\documentclass[12pt]{article} 
\usepackage[hyperfootnotes=false]{hyperref}
\usepackage{epsfig}
\usepackage{amsmath}
\usepackage{amssymb}
\usepackage{graphicx}
\usepackage[dvipsnames]{xcolor}
\definecolor{db}{rgb}{0,0,0.7}
\definecolor{dg}{rgb}{0,0.65,0}
\definecolor{pi}{rgb}{1,0.2,0.2}
\usepackage{slashed}
\usepackage{float}
  \newlength{\abstractwidth}
\usepackage{comment}
 \usepackage{mciteplus} 
\usepackage{subcaption}
 \usepackage{bbold}

  \newcommand{\be}{\begin{equation}}
  \newcommand{\bea}{\begin{eqnarray}}
  \newcommand{\eea}{\end{eqnarray}}
  \newcommand{\beq}{\begin{equation}}
  \newcommand{\ee}{\end{equation}}
  \newcommand{\eeq}{\end{equation}}

  \newcommand{\half}{{1\over 2}}

  \newcommand{\tl}{\hat{t}}

  	\newcommand{\slt}{SL(2) }
	\newcommand{\sltg}{SL(2)$_g$ }
  	\newcommand{\ads}{AdS$_2$ }
	\newcommand{\adsn}{AdS$_2$}
	\newcommand{\ncft}{NCFT$_1$ }
	\newcommand{\ncftn}{NCFT$_1$}
  	\newcommand{\q}{\mathfrak{q}}
	\newcommand{\eqn}[1]{\begin{equation}\begin{split} #1 \end{split}\end{equation}}
	
	\newcommand{\lp}{\left (}
	\newcommand{\rp}{\right )}

	\newcommand{\pa}[2]{\frac{\partial #1}{\partial #2}}
	\newcommand{\pd}{\partial}
	
	\newcommand{\inv}{^{-1}}
	\newcommand{\tr}{\text{tr} \,}
	
	\newcommand{\hf}{\frac{1}{2}}

	\newcommand{\lb}{\left [}
	\newcommand{\rb}{\right ]}
	\newcommand{\bra}[1]{\left \langle #1 \right|}
	\newcommand{\ket}[1]{\left| #1 \right \rangle}
	
	\newcommand{\ev}[1]{\left \langle #1 \right \rangle}

\def\la{\label}

\def\32{{3 \over 2 } }

  \def\ba{\begin{eqnarray}}
  \def\ea{\end{eqnarray}}

 \def\simleq{\; \raise0.3ex\hbox{$<$\kern-0.75em
      \raise-1.1ex\hbox{$\sim$}}\; }
 \def\simgeq{\; \raise0.3ex\hbox{$>$\kern-0.75em
	\raise-1.1ex\hbox{$\sim$}}\; }

\def\nref#1{(\ref{#1})}

\begin{document}

\begin{titlepage}
 % \rightline{}
  \bigskip

  \bigskip\bigskip

  \bigskip

\begin{center}
 
\centerline
{\Large \bf {Symmetries Near the Horizon }}
 \bigskip

 \bigskip
%\centerline
{\Large \bf { }} 
    \bigskip
\bigskip
\end{center}

  \begin{center}

 \bf {Henry W. Lin$^1$, Juan Maldacena$^2$, and Ying Zhao$^2$ }
  \bigskip \rm
  
\bigskip
 $^1$Jadwin Hall, Princeton University, Princeton, NJ 08540, USA\\

 \rm 
 \bigskip
 $^2$Institute for Advanced Study,  Princeton, NJ 08540, USA  
\rm
 \bigskip

 % \bf {Write authors  }
  \bigskip \rm
\bigskip
 
 %   Institute for Advanced Study,  Princeton, NJ 08540, USA  \\
\rm

\bigskip
\bigskip

% \vspace{2cm}
  \end{center}

 \bigskip\bigskip
  \begin{abstract}
We consider a nearly-AdS$_2$ gravity theory on the two-sided wormhole geometry. 	  We construct three gauge-invariant operators in N\ads which move bulk matter relative to the dynamical boundaries. In a two-sided system, these operators satisfy an  \slt algebra (up to non perturbative corrections). In a semiclassical limit, these generators act like \slt transformations of the boundary time, or conformal symmetries of the two sided boundary theory. These can be used to define an operator-state mapping. 
	A particular large $N$ and low temperature limit of the SYK model has precisely the same structure, and this construction of the exact generators also  applies. We also discuss approximate, but 
	simpler,  constructions of the generators in the SYK model. These are closely related to the ``size'' operator and are connected to the maximal chaos behavior captured by out of time order correlators. 
	
	%We consider a nearly-AdS$_2$ gravity theory on the two-sided wormhole geometry. We construct three gauge-invariant operators in N\ads which move bulk matter relative to the dynamical boundaries. In a two-sided system, these operators satisfy an  \slt algebra (up to non perturbative corrections). In a semiclassical limit, these generators act like \slt transformations of the boundary time, or conformal symmetries of the two sided boundary theory. These can be used to define an operator-state mapping. 
	%A particular large $N$ and low temperature limit of the SYK model has precisely the same structure, and this construction of the exact generators also  applies. We also discuss approximate, but simpler, constructions of the generators in the SYK model. These are closely related to the ``size'' operator and are connected to the maximal Lyapunov exponent for the 4-pt functions. 
	  
 \medskip
  \noindent
  \end{abstract}
\bigskip \bigskip \bigskip

\vspace{1cm}

\vspace{2cm}

  \end{titlepage}

  %  \starttext \baselineskip=17.63pt \setcounter{footnote}{0}
   \tableofcontents

 % \sc
   %\section*{STUFF TO DO}
   %\begin{enumerate}
%	   \item check all formulas that need to be checked
%	   \item finish conclusion
%	   \item size section needs to be streamlined
	   %\item Finalize appendices. Straighten out some conventions.
 %  \end{enumerate}

\section{Introduction and motivation} 

Any black hole with finite temperature has a near horizon geometry that can be approximated by flat space. The boost symmetry of this flat space region  corresponds to the full modular Hamiltonian of the outside region of the black hole, and it is an exact symmetry of the full wormhole geometry. The two translation symmetries of this flat space region are 
 more mysterious. It is important to understand them because they can take matter into the black hole interior.  In this paper,  we construct explicitly these symmetries for nearly AdS$_2$ gravity and also for the related SYK model.

Nearly \ads (N\adsn) gravity \cite{Almheiri:2014cka,Jensen:2016pah,Maldacena:2016upp,Engelsoy:2016xyb} captures 
the gravitational dynamics of near extremal black holes after a Kaluza-Klein reduction. The important gravitational mode is non-propagating and can be viewed as living at the boundaries of the nearly 
AdS$_2$ region. The action of these boundary modes is universal and can be written in terms of  a Schwarzian action for
a variable that can be viewed as a map from the boundary proper time to a time coordinate in a rigid AdS$_2$ spacetime. 
A similar mode appears in the description of Nearly CFT$_1$ (\ncftn) quantum systems, such as the SYK model 
\cite{Sachdev:1992fk,KitaevTalks,Kitaev:2017awl,Maldacena:2016hyu}, which exhibit nearly conformally-invariant correlation functions at relatively low energies.

In these systems, there is  an approximation where matter appears to move in a rigid AdS$_2$ background geometry displaying an $\widetilde{\mathrm{SL}}(2,\mathbb{R})$ isometry group\footnote{In Euclidean signature, the isometry group is PSL$(2,\mathbb{R})$. In this paper, we are mostly concerned about the algebra and not the group.}, henceforth denoted by \slt\!\!.  This approximation becomes 
 better and better as the boundaries are further and further away. However, this does not obviously translate into a physical symmetry, since only the relative position between the boundaries and the bulk matter is physical. 
 Nevertheless, we will find three \slt generators that act on the full physical 
 Hilbert space of the system. These generators obey an exact \slt algebra, but they do not commute with the Hamiltonian. 
However,  they have a relatively simple behavior under Hamiltonian evolution, which can be used to define 
 ``conserved'' charges through a more subtle construction. 
 
 It is convenient to describe the N\adsn/\ncft system in terms of an extended Hilbert space with a gauge constraint. 
 The extended Hilbert space factorizes into three pieces: two systems describing the two boundaries of the eternal black hole and a system describing the bulk matter fields. Each of them can be viewed in terms of particles moving on an exact AdS$_2$ spacetime \cite{Kitaev:2018wpr,Yang:2018gdb}. The physical Hilbert space is obtained by imposing an \sltg gauge
  constraint that sets to zero 
 the overall \sltg charges of the three systems. 
 This constraint imposes that only the relative position between the two boundaries, or between the boundary and the bulk matter, are physical. 
 In this paper we discuss physical \slt generators which are invariant under the \sltg gauge symmetry. 
 It is important not to confuse these two \slt groups, the gauge one and the physical one. This paper is about they physical one. Our construction will define these physical 
 generators relative to boundary positions in such a way that they 
 are invariant under the gauge symmetries. 
 Due to the fact that they involve the boundary positions, they are not conserved under time evolution, since
 the boundary positions change in time. However, the dynamics of the boundary positions is integrable 
 \cite{Bagrets:2016cdf,Bagrets:2017pwq,Stanford:2017thb,Mertens:2017mtv,Kitaev:2018wpr,Yang:2018gdb}, and one could use this fact  to define ``conserved'' charges by simply ``undoing'' the boundary evolution.

Our discussion is exact in a scaling limit where we go to low temperatures, but we scale up the size of the black hole
so that we keep fixed the coupling of the Schwarzian mode, or the quantum gravitational effects in AdS$_2$. 
This is a limit where the near extremal entropy $\Delta S = S -S_e$ is kept fixed\footnote{For a four dimensional charged near extremal black hole this is $\Delta S \propto r_e^3 T/l_p^2$, where $r_e$ is the extremal radius. We take $\Delta S$  fixed with $r_s \to \infty$, $T\to 0$.}.  
  In SYK variables this is the limit $N\to \infty$, $\beta J \to \infty$ with $N/(\beta J)$ fixed. We have not 
  included finite $\beta J$ effects or finite $N$ effects. 
  The generators we define involve variables,  such as the distance between the two boundaries, which are well defined in the gravity theory, in the scaling limit we define, but are not expected to make sense when non-perturbative 
  effects are taken into account. In particular, they are not expected to make sense for finite $N$ in the SYK model. This should not be surprising since unitary \slt representations are infinite dimensional.
  %In fact, our generators imply that there is an infinite number of states for a wormhole geometry, since \slt representations have an infinite number of states. This is clearly incompatible with a finite $N$. 
  %For this reason we expect that when the extrema$S_0$ is finite, there should be corrections.  or finite
  %entropy situation. 
  However, we also relate the generators we defined to other operators which are well defined for finite $N$, but   
  agree with the generators in the semiclassical limit. 
  This allows us to identify operators in both a gravity theory and the SYK model which approximately obey an
  \slt algebra and should be identified with the symmetries of AdS$_2$. These approximate symmetries behave as 
  $SL(2)_u$ transformations of the physical boundary time of  a pair of  \ncftn%
  s, and we give an approximate state-operator map that organizes the \ncft Hilbert space into primaries and descendants, in analogy with higher dimensions. 
  
  These generators are connected to the operators that generate traversable wormholes \cite{Gao:2016bin}. These move matter from one side of the horizon to the other. In fact, the approximate expression for the global time translation operator is essentially the same as the coupled Hamiltonian in \cite{Maldacena:2018lmt}. 
  These approximate generators also make contact with another approach for describing bulk motion via the 
  ``size'' operator in  \cite{Susskind:2018tei,Brown:2018kvn,Qi:2018bje}.  So the discussion in this paper explains why such operators 
  act like approximate \slt isometries in the bulk. 
  %As an application of the approximate \slt generators, we give a simple derivation of the maximal Lypanuov exponent.
  We also point out that the structure of the approximate generators is similar
  to the structure of the exact generators in the case of higher dimensional conformal field theories in Rindler space.

	{\bf Outline}. 
%	\HL{rewrite}		
%	WRITE OUTLINE ONCE WE DECIDE ON THE STRUCTURE OF THE PAPER...
	In section {\bf two}, we review nearly AdS$_2$ gravity at low energies in the embedding space formalism. The Hilbert space of the system consists of two boundary modes plus arbitrary matter, with an overall SL(2) gauge constraint. We briefly review how this structure also emerges in the SYK model. 

	In section {\bf three}, we construct the generators which satisfy an exact SL(2) algebra. Although the quadratic Casimir commutes with the usual Hamiltonians $H_l $ or $H_r$, the individual generators do not. We nevertheless explain how to obtain conserved charges.

	In section {\bf four}, we consider the charges in the semi-classical limit. In this limit, the generators can be viewed as conformal symmetries of the boundary time. We show that the Hilbert space organizes into primaries and descendants and give a state-operator correspondence analogous to the higher dimensional versions.

	In section {\bf five}, we show how to use these charges to explore bulk physics. We comment on drama near the inner horizon. We also discuss applications of our construction to previous work. Our charges are closely related to the coupled Hamiltonian in  \cite{Maldacena:2018lmt}. 
	%Another consequence of our construction is that it provides a more algebraic derivation of the maximal exponential growth of the OTOC. %Our construction is also closely related to chaos.

	In section {\bf six}, we explain how these approximate charges can be realized in the SYK model.  We also relate
	 our charges to ``size'' in SYK \cite{Roberts:2018mnp, Brown:2018kvn, Qi:2018bje}.
	We note that the generators written in terms of the microscopic variables has an analogous form in higher dimensional CFTs.

	In section {\bf seven}, we discuss some issues and draw conclusions.

	In the {\bf appendices}, we explain how to construct gauge-invariant SO(3) generators in a rather pedestrian system involving two non-relativistic particles on a sphere, plus some arbitrary matter, with an overall angular momentum gauge constraint. We also explain how to compute commutators in both the canonically quantized Schwarzian theory and its linearized cousin. We also discuss an alternative to the embedding space formalism which uses \slt spinors instead of the vectors. Finally, we comment on a modified eternal traversable wormhole where the oscillation frequency of the Schwarzian mode is very large. % extremely% heavy.

%	In section two, we first discuss how given two points in \ads, we may geometrically construct three preferred Killing vectors, which correspond to time translation, spatial translation, and Lorentz boost. We then realize these symmetries in a simple model of 2-dimensional quantum gravity, the Jackiw-Teitelboim (JT) model with arbitrary bulk matter. 

	%In section three, we analyze the action of these charges on states. We show that in the semi-classical limit, these charges act on states in a way that is highly analogous to the state-operator correspondence in higher dimensions. This allows us to organize states in the physical Hilbert space in terms of primaries and descendents.

	%In section four, we discuss how these generators might be explicitly constructed in a dual holographic theory. This involves translating the expressions for the charges into matter correlation functions. We show how this can be done explicitly in the SYK model.

	{\bf Notation}.
	In most of this paper we work in units, where in the SYK language $\alpha_s N /{\cal J } =1$, or in 4d near extremal 
	charge black hole language, ${ r_e^3 \over G_N} =1$, where $r_e$ is the extremal radius. Such factors can be restored by dimensional analysis. 
	
	%In SYK language, we work in units where ${\mathcal J}/N = 1$. One can convert to SYK units by dimensional analysis. For example, the statement $\beta \ll 1$ should be translated in SYK units to $\beta \mathcal J \ll N$.

	%we construct the charges in terms of the microscopic variables of the SYK model.
%	 Since the Hilbert space is finite, and all non-trivial unitary representations of \slt are infinite, these charges can at best form an approximate \slt algebra at finite $N$. 
	
%	In section four, we compute the action of these charges on low-energy states. Here our analysis applies to both SYK and gravity. The charges act on primary and descendant states in a way that is highly analogous to higher dimensional CFTs.
	
	%In section five, we discuss the connection of our construction to chaos and make some additional comments.

%\section{Charges in Nearly $AdS_2$ gravity }
  
 % \la{SecAdST}
 
 \section{Review}
   
  \subsection{Review of the symmetries of $AdS_2$}
  \la{ReviewAdS}
  
\begin{figure}[ht]
\begin{center}
\includegraphics[scale=1]{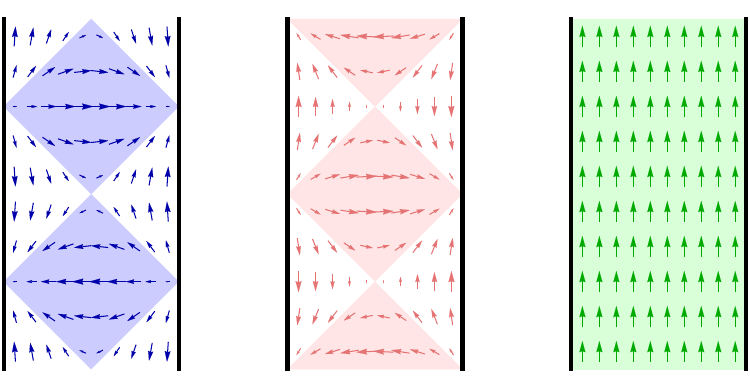}
\caption{ The three Killing vectors: {boost \color{db} $B$ (blue)}, {momentum \color{pi} $P$ (pink)}, and {\color{dg} global energy $E$ (green)} given in \nref{Engen} in the coordinate system \nref{Coord}. The shaded regions delineate different orbits of the symmetries.  }
\label{SymmetriesCoordinates}
\end{center}
\end{figure}

In the embedding space formalism, \ads is the universal cover of the surface defined by
\eqn{Y . Y = \eta_{ab} Y^a Y^b =   -(Y^{-1}) ^2 - (Y^0)^2 + (Y^1)^2 =-1.\la{embedads}} 
From this definition, it is clear that \ads has an $SO(2,1) \simeq SL(2,R)$ symmetry generated by
\eqn{Q^a = \half \epsilon^{abc} J_{bc}
~,~~~~~~~~~~~
% \\
J_{ab} = - i Y_a \pa{}{Y^b} + iY_b \pa{}{Y^a}.}
These generators satisfy the algebra 
\eqn{[Q^a, Q^b] = i \epsilon^{abc} \eta_{cd} Q^d.}
%The bifurcation point of \ads is at $Y_0 = Y_1=0$ and $Y_{-1} = 1$. Near this point, the three generators act as a Lorentz boost $B \approx i Y_0 \pd_1 - i Y_1 \pd_0$, momentum $P \approx i \pd_1$, and energy $E = i \pd_0$. Hence,
%
%\eqn{
%	J_{0,1} &\approx 0\\
%	J_{-1,1} &\approx i \pd_1\\
%	J_{-1,0} &\approx i \pd_0 
%}
To see how these generators act on \ads more explicitly, we can solve the constraint \ref{embedads} using global coordinates:
\eqn{Y = (Y^{-1},Y^{0},Y^{1}) = \lp {\cos T \over \sin \sigma},{\sin T \over \sin \sigma}, {-1 \over \tan \sigma}\rp, \quad \sigma \in [0,\pi]. }
Then, the Killing vectors
\eqn{	
B &= Q^{-1} = J_{0,1} = i \lp  - \cos T \cos \sigma \pd_T + \sin T \sin \sigma\pd_\sigma\rp,\\
P &= Q^0 = -J_{-1,1} = -i \lp \sin T \cos \sigma \pd_T + \cos T \sin \sigma \pd_\sigma\rp, \\
E &= Q^1 =J_{-1,0} = i \pd_T. \la{Engen}
}
Note that $P = -i [ B, E]$. 
Near the bifurcation point $T=0, \sigma = \pi/2$ these symmetries act as time translation/energy $E$, spatial translation/momentum $P$, and boost $B$. See figure \ref{SymmetriesCoordinates}. Of course, the algebra is SL(2),  
not Poincare, so that $[E,P]=iB$. 

We can choose coordinate systems \footnote{The second coordinates describe a Friedmann-Lemaitre-Robertson-Walker (FLRW) cosmology.} that simplify the action of these generators
\begin{subequations}
	\begin{align}		
	&{\rm Rindler: } &ds^2 &= { - dt^2 \sinh^2 \rho + d\rho^2 },&\rho &\in [-\infty,\infty], & B &= i \partial_t
\cr 
	&{\rm FLRW: } &ds^2 &= { - d\tau^2 + \sin^2 \tau dx^2 },  &\tau &\in [ 0 , \pi ], & P &= - i \partial_x 
\cr 
\la{Coord}
	&{\rm Global: } &ds^2 &= { - dT^2 + d\sigma^2 \over \sin^2 \sigma }, &\sigma &\in [ 0 , \pi ], & E &= i \partial_T 
\end{align}
\end{subequations}

%The time translations in various coordinate systems correspond to various combinations of these charges
%
%These three generators $Q^a$ (in some basis) also correspond to the time translation generators in three different coordinate systems:
%\bea\la{Coord}
%  & {\rm Rindler: } ~~~ds^2 = { - dt^2 \sinh^2 \rho + d\rho^2 } ~,~~~~~~~~~~~~&B = i \partial_t
%\\
%  &{\rm FLRW:} ~~~~~~ds^2 = { - d\tau^2 + \sin^2 \tau dx^2 } ~,~~~~\tau\in [ 0 , \pi ]~,~~~~~~ &P = - i \partial_x 
%\\
%  &{\rm Global:} ~~~~ds^2 = { - dT^2 + d\sigma^2 \over \sin^2 \sigma } ~,~~~~~~~~~~\sigma \in [ 0 , \pi ]  ~,~~~~~~&E = i \partial_T ~
%\\
%&{\rm Poincare:} ~~ds^2 = { - dt_P^2 + dz^2 \over z^2 } ~,~~~~~~~~~~~~~~
%E +B = i \partial_{t_P} 
%\eea

Notice that, given a vector $W^a$, we can assign a charge $Q_W = W_a Q^a$. This charge
has the property that it leaves the bulk point $Y^a \propto W^a$ fixed\footnote{  If the vector $W^a$ is 
spacelike, then we will not have any fixed point in $AdS_2$. An example is the generator $E$ in \nref{Engen}, 
see figure \ref{SymmetriesCoordinates}. }. This is basically just the familiar fact that a rotation about some axis fixes the axis.

Points at the boundary are naturally described in terms of projective coordinates, 
%If a particle approaches the conformal boundary of \ads, $Y$ becomes nearly null in embedding space. 
%In such situations, it is often convenient to describe the trajectory of the boundaries using projective coordinates
 $\tilde X^a$ with the constraint $\tilde X . \tilde X =0$ and the 
 identification $  \tilde X^a \sim \lambda \tilde X^a$. If we have a charge associated 
 to the vector $W^a$, $Q_W = W.Q$, 
  then this charge will leave invariant the boundary points that are light-like 
  separated from $W^a$, $W.\tilde X =0$. 

A particle moving in \ads can be described by a trajectory $Y^a(u)$ constrained to live on the surface $Y . Y = -1$. Since $Y$ is a vector, $[Q^a,Y^b] = i \epsilon^{abc} Y_c$. 
For a standard massive particle, of mass $m$, the charges are given by 
\be
Q^a = m \epsilon^{a}_{~bc} Y^b \dot Y^c ~,~~~~~~~~Y.Y=-1 ~,~~~~~ \dot Y . \dot Y =-1
\ee
If the particle is also charged under an electric field that is uniform in $AdS_2$, then the 
charges are 
\be \la{QYch}
Q^a =  m \epsilon^{a}_{~bc} Y^b \dot Y^c  - \q Y^a ~,~~~~~~~~Y.Y=-1 ~,~~~~~ \dot Y . \dot Y =-1
\ee
The charges $Q^a$ are conserved, and the particle trajectories are given by 
 $Q.Y = \q $.
  
Alternatively we can say that if we have a particle moving in $AdS_2$, 
 % a relativistic particle in embedding space, first ignoring the constraint \ref{embedads}. 
 its Hilbert space has operators satsifying
\eqn{
[Y^a,Y^b] &= 0,\\
[Q^a,Q^b] &= i \epsilon^{abc} \eta_{cd} Q^d,\\
[Q^a,Y^b] &= i \epsilon^{abc} \eta_{cd} Y^d.\la{poincare}
}
This is the Poincare algebra $R^{2,1} \rtimes SL(2,R)$. The Casimirs of this algebra are %$SO(2,1) \simeq SL(2,R)$. The Casimirs of this algebra are
\eqn{r^2= Y . Y, \quad \q = Y . Q.}
The values of these Casimirs are inputs of the physical theory. For example, for a spin-less particle freely propagating in \ads we have $r^2 = -1, \q =0$.
From this point of view $\q$  is the spin of the particle. 
 
%Points at the boundary are naturally described in terms of projective coordinates, 
%If a particle approaches the conformal boundary of \ads, $Y$ becomes nearly null in embedding space. 
%In such situations, it is often convenient to describe the trajectory of the boundaries using projective coordinates
% $\tilde X^a$ with the constraint $\tilde X . \tilde X =0$ and the 
% identification $  \tilde X^a \sim \lambda \tilde X^a$. 
%This is particularly relevant for this paper, since in the Schwarzian limit of Jackiw-Teitelboim gravity, the left and right dynamical boundaries of Nearly \ads (N\ads\!\!) approach the conformal boundaries, and we parameterize their positions using null vectors $X_l(u_l)$ and $X_r(u_r)$.

%These various coordinate times are related to $X$ via
%\bea  \la{BdyTimes}
%& ~&e^{ i T_r}  = X^{0} + i X^{-1} ~,~~~~{\rm for } ~~~X^1 =1 ~,~~~~~e^{ i T_l} = X^{0} + i X^{-1} ~,~~~~{\rm for } ~~~ X^1=-1
% \cr
%& ~&   { X^{-1} \over X^{0} + X^1 } = t_P ~,~~~~~~~~~~~~ e^{ t} = X^{1} + X^{-1} ~,~~~ {\rm for} ~~X^{0}=1
%  \eea
%We will mostly use the boundary Rindler time $t$ in this paper.
%It is often convenient to use null coordinates $Y^\pm  = Y^1 \pm Y^{-1}$. In these coordinates, the embedding space metric is 
%\eqn{ V. W = \hf \lp V^+ W^- + V^- W^+\rp - V^{0} W^{0}. }
%For the left boundary, $X = (X^+, X^-, X^0) \propto (e^t, e^{-t}, 1)$, so the symmetries act on the boundary Rindler time via 
%\eqn{Q_n: \, t(u) \to t(u) + \epsilon e^{n t}. }
%Continuing to Euclidean time $t = i\tau$, these are just the symmetries of a CFT$_1$ on the thermal circle.

If we have quantum fields moving on $AdS_2$ the charges can be written in terms of the stress tensor and the associated Killing vector 
\be \la{CharKil}
Q_\zeta = \int_{\Sigma}  n^\mu T_{\mu \nu } \zeta^\nu 
\ee
where $\zeta^\mu$ are each of the Killing vectors in \nref{Engen}. 
These charges are constant and independent of the spatial slice $\Sigma$  used to evaluate them, if the fields obey appropriate reflecting conditions at the AdS$_2$ boundary.

\subsection{Review of the nearly-$AdS_2$ gravity theory} 
\la{ReviewJT}

We will be considering the JT theory coupled to matter as 
\cite{Jackiw:1984je,Teitelboim:1983ux,Henneaux:1985nw,LouisMartinez:1993eh}
\be \la{JTAct}
S = \phi_0 \left[ \int R - 2 \int K \right]  + \int \phi( R+2)  - 2 \phi_b \int K + S_m[ g_{\mu \nu},\chi]
\ee
where we have also indicated the boundary terms. 
 The first term is topological and only contributes to the extremal entropy. 
We have also assumed that the matter couples to the metric but not to $\phi$. 
We will also assume that the boundary is very far away so that matter effectively feels as if it was in exactly $AdS_2$ space.   This is sometimes called the ``Schwarzian" limit because 
in this case the boundary dynamics is governed by  \cite{Jensen:2016pah,Maldacena:2016upp,Engelsoy:2016xyb}
\be \la{SchA}
S_{\rm Sch}[t] = -      \int du \, \{ e^{t(u)} , u \} ~,~~~~~~~~\{ f, u \} = 
	{ f''' \over f'} - { 3 \over 2} { {f''}^2 \over {f'}^2}  ~,~~~~d\tau_p = 2 \phi_b \, du
	\ee
	where $u$ is a rescaled version of proper time $\tau_p$,   and 
	$t$ can be viewed as the Rindler time $t$ in \nref{Coord} near the boundary. 
	We can view the curve $t(u)$ as parametrizing the position of the boundary. The action \nref{SchA} captures
	a gravitational degree of freedom that we can view as living on the boundary\footnote{
	 This should not be confused with a possible holographically dual boundary quantum mechanical theory, which would 
	 describe the full system.}.

We will consider spacetimes describing a two sided eternal 
black hole, so that we have two boundaries and two variables $t_r$, $t_l$, each with the action \nref{SchA}. 
The dynamics of the full system \nref{JTAct} reduces to the dynamics of three decoupled systems connected 
only by an overall \sltg constraint
 \be  \la{ActThree}
	 S = S_{\rm Sch}[t_r] + S_{\rm Sch}[t_l] + S_m[g_{\mu \nu} , \chi] 
	 \ee
 These three decoupled systems are the following. 
First we have the matter which lives in exactly $AdS_2$ space and has \sltg charges $Q^a_m$. Then we have the right and left boundaries. In this limit,  they are not 
directly coupled to each other or to the matter. 
 However, in N\ads gravity, an overall \sltg 
 transformation is a redundancy of our description.
  Hence the physical Hilbert space is \cite{Jensen:2016pah,Maldacena:2016upp,Engelsoy:2016xyb}
 \be
{\cal H}_{\rm Physical} = ( {\cal H}_l \times {\cal H}_{matter} \times {\cal H}_r)/SL(2)_g ~,~~~~~~~~~~ Q^a_l + Q^a_m + Q_r^a =0  \la{Constr}
 \ee
 where the charges $Q^a$ are the \sltg charges of each of the systems. 
Physically, this says that only the relative positions of the matter and the boundaries
matter. As pointed out in \cite{Yang:2018gdb}, we can view it as ``Mach's" principle, where the boundaries are the ``distant stars''. These are part of the usual constraints of general relativity. 

One can find explicit expressions for the \sltg charges of the right and left boundaries
by using the Noether procedure on \nref{SchA} \cite{Maldacena:2016upp}
\def\tr{{t_r}}
\def\tl{{t_l}}
\bea
  Q^{-1}_r & = & { \tr''' \over \tr'^2 }  - { \tr''^2   \over \tr'^3 } - \tr'
  \cr 
   Q^+_r & = & e^{  \tr} \left[ { \tr''' \over \tr'^2 } - { \tr''^2 \over \tr'^3 } - { \tr'' \over \tr'}  \right]\cr
  Q^-_r & = &  e^{ - \tr} \left[ -{ \tr''' \over \tr'^2 } + { \tr''^2 \over \tr'^3 } - { \tr'' \over \tr'}  \right]
 .\la{qs}
  \eea
where $Q^\pm = Q^0 \pm Q^1$. 
  The left side charges may also be obtained by analytic continuation $Q_l \to - Q_r$ with
   $\tl = -\tr + i \pi$, $u_l \to - u_r + i$(constant). We are defining $u_l$ and $t_l$ so that they go 
   forwards in time
   in the thermofield double interpretation. 
   \bea  
  Q^{-1}_l & = & -{ \tl''' \over \tl'^2 }  + { \tl''^2   \over \tl'^3 } + \tl'\cr 
  Q^+_l & = & e^{-  \tl} \left[ { \tl''' \over \tl'^2 } - { \tl''^2 \over \tl'^3 } + { \tl'' \over \tl'}  \right] %\sim   e^{  - u} ( \epsilon''_l + \epsilon'''_l ) 
  \cr
  Q_l^- & = & e^{  \tl} \left[ -{ \tl''' \over \tl'^2 } + { \tl''^2 \over \tl'^3 } + { \tl'' \over \tl'}  \right] %\sim  e^{ u} ( - \epsilon''_l + \epsilon'''_l ) 
  %\sim 1 + \epsilon'_l - \epsilon'''_l 
 \la{qsl} \eea 
%  I CHANGED THE LEFT CHARGES, CHECK. 
  One can check  that $ \half Q_r . Q_r = \{e^{-\tr}, u\} $. 
  %, where $E$ is the energy associated to the   action \nref{SchA}.
  %, where $\{f,u\}$ is the Schwarzian. %,  X_r. X_r =0.$

 For our subsequent discussion it is convenient to write a nicer expression for the 
 boundary position so that its \slt transformations properties are more manifest. 
   The dynamics of the   boundary is 
closely related to the dynamics of a charged massive particle, or a particle with spin, 
in the limit that the mass and the charge (or spin) becomes both very large, while 
keeping the total \slt charges $Q^a$ finite \cite{Kitaev:2018wpr,Yang:2018gdb}. 
We have \nref{poincare} with $\q = 2 \phi_b$. 
In this case, the coordinates $Y^a$ become very large because we approach the boundary. 
So it is convenient to define rescaled coordinates $X^a$ via 
\eqn{X^a_r = {Y_r^a\over Y_r . Q_r} = {Y_r^a \over \q} ~,~~~~~~~X_r.Q_r=1~,~~~~~~ \quad ~~~~X_r^2 \to 0 ~ }
So, from the point of view of \nref{poincare} we have $r^2 = 0$, $\q_x = 1$ for the variable $X_r^a$.
For $X_l$ we get  $ \q_x=-1$.  
We can also rescale proper time by the same factor so that now we obey $\dot X_r .\dot X_r =-1$. 
In terms of our previous variables these can be written as 
\eqn{X_r &= \lp X^{-1} ,X^+,X^- \rp =  \lp {1 \over \tr'},{e^{\tr} \over \tr'},-{e^{-\tr} \over \tr'} \rp, ~~~~~~~~ X^\pm \equiv X^0 \pm X^1 \la{Xfromt}
	\cr
	X_l &=  \lp  {1 \over \tl'},- {e^{-\tl} \over \tl'},{e^{\tl} \over \tl'} \rp.
	}
	We can check that $-X_l . Q_l = X_r . Q_r = 1$.  
	In Appendix \ref{appendixCan}, we verify that in the canonically quantized Schwarzian theory, the above operators satisfy the Poincare algebra \nref{poincare} with the appropriate Casimirs $r^2 =0$, $\q_x = \pm 1$. 
	(In appendix \ref{Spinor} we give an alternative description in terms of spinors.) 
	In addition, the Hamiltonian corresponding to the Schwarzian action \nref{SchA} is 
	\be
	2H_r = -  Q_r^2 =  \ddot{X}_r^2. \la{HamQsq} 
	\ee
		Using \nref{poincare} this gives us the quantum mechanical relation
	\be 
	\dot X_r^a = i [H_r,X_r^a] = - { i \over 2} [ Q_r^2 , X^a_r] = \half  \epsilon^{a}_{~bc}  (X_r^b Q_r^c + Q_r^b X_r^c) \la{dotX}.
	\ee
	Taking a second derivative we get the operator equations
\bea \la{ChargXr}
Q_r^a &= \ddot X_r^a -  H_r X_r^a - X^a_r H_r  ~,~~~~~X_r . Q_r=1 ~,~~~~~\dot X_r^2 =-1,
 \\ 
 Q_l^a &= - \ddot X_l^a +  H_l X_l^a + X^a_l H_l  ~,~~~~~X_l.  Q_l=-1 ~,~~~~~\dot X_l^2 =-1 ~,~~~~\la{ChargXl} \eea
 where the difference in signs is due to the difference in sign of $\q$ for the left boundary. We may also use the algebra to compute commutators between $X, \dot X, \ddot{X}$. For example,
\be
[ \dot X_r^a ,  X_r^b] = - i X_r^a X_r^b   \la{dotXX}
\ee

Using these coordinates it is also possible to write the correlation functions of operators dual to 
matter fields in the bulk. If we have a massive field in the bulk giving rise to an operator of dimension
$\Delta$, then its left right correlator is given by 
\be \la{CorrLR}
\langle O(u_l) O(u_r) \rangle \propto { 1 \over (- 2 X_l(u_l).X_r(u_r) )^{\Delta } } =\left( { t_l'(u_l) t_r'(u_r) 
\over 4 \cosh^2({ t_l(u_l) + t_r(u_r) \over 2 } ) } \right)^\Delta 
\ee
where we used \nref{Xfromt}. 
We get a similar formula for  correlators on the same side.

\subsection{Review of SYK } 
\la{RevSYK}

The SYK model contains $N$ Majorana fermions with random interactions affecting $q$ fermions at a time, $q=4,6,\cdots$
\cite{Sachdev:1992fk,KitaevTalks,Kitaev:2017awl}. 
In the large $N$ limit, one can write down an effective action in terms of a bilocal field $G(u_1,u_2)$, which becomes
equal to the average two point function once we impose the equations of motion. 
At low energies this action becomes nearly reparameterization-invariant, except for a low action reparametrization mode (or soft mode),  which 
has a Schwarzian action with  an  overall coefficient scaling as $N/J$, with $J$ an energy scale setting the strength of the interactions of the original model.   

In more detail,  we start with a scaling solution 
\be \la{Unpe}
G_0(t_1,t_2) \propto |t_1 -t_2|^{-2\Delta}.
\ee
The soft mode corresponds to functions $G$ obtained by a reparametrization of \nref{Unpe}, 
$G =  [f'(u_1) f'(u_2)]^{\Delta} G(f(u_1) , f(u_2) ) $. 
We can also generate new configurations by 
  having fluctuations $\delta_\perp G(t_1,t_2)$ which lie in the directions orthogonal to the soft mode. 
For now the coordinates $t_1$ and $t_2$ are some coordinates that appear in the solution of the low energy equations of
the SYK model and are defined by the form of the unperturbed solution \nref{Unpe}. 
We introduce the soft mode by writing the full physical $G$ as \cite{Kitaev:2017awl}
 \be \la{Gfull}
 G(u_1,u_2)  = [f'(u_1) f'(u_2)]^{\Delta}   G_0(f(u_1),f(u_2)) +  [f'(u_1) f'(u_2)]^{\Delta} \delta_\perp G(f(u_1), f(u_2) )  
\ee
This can be viewed as   parametrization the full space of functions $G$. Namely, 
  we think of the 
integration variables as $f(u)$ and $\delta G_\perp (t_1,t_2)$.   
Inserting this expression into 
the SYK action, and taking a low energy limit,  we find that the full (Euclidean) action becomes 
\be \la{SYKActT}
S = S[ G_0(t_1,t_2) + \delta_\perp G(t_1,t_2) ]  - { N \alpha_S \over   {\cal J } } \int du \{ f, u \}
\ee
where we used   the approximate reparametrization invariance of the low energy action.  This means that 
the first term in \nref{SYKActT} is independent of $f$. All the dependence on $f$ is in the second term of \nref{SYKActT} and it comes from a small violation of the reparametrization symmetry \cite{Kitaev:2017awl}.  
To evaluate the path integral, one should sum over different $f$ and $\delta_\perp G$. An important point is that in this parametrization, we have an \sltg symmetry  
 \be \la{SLTg}
 f \to { a f + b \over c f + d} ~,~~~~~~~~~~ \delta_\perp G(t_1, t_2 ) \to { 1 \over \left[ ( a- c t_1)( a- c t_2)\right]^{2\Delta} } \delta_\perp G\left( { d t_1 - b \over -c t_1 + a } , { d t_2 - b \over -c t_2  + a } \right)
 \ee
 The arguments of $\delta_\perp G$ are transforming in the inverse way than $f$ so that the second term in 
 \nref{Gfull} remains invariant. The first term in \nref{Gfull} also remains invariant under this transformation. 
 Therefore \nref{SLTg} is a redundancy in our parametrization of the space of $G(u_1,u_2)$ \nref{Gfull} 
 and we should demand that
 everything is invariant. 
 Note that when we write the action as the sum of two terms such as in \nref{SYKActT} (or three terms if we wrote the 
 Lorentzian action for the thermofield double), then the \sltg symmetry will act in a non-trivial way on the variables of each term. In particular, the \sltg action transforms $\delta_\perp G(t_1,t_2)$, as in 
 \nref{SLTg}. So, even though the two terms of the action \nref{SYKActT} are decoupled, they become connected by the 
 total \sltg constraint.  
 
 The conclusion is that in the SYK model we have a structure which is similar to the one we had in nearly AdS$_2$ 
 gravity. We have three separate systems connected by an overall gauge constraint. The Schwarzian parts are identical 
 to what we had in gravity. But the analog of the matter action $S_m[g_{\mu \nu} , \chi]$ is the first term in 
 \nref{SYKActT}. It is independent of the Schwarzian variables, but its variables transform nontrivially 
 under $SL(2)_g$. 
%
 % The Hilbert space of the conformal fluctuations can be characterized in terms of representations of 
%$SL(2)$, but we do not have a very explicit form for it (see e.g. \cite{Gross:2017aos}). 
%Nevertheless, if we consider the thermofield double, we see that the SYK model has a structure similar to \nref{Constr}, where 
%${\cal H}_{\rm matter} $ can be viewed as the fluctuations $\delta_\perp G$ in the directions which are not reparametrizations. 
% In addition, we could also include insertions of single fermions, or other fluctuations that are not $O(N)$ invariant. 

\section{Exact generators } 

\subsection{Construction of gauge invariant \slt generators} 
\la{ExactGenCons}

In N\ads gravity, bulk matter ``feels'' as if it was moving in empty $AdS_2$. This suggests that
we could define $SL(2)$ generators that move the matter. Naively these would be $Q^a_m$. However,
these are not physical   because they are not invariant under the $SL(2)$ gauge symmetry. Said slightly differently, once we go from quantum field theory on a fixed background to quantum gravity, we must gravitationally dress all observables. Since the metric of NAdS$_2$ is essentially rigid, the dressing should involve the boundary degrees of freedom.

\begin{figure}[ht]
\begin{center}
\includegraphics[scale = 0.55]{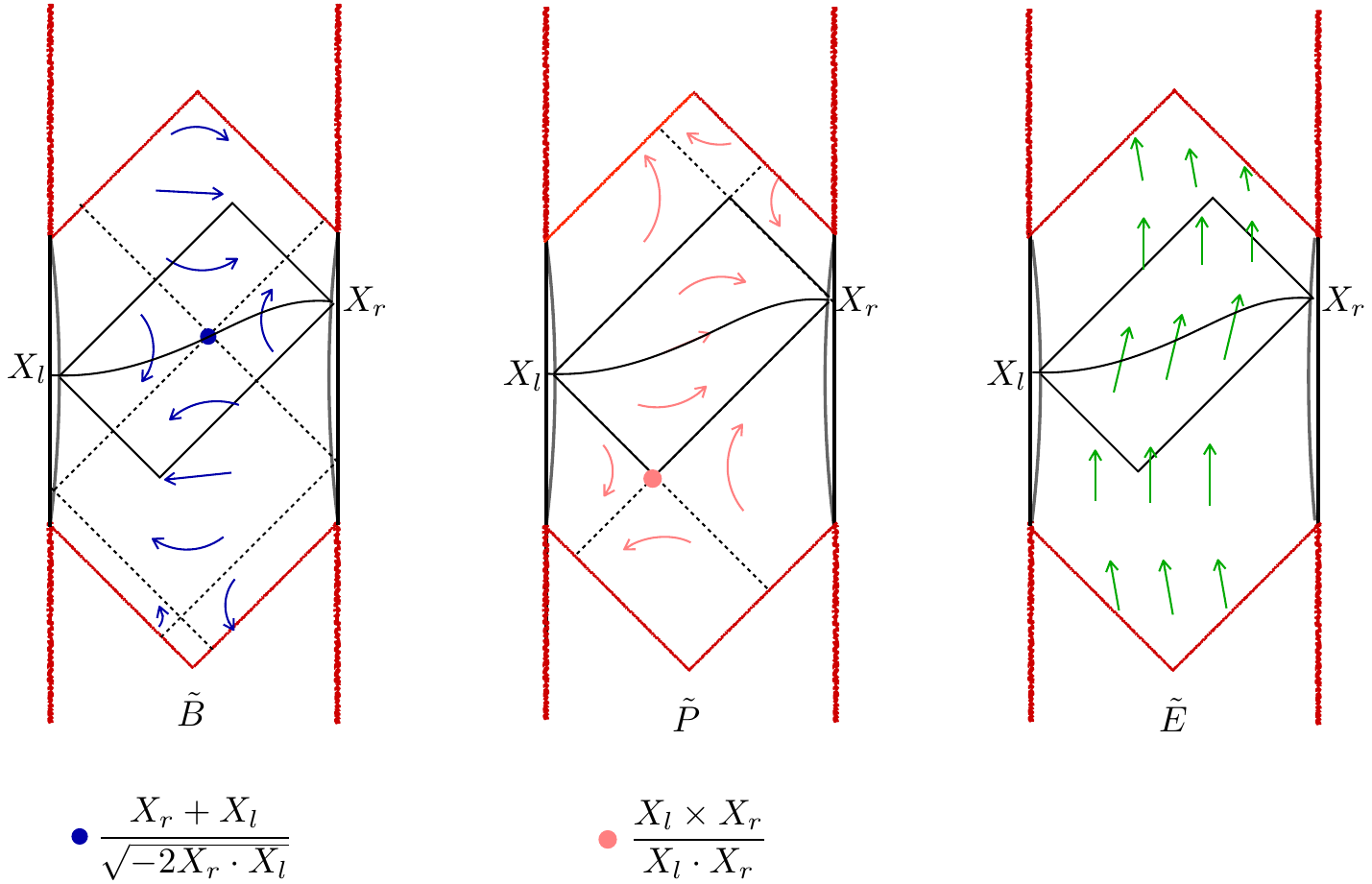}
\caption{Geometrical action of the gauge invariant charges $\tilde P, \tilde B, \tilde E$. The points which are fixed by the symmetry generators are also indicated.}
\label{SymmetriesWBoundary}
\end{center}
\end{figure}

%However, we can use the boundary positions to define invariant charges. 
For example, given two 
boundary positions $X_l^a$ and $X_r^a$ we can define the vector 
$W^a = \epsilon^{abc} X_{lb} X_{rc}$ and the generator 
\be \la{MomInv}
 G^0 ={\tilde P} = { \epsilon_{abc} Q_m^a X_{l}^b X_{r}^b  \over X_l . X_r } 
\ee
where we introduced two different notations for the generator. 
This generator leaves the boundary points $X_l$ and $X_r$ invariant. It is a translation in the bulk 
along the geodesic that joins these two boundary points, see figure \ref{SymmetriesWBoundary}. 
In addition, we have normalized it so that 
it generates translations by a ``unit'' proper distance in the bulk.
In the case that $X_l$ and $X_r$ correspond to the points with $T=0$ and $\sigma =0,\pi$ in 
\nref{Coord} get we the generator $P$ in \nref{Engen}. For general positions for $X_l$ and $X_r$ we 
get a linear combination of the generators in \nref{Engen}. A nice feature of \nref{MomInv} is that 
it is invariant under the $SL(2)$ gauge transformations.  
Another nice feature of \nref{MomInv} is the fact that it acts within the so called ``Wheeler-de-Witt''
patch, which is the set of points that are spacelike separated from both boundary points, $X_l$ and 
$X_r$, see figure \ref{SymmetriesWBoundary}. 

We can now wonder whether we can define two other generators in a similar way. Natural candidates are 
\be \la{OtherTwo}
G^1 + G^{-1} = {\tilde E + \tilde B} = - 2 { Q_m^a X_{la} \over \sqrt{ - 2 X_l . X_r } } ~,~~~~~~~~
G^1 - G^{-1} =  {\tilde E  -\tilde B} = 2 { Q_m^a X_{ra} \over \sqrt{ - 2 X_l . X_r } } 
\ee
These are generators which leave one of the points fixed, ($X_l$ for the first, and $X_r$ for the second). 
They do not act within the Wheeler de Witt patch, and can map points inside to points outside, see figure 
\ref{SymmetriesWBoundary}. 
%We can also define a linear combination that is more similar to the ones in \nref{Coord}, 
%\be Q^{\tilde E} = \half ( Q^{\tilde E + \tilde B} + 
%Q^{\tilde E - \tilde B} ) ~,~~~~~~~~~~
%  Q^{\tilde B} =  \half ( Q^{\tilde E + \tilde B} - 
%Q^{\tilde E - \tilde B} )
%\ee 
These generators have been defined so that they obey the same algebra as the generators in \nref{Engen}, but
are defined  relative to the two boundary positions. 
Notice that they involve a matter operator, $Q_m^a$ and operators of the boundary systems $X_{l,r}^a$. 
Since they are gauge invariant, they map physical states to other physical states. 

We can think of the generators ${\tilde B}$ as defined by the following procedure. Imagine that 
have have two points $X_l$ and $X_r$ that are very far away, but not yet at the boundary. Then we 
join them by a geodesic and determine their midpoint. Then ${\tilde B}$ is the boost around this 
midpoint. Then the third generator, ${\tilde E}$, results from commuting the previous ones and 
gives a generator that locally looks like a time translations around the midpoint, see figure \ref{SymmetriesWBoundary}. 
These are time translations locally orthogonal to the geodesic joining $X_l$ and $X_r$. 

Acting on a state with given boundary coordinates $X_l$ and $X_r$, this state moves the matter around 
leaving the boundary points fixed. The resulting time evolution of $X_l$ and $X_r$ can be changed by
the action of these generators, but not their instantaneous positions. 

We have found the action of a physical $SL(2)$ symmetry on the physical Hilbert space. In particular
this means that the physical Hilbert space is infinite dimensional due to the matter degrees of freedom, 
and their descendants. 

 In this discussion, we have neglected the possibility of topology changes, such as the ones in \cite{Saad:2019lba}, 
since we assumed that the topology is essentially a strip. Therefore we are assuming that $\phi_0$ in 
\nref{JTAct} is very large so that topology changes are highly suppressed. It would be interesting to understand how other topologies change the picture; presumably it should be related to cutting off the algebra to  a finite dimensional Hilbert space.

An alternative way to describe this same construction is to say that we have defined three vectors 
$e^{A }_{\,a}$, where $A$ is an index running over the three vectors, and then we defined three gauge invariant
generators 
\be
G^A = e^{A }_{\, a} Q^a_m.
\ee
The three vectors were the ones in \nref{MomInv} \nref{OtherTwo}, i.e., 
\eqn{
\la{evec}
&e^0_{\,a} =  { \epsilon_{abc} X_{l}^b X_{r}^c   \over X_l . X_r } ~,~~~~~~e^{-1}_{\, a} =- { 1 \over \sqrt{-2 X_l.X_r } } (X_{r a } + X_{la } ) ~,\\
&e^{1}_{\, a} = { 1 \over \sqrt{-2 X_l.X_r } } 
(X_{r a } - X_{la } ).
}
The $G^A$ also obey the $SL(2)$ algebra, $[G^A,G^B] = i \epsilon^{ABC} G_C$, 
 due to the properties of $e^{A}_{\, a}$ and the 
commutation relations of $Q^a_m$. 
We can also write the matter Casimir 
\be \la{Casimir}
C \equiv  G^A G^B \eta_{AB} = Q_m^a Q_m^b \eta_{ab}  =\tilde E^2 - \tilde B^2 - \tilde P^2 = E_m^2 - B_m^2 - P_m^2
\ee
which is \slt gauge invariant and commutes with the Hamiltonian. 

As a side comment, we may preserve the algebra by rescaling $E, B, P$ by a factor depending on $X_l . X_r$ if we also rescale $\eta$ by a compensating factor.

\subsubsection{Writing the charges purely in terms of boundary quantities} 

We can use the fact that $Q^a_m = - (Q^a_r + Q^a_l)$, \nref{Constr}, together with 
%\nref{ChargXr} \nref{ChargXl}
\nref{dotX},  to write 
\bea 
 G^0={\tilde P } &=& -{ \epsilon_{abc} X_{l}^b X_{r}^c ( Q^a_r + Q^a_l ) \over X_l . X_r}  =
 { \dot X_l . X_r - X_l \dot X_r \over X_l . X_r }   
\cr \la{MomChaXX}
\tilde P &=& 
( \partial_{u_l} - \partial_{u_r} ) \log[ - 2  X_l . X_r ] 
=(\partial_{u_l}  - \partial_{u_r} ) \ell  
\\ \la{SYKNERN}
\tilde P &=& { \alpha_S N \over {\cal J } }(\partial_{u_l}  - \partial_{u_r} ) \ell = { r_e^3 \over G_N }  (\partial_{u_l}  - \partial_{u_r} ) \ell
\eea
where we noted that $\log[ - 2 X_l . X_r]$ is the regularized distance between the two boundaries, in units
of the radius of AdS$_2$. 
 By 
``regularized" we mean that we have subtracted an infinite additive constant
 to the actual proper distance\footnote{This infinite constant is
independent of time and independent of the  $X_l$ or $X_r$ variables.}. 
In \nref{SYKNERN} we have restored the constants that we had set to one for the SYK case or the 4d
near extremal charged black hole. 

Notice that,  due to \nref{dotXX}, the numerator commutes with the denominator, even though each term 
in the numerator does not commute with the denominator. 
In this formula, \nref{MomChaXX}, 
 we see that the total momentum is expressed purely in terms of boundary quantities, or gravitational 
quantities\footnote{Note that we are talking about the boundary gravitational degrees of freedom, and not the
holographically dual boundary quantum mechanical theory.}. In addition, this generator can  
 be interpreted as the matter momentum
in a frame set by the boundary positions. 
%  The logarithm in \nref{MomChaXX} is a regularized proper distance between the two boundaries. 

Notice that \nref{MomChaXX} is a rather pleasing expression because it can be interpreted as saying that
the momentum of matter is minus the  momentum of the left plus right boundaries. Namely, if we choose
a coordinate $x$ along the geodesic connecting the two boundaries, then the distance is 
$\ell = x_r - x_l  $ and the momentum is
\be
\tilde P =( \partial_{u_l} - \partial_{u_r} ) 
\left[x_r(u_r) -  x_l(u_l)\right]  = - ( \dot x_r + \dot x_l  ) .
\ee
This is saying that the matter momentum is minus the sum of the momenta of the boundary particles. 
 We get the naive expression for the momentum of the boundary particles 
 because the term involving $\q$ in \nref{QYch} drops out when we contract 
 $\vec Y_l \times \vec Y_r$ with $\vec Q$. So   we 
 get the same result as for an ordinary massive particle.  

 Note that in writing \nref{MomChaXX} we assumed a particular form for the Hamiltonian that generates the 
 $u$ dependence. In particular, we have assumed that we have a decoupled evolution, by $H_l$ and $H_r$ 
 in \nref{HamQsq}. In contrast, the expressions \nref{MomInv} \nref{OtherTwo} did not use the form of the Hamiltonian and are valid more generally (for example we could have a small coupling between the left and right sides). We can obtain expressions that are more generally valid by writing $Q^a_l$ and $Q^a_r$ in terms of the boundary positions and their conjugate momenta, see appendix \ref{appendixCan} and \nref{PQCharges}.

We can also consider the expressions for the other generators. Again, we start from 
\nref{OtherTwo} and we express the matter charges in terms of the left and right charges, and
use \nref{ChargXr} \nref{ChargXl} to obtain  (ignoring operator ordering issues) 
 \eqn{ G^1 ={\tilde E}&=   \lp - 2 X_l . X_r  \rp^{1/2} \lp H_l + H_r + { 1 \over  X_l . X_r}-  { \ddot X_l . X_r +  X_l . \ddot X_r\over 2 X_l . X_r} \rp, \\
	     G^{-1} = {\tilde B}  &= - \lp -2 X_l . X_r  \rp^{1/2} \lp H_l - H_r + {X_l. \ddot X_r - \ddot X_l . X_r \over 2 X_l . X_r} \rp. 
	\la{OtherTwoXX}     }
%We also see that we can express these in terms of distances between the left and right sides. 
We can also express the Casimir \nref{Casimir} in terms of purely boundary quantities. 
Of course, these expressions depend on {\it both} boundaries. 

These observables can be expressed in terms of energies and distances between left and right sides. We will discuss how to measure distances in Section \ref{DistMeas}. 
% We will also give an interpretation of equation \nref{OtherTwoXX} as a modified eternal traversable wormhole in Section \ref{EvolvChar}.

%And the distance between both boundaries is a subtle
%observable. 
 %\bea
%G^A G^B \eta_{AB} &  =& - 2 (H_l + H_r) + 2 Q_l . Q_r  
%\cr
% & =& -2 ( H_l + H_r) - 2 \ddot X_l .
% \ddot X_r + 4 \ddot X_r . X_l H_l + 4 \ddot X_l . X_r H_r - 8 H_l H_r X_l . X_r 
% \notag \eea
% I AM NOT SURE THIS EQUATION IS ILLUMINATING... 

\subsection{Defining ``conserved'' charges } 
\la{ConsCharEx}

The generators we have defined above act on the physical Hilbert space but they do not commute
with the Hamiltonians of the system, $H_l$ or $H_r$. Therefore we cannot call them conserved quantities. 
(Of course, the Casimir \nref{Casimir} is indeed conserved.) 
However, one feature of the gauge-non-invariant matter 
charges is that they are conserved $[H_{l,r},Q_m^a] = 0$ in the unphysical Hilbert space. 

The charges we defined depend on the left and right times through the boundary positions $X_l(u_l)$ and
$X_r(u_r)$. Then the charges in \nref{MomInv} \nref{OtherTwo} depend on the two times $G^A(u_l,u_r)$. 
However, the dynamics of the left and right boundaries is solvable as a quantum mechanical theory. 
This means that the change in the charges follows a reasonably predictive pattern. In particular, 
we would obtain time independent expressions for the generators by solving the boundary dynamics so that
we can work with $X_l(0)$ and $X_r(0)$. 
Therefore we can simply say that the ``conserved'' charges are 
simply $G^A(0_l,0_r)$ where we have set both times to zero. 
Now, this looks like we are cheating since we can always define a conserved quantity by undoing the time evolution. However, in this case,   the statement has non-trivial content because
we {\it only} have to undo the evolution of the boundary mode, the Schwarzian degree of freedom. In 
particular, we are not undoing the evolution of matter, which could be a complicated self interacting
theory. 
In addition, in the  classical limit, we can undo the classical evolution of the boundary theory in 
a simple way. In principle, we can also express $G^a(0,0)$ in terms of the correlators at zero as in 
\nref{MomChaXX} \nref{OtherTwoXX}. 

Formally we can write down 
\be \la{Gzz}
G^A(0,0) =  e^{-i (H_l u_l + H_r u_r)} G^A(u_l,u_r) e^{i (H_l u_l + H_r u_r)} = 
 \Lambda^{A}_{~B}  G^B(u_l,u_r) 
\ee
with 
\be
\Lambda^A_{~B} = e^A_{\, a }(0 ,0)  \left( e^{-1}(u_l,u_r) \right)^{a}_{~B} =  e^{A}_{\,\, a}(0,0) \eta^{ab} e^C_{\,\,b}(u_l,u_r) \eta_{CB}
 \la{LambdaEq}.
\ee
This expression for $\Lambda$  involves the quantum operators 
$X_{l,r}$ evaluated at zero and also evaluated at $u_l$, $u_r$, so it is
a rather complex expression in the quantum Schwarzian theory. 
Note that the operator $X_r(0)$ can be expressed explicitly in terms of operators at time 
$u_l, ~u_r$ by using the propagators for the Schwarzian theory \cite{Yang:2018gdb,Kitaev:2018wpr}. 
Unfortunately, the operators are not diagonal in the $X_r(u_r)$ basis, so it is hard to express them in terms of correlators at time $u_r$, and we will not attempt to do it here. 
 
Let us mention that even the standard 
expressions for the matter charges \nref{CharKil} involve some explicit time dependent 
 expressions, since (some of) the Killing vectors depend
explicitly on time. In \nref{CharKil}, this dependence is very simple. In our 
problem the time dependence is a bit more complicated, but in principle solvable.

One case where the dynamics can be solved simply is the classical limit, as we will see in \nref{MatrLa}.
%  In this case, 
% we can write $\Lambda$ (and therefore $G$) explicitly using \nref{LambdaEq} which becomes
%\eqn{\Lambda^A_{\,\,B} = \left(
%\begin{array}{ccc}
%1 & 0 &0 \\ 0& \cosh \gamma & \sinh \gamma   \\
%0 & \sinh \gamma & \cosh \gamma    \\
%\end{array}   
%\right) ~,~~~~~~~ \gamma \equiv { 2 \pi (u_r - u_l) \over \beta}
%}
%CHECK THE SIGN OF GAMMA. \HL{I got the same sign} 
%where $\beta$ is the temperature of the black holes. 
%So, if we add quantum matter to this wormhole we can get approximate expressions for $G(0,0)$ using 
%\nref{Gzz}. 
%In this approximation, the fact that we need to multiply $G$ by some time dependent factors is
%similar to the time dependence of the killing vectors in \nref{CharKil}. 
%The classical limit is appropriate for relatively small $\beta$ and short enough times (with our 
%normalization of the action \nref{SchA}, this means $\beta \ll 1$, $\Delta u \ll 1$). 
%Beyond this regime we have to use the full quantum mechanical formulas. 

\section{Approximate expressions for the generators} 
 
  %=========================================
\subsection{The generators in the semiclassical limit}
%=========================================
  \la{SecLinearized}
 
It is instructive to consider the above construction in the semiclassical limit. So we start with 
 a two sided black hole solution with $\beta\ll 1$, 
\be \la{BkgSo}
 t_r = \tilde u ~,~~~~~~t_l = \tilde u ~,~~~~~~~~ \tilde u \equiv \mathfrak{s} u ~,~~~~~\mathfrak{s} \equiv  { 2 \pi  \over \beta } = { 2 \pi \alpha_S N \over \beta \cal J } = { 2 \pi r_e^3 \over \beta G_N } 
 \ee
% \be \la{BkgSo}
% t_r = \tilde u ~,~~~~~~t_l = \tilde u ~,~~~~~~~~ \tilde u \equiv 2\pi u/\beta  ~,~~~~~\mathfrak{s} = { 2 \pi \alpha_S N \over \beta \cal J } = { 2 \pi r_e^3 \over \beta G_N } 
% \ee
 where we have defined the coefficient of the Schwarzian action $\mathfrak{s}$, which has the interpretation of the $SL(2)$ spin of the state in the Schwarzian theory ($j = \half + i \mathfrak{s})$. It is also related to the near 
 extremal entropy, $S-S_0 = 2 \pi \mathfrak{s}$. We have also restored the constants we had 
 set to one for the case of SYK or
 4d near extremal charged black holes. 
 %\footnote{Note that when $\tilde u$ appears in a dimensionless expression, such as  $e^{\tilde u}$, we restore constants as $\tilde u = { 2 \pi u \over \beta } $. We are choosing units of energy so that $\mathcal{J} = \alpha_s N$.}.

   It is common in these discussions to keep two parameters $N$ and ${\cal J}$ as independent, and the fact that we have removed them completely might confuse some readers. Indeed they are independent parameters in a model such as $SYK$. However, all of our discussion is centering on the low energy regime and the coupling of the Schwarzian theory to the conformal sector. These parameters appear only in the overall coefficient of the Schwarzian action. We have rescaled the units of time $u$, so as to set this constant to one, see \nref{SchA}. This highlights the fact that there is only an overall lengthscale appearing the problem, and we have chosen units where this lengthscale is set to one. This is the timescale at which the Schwarzian theory becomes strongly coupled. In these units the limit $\beta \ll 1$ corresponds to the semiclassical limit of the Schwarzian theory. 
 We can restore the full dependence on $N/{\cal J}$ by restoring such constants by using dimensional analysis. For example, thinking of $\mathfrak{s}$ as the entropy we restore them as in \nref{BkgSo}. If we had $\beta$ appearing in a dimensionless quantity, such as $e^{ 2 \pi u/\beta  }$, then no further change is needed, since the rescaling of $u$ and $\beta$ cancel out. 
 The semiclassical limit corresponds to $\mathfrak{s} \gg 1$. 
 %\footnote{As we will see later, in SYK variables,  $\mathfrak{s} = { N \over \beta J}$. We are choosing units where $J=N$.}. 

 Inserting \nref{BkgSo} into the right-left charges \nref{qs} \nref{qsl} we find that $Q_r^a + Q_l^a=0$ as 
 expected. The only non-zero components of these charges are 
 $ Q^{-1}_l = -Q^{-1}_r = \mathfrak{s}$. 
 We now add a relatively small amount of bulk matter $Q_m$. By small we mean that the changes in the 
 boundary trajectories are small, 
 \be  \la{TimeExp}
 t_r = \tilde u + \epsilon_r(\tilde u) ~,~~~~~t_l = \tilde u + \epsilon_l(\tilde u ) 
 \ee
 with $\epsilon_{r,l} \ll 1$. In this case we can expand the charges $Q^a_r$, $Q^a_l$ in $\epsilon$. 
 In fact, it is convenient to expand the sum of the charges because this sum is then equated to 
 $Q^a_m = - (Q_l^a + Q^a_r)$. 
 \def\SemFactor{\mathfrak{s}}

 This gives 
 \bea
%q_m^a & =&  - ( Q_L^a + Q_R^a)
%\cr
Q_m^{-1} &\simeq &  \SemFactor \left[ \epsilon_r' - \epsilon_r''' - \epsilon_l' + \epsilon_l''' \right] 
~,~~~~~~~~{\rm where} ~~~~~ ' \equiv  { \partial_{\tilde u } } \la{Chexp}
\\ \notag
Q_m^0 + Q_m^1=Q_m^+ & \simeq &  \SemFactor \left[  e^{ \tilde  u_r} ( \epsilon''_r -\epsilon'''_r ) + e^{-\tilde u_l} (  -  \epsilon''_l - \epsilon'''_l ) \right] = \SemFactor \left[    \epsilon''_r(0) -\epsilon'''_r(0)    -  \epsilon''_l(0) - \epsilon'''_l(0)  \right] 
\cr
Q_m^0 - Q_m^1=Q_m^- &\simeq  & \SemFactor \left[ e^{-\tilde u_r} ( \epsilon''_r +\epsilon'''_r ) +e^{    \tilde u_l} (- \epsilon''_l + \epsilon'''_l ) \right] =\SemFactor \left[  \epsilon''_r(0) +\epsilon'''_r(0)   - \epsilon''_l(0) + \epsilon'''_l(0)  \right] 
 \eea
where the primes on $\epsilon_{l }(\tilde u_l)$ and $\epsilon_r(\tilde u_r)$ are derivatives with respect to $\tilde u_l$, $\tilde u_r$ respectively. 
These expressions are naively $u$ dependent, but the equations of motion for $\epsilon$ make sure that they are $u$ independent, and we have given 
the expressions for zero times. 
 Namely, from the conservation of energy, 
\be
\la{EnLin} 
H_r= { \SemFactor^2 \over 2   }  + { \SemFactor^2 } ( \epsilon_r' - \epsilon_r''') + \cdots 
~,~~~~~~
H_l= { \SemFactor^2 \over 2  }  + { \SemFactor^2 } ( \epsilon_l' - \epsilon_l''') + \cdots 
\ee
we get 
\be
\epsilon''''-\epsilon'' =0 ~,~~~~~\epsilon = \epsilon_{r,l}(\tilde u_{l,r}) 
\ee
which ensures that the right hand sides of \nref{Chexp} are all independent of time (as are the left hand sides). 

Inserting \nref{BkgSo} into \nref{Xfromt} and then computing the generators $G^A$ to zeroth order in 
$\epsilon$ we find 
\be \la{GenZer}
G^{A}(0,0) \simeq  Q^a_m ~,~~~~A=a.
\ee
This equation is valid in the gauge we used to write the background solution \nref{BkgSo}. 
In general we can also compute the generators $G^A$ at more general times. Using the classical evolution to 
evolve the vectors $e^A_{\, a}$ we can express them as a linear combination of the ones in \nref{GenZer}, see
\nref{LambdaEq}, 
\be
  G^A(0,0) \simeq \Lambda^A_{~B}G^B(u_l,u_r)  ~,~~~~~~
\ee
with 
\bea
\Lambda &=&  \left(
\begin{array}{ccc}
1 & 0 &0 \\ 0& \cosh \gamma & \sinh \gamma   \\
0 & \sinh \gamma & \cosh \gamma    \\
\end{array}   
\right). \left(
\begin{array}{ccc}
\cos \alpha  & -\sin \alpha  &0 \\ \sin \alpha & \cos \alpha & 0   \\
0 & 0 & 1  \end{array}   
\right)  ~,
\cr
 & ~&{\rm where  }  ~~~~~ \sin \alpha = 
\tanh{ \tilde u_l + \tilde u_r \over 2 } ~,~~~~~~  \gamma = {\tilde u_l - \tilde u_r \over 2} . 
\la{MatrLa} 
\eea
 This is reflecting the fact that when we pick arbitrary left and right times, in the classical limit, 
  the generators $G^A$ have been 
 rotated relative to the ones at zero times, see figure \ref{SymmetriesWBoundary}. 
 In this case, the time dependence of the generators is simple and can be extracted to define the time 
 independent generators $G^A(0,0)$. This is the classical version of the formula \nref{LambdaEq}. 
 
 It is also interesting to note that we can also obtain \nref{Chexp} by evaluating $G^A(u_l,u_r)$ using correlators, as
 in \nref{MomChaXX} and \nref{OtherTwoXX}. In detail,  what we have in mind is the following. Let us consider 
 \nref{MomChaXX}, for example. We express $X_l.X_r$ in terms of the boundary times as in \nref{CorrLR}. 
 We then expand the times as in \nref{TimeExp},  to  obtain 
 \be
 \tilde P(u_l , u_r) = (\partial_{u_l} - \partial_{u_r} ) \log[ - 2 X_l.X_r] = \SemFactor \left[ \epsilon''_r - \epsilon''_l  + \half ( \epsilon_l' - \epsilon_r') \tanh{ \tilde u_l + \tilde u_r \over 2 } 
 \right] .
 \ee
 In a similar way we get can get the other generators. Then applying the inverse of the matrix in \nref{MatrLa} 
 we get the generators $G^A(0,0)$ which are the ones in \nref{GenZer}, with the expression \nref{Chexp}.

 \subsection{The semiclassical   limit and $SL(2)$ symmetries of the physical boundary time}
 \la{SemSLTSym}
 
 In this semiclassical limit, we can think of the $G^A(0,0)$ as generating a symmetry that acts as ordinary 
 reparametrizations of $\tilde u$, generated by the infinitesimal transformations 
 \be \la{slure}
 {\rm SL(2)}_{u}: ~~~~\tilde u \to \tilde u +  \alpha_{-1}  + \alpha_+ e^{  \tilde u}  + \alpha_- e^{ - \tilde u } ~,~~~~~\tilde u = \SemFactor u ~,~~~~~~\SemFactor= { 2 \pi \over \beta} 
 \ee
 In order to make this manifest we will analyze how the charges $G^A(0,0)$ act on states created by the 
 insertion of operators in Euclidean time. In fact, we will discuss a state/operator map 
 for the nearly-$CFT_1$ that is dual to this gravity theory. 
 
 More precisely, we view the state at $u_l = u_r=0$ as created by Euclidean time evolution over a time 
 $\Delta u_e= \beta/2$  (or $\delta \tilde u_e = \pi $). This Euclidean evolution generates the empty wormhole. 
 We can then create excitations by acting by operators during the euclidean evolution period. To simplify the 
 notation we will denote by $\varphi$ the rescaled Euclidean time $\varphi \equiv { 2 \pi \over \beta} u_e$, 
 so that $\varphi \sim \varphi + 2\pi$. 
 
\begin{figure}[ht]
\begin{center}
\includegraphics[scale = .5, trim = 0pt 0pt 0pt 0pt]{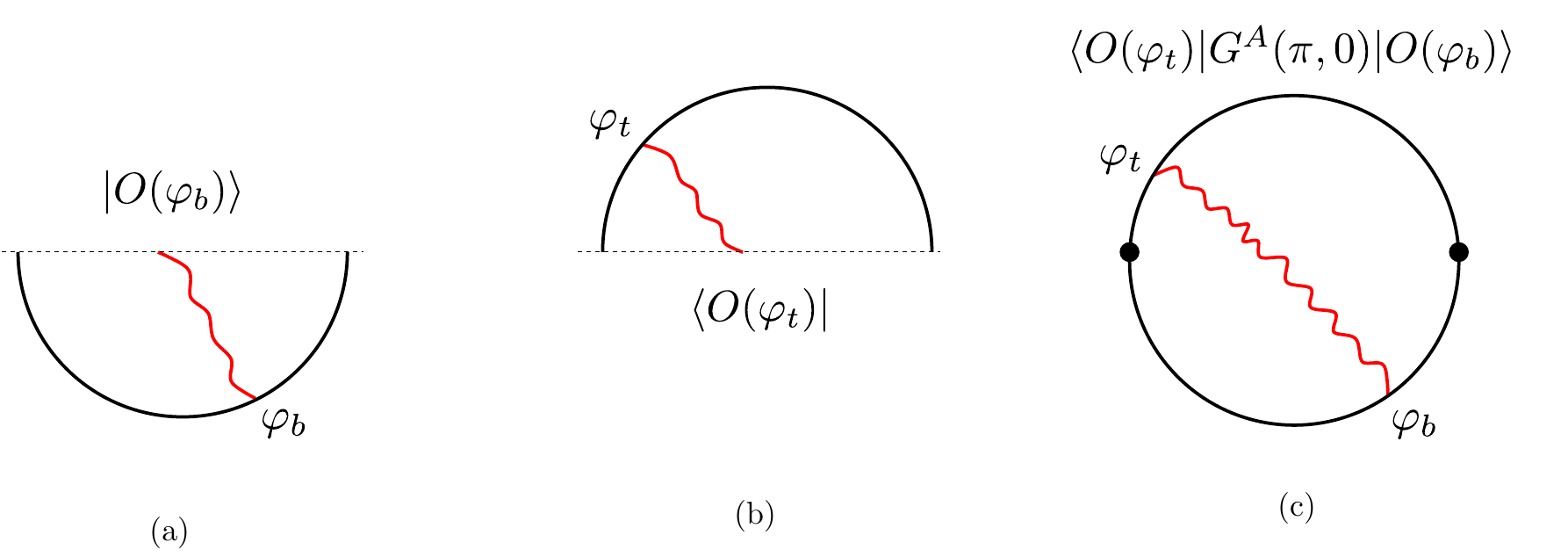}
%\end{figure}
%[scale = 1., trim = 0pt 0cm 0cm 0cm, clip=true]{Charges.pdf}
\caption{(a) By performing euclidean evolution over time $\beta/2$ and inserting an operator at some point during
the euclidean evolution we create a state. This defines a map between operators and states. These are states of a wormhole or states living in the Hilbert space of two copies of the dual boundary quantum system.  (b) The same for the bra. 
(c) We can take the inner product and add the action of a charge $G^A(\pi,0)$, represented by the black dots. In the semiclassical regime, these charges act as $SL(2)$ generators on the states or the operators.   }
\label{OperatorState}
\end{center}
\end{figure}

 We can now consider a general two point function between an operator inserted on the top half of the circle and 
 one on the bottom half
 \be \la{Corrtb}
 \langle O(\varphi_t) O(\varphi_b) \rangle = \langle O(\varphi_t) | O(\varphi_b) \rangle  \propto { \SemFactor^{2\Delta} \over
 \left[ \sin { \varphi_t - \varphi_b \over 2} \right]^{ 2 \Delta} } 
 \ee
 We can view this as the overlap of two states. One is a state that is obtained by doing the path integral over
 the bottom half and the other is the one obtained by doing the path integral over the top half. 
 This defines an operator/state map. See figure 
 \ref{OperatorState}.
 
 We will now act with the charges $G^A(0,0)$, or in this case $G^A(\pi, 0)$ and will demonstrate that they act 
 as expected on the states created by these operator insertions, in other words 
 \be \la{SLAction}
 \langle O(\varphi_t) | G^A(\pi, 0) |O(\varphi_b) \rangle = 
 \left[ \zeta^A(\varphi_b) \partial_{\varphi_b} + \Delta (\partial_{\varphi_b} \zeta^A(\varphi_b) ) \right]\langle O(\varphi_t) | O(\varphi_b) \rangle
 \ee
 where $\zeta^A$ is a linear combination of the vectors generating the infinitesimal $SL(2)$ reparametrizations 
 \nref{slure}, see \nref{CharEps}. This is physically saying that the action of $G^A(\pi,0)$ is acting with an infinitesimal reparametrization on
 the bottom part, see figure \ref{OperatorState}(c). 
 We can equally view it as acting with (minus)  the reparametrization on the top part, since
 acting with the reparametrization both on the top and bottom leaves the correlator \nref{Corrtb} invariant.  
 
\begin{figure}[ht]
\begin{center}
\includegraphics[scale = .5, trim = 0pt 0pt 0pt 0pt]{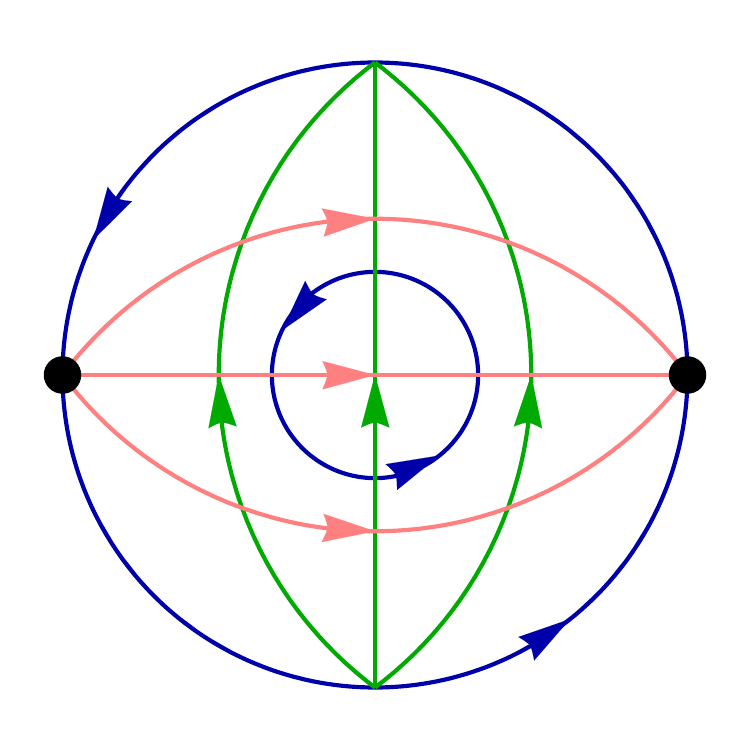}
%\end{figure}
%[scale = 1., trim = 0pt 0cm 0cm 0cm, clip=true]{Charges.pdf}
\caption{ Geometric action of the generators in Euclidean \ads. Here the charges are inserted at the black points,  at  $\varphi = 0$ and $\varphi=\pi$.  {\color{db} Blue} lines follow the boost Killing vectors {\color{db} $\tilde {B}$} ; {\color{pi} pink} lines, the momentum ${\color{pi} \tilde{P}}$ ; and {\color{dg} green} lines, the global energy vectors ${\color{dg} \tilde{E}}$. }
\label{EuclSymm}
\end{center}
\end{figure}

 We will demonstrate \nref{SLAction} as follows. First we write down the three  generators and their associated 
 vectors. 
 \be \la{CharEps}
 \begin{array}{lll}	&\tilde{B}  \simeq \SemFactor\left[   \epsilon'(0)  +\epsilon'''(0) - \epsilon'(\pi) - \epsilon'''(\pi) \right]  ~,~~~~~~
  & \zeta^{\tilde B}  = 1 \\
 	&{\tilde{P}}  \simeq  \SemFactor \left[  \epsilon''(0)  + \epsilon''(\pi)\right]  ~,~~~~~~~ &\zeta^{\tilde P} = -\sin \varphi 
 	\\  
 	& {\tilde{E}} \simeq \SemFactor \left[  \epsilon'''(0)   + \epsilon'''(\pi)\right] ~,  ~ ~~~~~ &\zeta^{\tilde E} = \cos \varphi   
\end{array}
\ee
These are the expressions appropriate for Euclidean time\footnote{
 Relative to \nref{Chexp} we have flipped the signs 
of $\epsilon_l$ and of $u_l$. This arises due to a different definition of the left time. In addition, when we go from Lorentzian to Euclidean time 
we need to say that $\epsilon \to i \epsilon$, $u \to i u$. We also removed an extra $i$ in  $\tilde P$. }. 
We can picture the geometric action of these generators as in figure \ref{EuclSymm}.

\begin{figure}[H]
\begin{center}
\includegraphics[scale = .6, trim = 0pt 0pt 0pt 0pt]{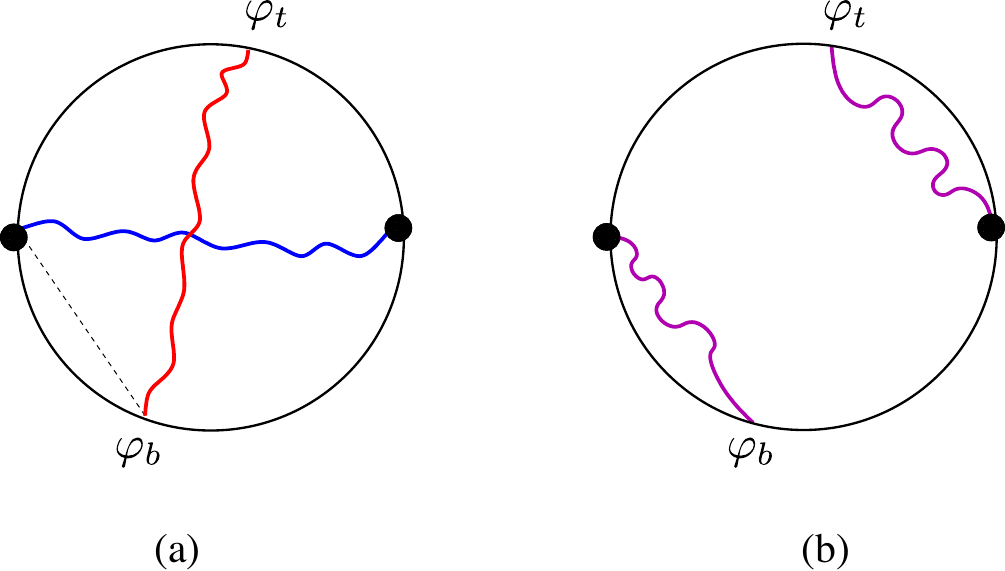}
%\end{figure}
%[scale = 1., trim = 0pt 0cm 0cm 0cm, clip=true]{Charges.pdf}
\caption{ (a) We consider the action of the charges. We have matter fields propagating from the bottom to the top indicated in red. These cause some backreaction on the boundary positions. These are summarized by the coupling to $\epsilon$ at the insertion points of operators. The definition of the charges involves computing a distance, which implicitly, or more explicitly (in the generators $\hat G^A$ in \nref{BoostSum}, \nref{MQEn},\nref{POper}), involve the propagation of other matter fields.  The interesting terms come from correlators between these $\epsilon$ insertions. We only have two point functions of $\epsilon$, only one of which is indicated in the diagram by a doted line. 
Other diagrams contain a dotted line between black points and  $\varphi_{t,b}$.  (b) In some specific models we might get contractions between the fields
in the definition of the charges and the insertions. We want to suppress this type of diagrams. They are indeed suppressed relative to those in (a) in the SYK model.   }
\label{Charges}
\end{center}
\end{figure}

Notice that in this classical limit $\tilde B$ is proportional to $H_r - H_l$, see \nref{EnLin} \nref{Chexp} and it generates
shifts in $\varphi$. 
Now, in order to evaluate the left hand side of \nref{SLAction} we will use the first order expression 
for $G^A$ in \nref{CharEps}. We then also expand the correlators to first order in $\epsilon$. In other words, 
we write the  correlators as in \nref{CorrLR} and expand the times as in \nref{TimeExp} to obtain 
\be
\langle O(\varphi_t) O(\varphi_b) \rangle \to \langle O(\varphi_t) O(\varphi_b) \rangle\left\{ 1 +    \Delta
 \left[ \epsilon_t' + \epsilon_b' - { ( \epsilon_t - \epsilon_b ) \over   \tan{ \varphi_t - \varphi_b \over 2 }  } 
 \right] \right\}
\ee
where the subindices of $\epsilon$ indicate where they are evaluated, $\epsilon_t = \epsilon(\varphi_t)$, etc. 
 Then the computation of \nref{SLAction} boils down to a computation in the linearized 
 Schwarzian theory with action 
 \be
 \la{SchLin}
 S_E = { \SemFactor} \int d \varphi \half ( {\epsilon''}^2 - {\epsilon'}^2 ) + \cdots
 \ee
 We see that the classical limit is indeed large $ \SemFactor $.  
  The propagator associated to this action is \cite{Maldacena:2016upp}
 \be \la{Propa}
 \langle \epsilon(\varphi) \epsilon(0) \rangle = { 1 \over \SemFactor } \left[ 
 G( |\varphi|) + a + b \cos \varphi \right] 
  ~,~~~~G(\varphi) \equiv - { (  \varphi -\pi)^2 \over 4 \pi } + { ( \varphi - \pi) \over 2 \pi}  \sin \varphi
\ee
 where $a$ and $b$ are constants that drop out when we compute $SL(2)$ gauge invariant quantities, such as the ones 
 we are computing. So we can set them to zero. 
For example, to compute an insertion of $\tilde P$ we need to compute the correlator 
 \be
 {\langle O(\varphi_t) | \tilde P | O(\varphi_b) \rangle  \over \langle O(\varphi_t) | O(\varphi_b) \rangle} = \Delta 
 { \SemFactor }\left\langle 
 \left[ \epsilon_t' + \epsilon_b' - { ( \epsilon_t - \epsilon_b ) \over   \tan{ \varphi_t - \varphi_b \over 2 }  } 
 \right]  \left[ \epsilon''(0) + \epsilon''(\pi) \right] \right\rangle
 \ee
 Using the propagator \nref{Propa} we find that this is equal to the expression we need to generate the
 right hand side of \nref{SLAction}. In other words, it is 
 \be \la{PMel}
   {\langle O(\varphi_t) | \tilde P | O(\varphi_b) \rangle  \over \langle O(\varphi_t) | O(\varphi_b) \rangle} = \Delta \left[ { 1 \over \tan { \varphi_t - \varphi_b \over 2 } } \zeta^{\tilde P }(\varphi_b)   + \left(\zeta^{\tilde P } (\varphi_b) \right)' \right]  ~,~~~~\zeta^{\tilde P } = - \sin \varphi 
  \ee
  with $\zeta^{\bar P}= -\sin \varphi$, as in \nref{CharEps}.  The diagrams we need to compute can be seen in 
  figure \ref{Charges}(a).  For $\tilde E$ and $\tilde B$ we also get results consistent with \nref{SLAction}, \nref{CharEps}. 

 In computing the matrix elements of $G^A$, we ignored 1-loop corrections to the 2-pt function. This is justified because such corrections actually cancel, since the zero-th order term in $G^A$ is actually zero. We also ignored the 1-loop correction to $G^A$ itself. If we use the exact charges, this too must vanish because $G^A$ exactly annihilates a state without matter.
However, if we used approximate expressions (as we will discuss in Section \ref{OtherSemExp}) for $G^A$, one should in principle subtract off these contributions order by order in perturbation theory, see equation \nref{MQEn}.

 As a more specific example we can consider the expectation values of all three generators on a state created by inserting the operator in Euclidean time at $\varphi_b <0$, see figure \nref{ParticleRest}: 
 \begin{figure}[H]
\begin{center}
\includegraphics[scale = .5, trim = 0pt 0pt 0pt 0pt]{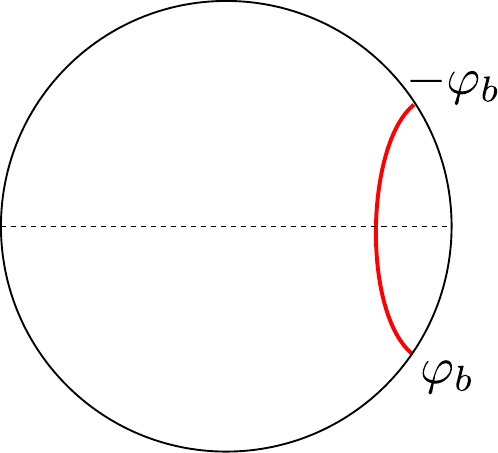}
%\end{figure}
%[scale = 1., trim = 0pt 0cm 0cm 0cm, clip=true]{Charges.pdf}
\caption{Inserting an operator in Euclidean time creates a particle at rest.   }
\label{ParticleRest}
\end{center}
\end{figure}
 We simply evaluate expressions like \nref{PMel} setting
 $\varphi_t = - \varphi_b$ and we obtain 
 \be \la{ExpRest}
 \langle \tilde P \rangle =  0 ~,~~~~~~~\langle \tilde B \rangle = { \Delta \over -\tan \varphi_b } ~,~~~~~~
 \langle \tilde E \rangle = { \Delta \over -\sin \varphi_b }
 \ee
 We should think of a state which contains a particle at rest on the initial slice. At small $|\varphi_b|$, the particle is located 
 at a propert distance of the order $-\log(-\varphi_b)$ from the horizon, and the redshift difference between the horizon and its position if of order $1/(-\varphi_b)$. See figure \ref{ParticleRest}.

The conclusion of this discussion is that around these classical states, the exact generators 
  $G^A(0,0)$ are acting as $SL(2)$ generators transforming the boundary time. 

  The primary states and their descendants defined by the state-operator correspondence are eigenstates of $\tilde{E}$ in the semiclassical approximation, but this is not expected to be exact in the Schwarzian theory. Presumably the exact eigenstates of $\tilde{E}$ could be obtained by smearing the primary in some suitable fashion.

 In previous sections we have seen that $G^A$ maps physical states to physical states. We have seen here that
 this map changes states as we expect from $SL(2)_u$ symmetries of the boundary time $u$ \nref{slure}. In other words, 
 the generators $G^A$ that are always well defined, become the $SL(2)_u$ generators of a \ncft in this limit. 
 This correspondence is not expected to hold away from the semiclassical limit. In fact, the boundary 
 dynamics is not invariant under $SL(2)_u$. But this is a good approximate symmetry in this classical limit. 
 Note that the semiclassical limit is really hardwired in our description of the symmetry itself, since the 
 action of the approximate symmetry depends explicitly on $\beta$ \nref{slure}. This represents a state dependence of
 the symmetry action. And it is reflected in the dependence of the generators on $\beta$ \nref{CharEps} (and will be
 more explicitly seen below). Furthermore, in section \ref{InsertEarly}, 
 we will see that, as we insert matter at early lorentzian times, this 
 semiclassical picture also breaks down. 

 Finally, we would like to caution the reader that this physical $SL(2)_u$ should not be confused with the exact gauge symmetry $SL(2)_g$ that has previously been discussed (and which we review in Section 2.3). The $SL(2)_u$ generators act on physical states $\ket{O(u)}$ and give new physical states, e.g., $\tilde{B} \ket{O(\phi_b)} \approx \ket{\pd O(\phi_b)}$.
 \subsubsection{Inserting matter at early lorentzian times} 
 \la{InsertLor}
 
 In the previous section we have discussed that inserting operators in euclidean time gives us states at $u_l = u_r=0$ 
 that contain bulk excitations, and we 
 explained how to read off the $SL(2)$ charges of these states 
 in the semiclassical limit. 
 
 Of course, these formulas also work in Lorentzian signature. More specifically, imagine that 
 we start with the thermofield double state at early times, say 
 $u_l=0, ~u_r \ll 0$, which we can obtain by evolving the TFD state backwards in time on one of the sides. 
 We then insert an operator at time $u_{r0} < 0$, and evolve up to $u_r=0$. See figure \ref{Lorentzian}(a).
 We will need to slightly smear it 
 in order to create a relatively low energy state that can be described within the conformal regime. 
   This will also create a matter state inside the wormhole. We can find the $SL(2)$ transformation properties of
   this state by acting with the charges, and we will obtain the expected action, as indicated in \nref{SLAction}. 
   It is interesting that now some of the ``conformal Killing vectors"  have an exponential depedence on time, 
   \be
   \zeta^{\tilde B} = 1~,~~~~~~\zeta^{\tilde P} = \sinh{ 2 \pi  u \over \beta }  ~,~~~~~~~\zeta^{\tilde E } = \cosh{ { 2 \pi u \over \beta}} 
   \ee
   This implies that if we insert the same operator $O$, earlier and earlier in time, we will get exponentially growing
   values for its energy and its momentum, from \nref{SLAction}.  
   At least this is true as long as these semiclassical expressions hold. 
   It turns out that for early times, times larger than the scrambling time, $u_{\rm scr} ={ \beta \over 2 \pi }   \log \SemFactor $, it becomes important to take into account the backreaction beyond the leading order in $\epsilon$. 
   
   \subsubsection{Inserting matter beyond the scrambling time and corrections to the semiclassical limit } 
   \la{InsertEarly}
   
  \begin{figure}[ht]
\begin{center}
\includegraphics[scale=.45]{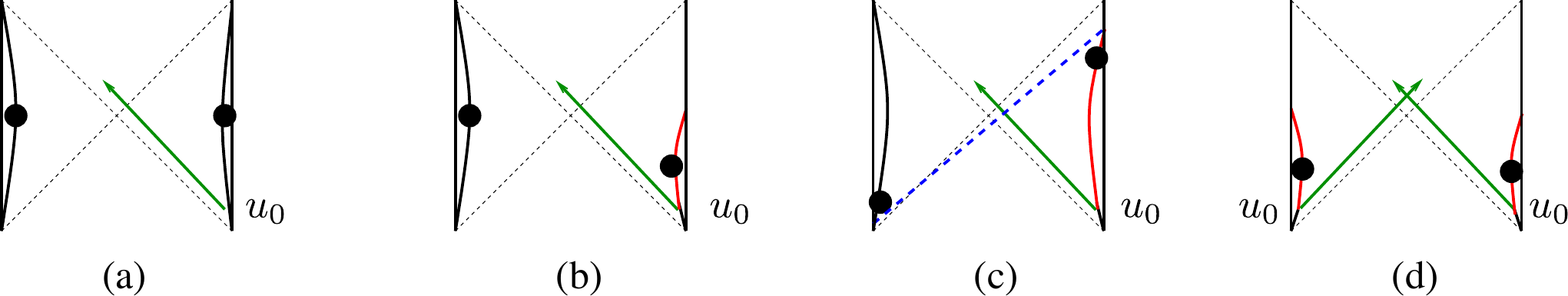}
 \caption{(a) We insert matter at some early Lorentzian time $u_0 < 0$ smaller than  the scrambling time so that the 
 backreaction on the boundary trajectory is small. (b) Same insertion but beyond the scrambling time. There is a large
 change in the boundary trajectory. (c) The same as (b) but after an overall boost (a gauge transformation) that explains better why there is a maximum momentum. Here we have kept the point $u_0$ fixed in the figure so that increasing it 
 corresponds to moving the left and right dots (the times where we evaluate the charges) down and up. 
  The point is that the boost angle between the green line (the matter we inserted) and the blue dotted line is finite. The blue dotted line   represents that highest boost angle for a geodesic joining the two black dots at late times.   (d) We insert matter on both sides. In this case, one can show that the energy continues to increase as we take 
  $u_0$ beyond the scrambling time.  }
\label{Lorentzian}
\end{center}
\end{figure}

 We have seen in the previous subsection that if we insert some mode of energy $\omega$ at some early time
 $u_0$, then its charges evaluated at $u_l = u_r= 0$ grow exponentially as $ \beta \omega e^{ -\tilde u_0}$  ($\tilde u_0 = 
 \SemFactor u_0$ is negative). 
Then, even if $\SemFactor \gg 1$, there can be a time when the simple small $\epsilon$ approximation breaks down.
The expansion parameter is really $\beta \omega e^{-\tilde u_0}/\SemFactor $, and the small $\epsilon$ approximation breaks down 
when this is of order one. This is the so called scrambling time \cite{Sekino:2008he}. The picture is that, by this time, the excitation 
has an order one commutator with any other simple excitation. 
Now, our basic expressions for the generators are exact and can be evaluated beyond the scrambling time. 
In this section we sketch the results for the exact generators \nref{MomInv} \nref{OtherTwo} when we go beyond the 
scrambling time. We will work in the large $\SemFactor$ approximation, and for simplicity 
 we will further assume that $\omega/\SemFactor \ll 1$ and $|u_0| \ll 1$ but we will work exactly in 
\be \la{alphsd}
 \alpha \equiv {\beta \omega \over \SemFactor } e^{ - \tilde u_0 }  ~,~~~~~~\tilde u_0 = \SemFactor u_0 = { 2 \pi u_0 \over \beta } 
 \ee
In this regime,  we can find the correction to the classical trajectory and compute the generators, see figure 
\ref{Lorentzian}(b). We find that the generators are equal to, see appendix \ref{ScramDet}, 
% \HL{These formulas need to be fixed!} 
\be \la{NonLinG}
 \tilde P \sim  \tilde E \sim  2 { \SemFactor   } { \hat \alpha \over 1 + \hat \alpha } ~,~~~~~~ \tilde B \sim 0 ~,~~~~~~
 \hat \alpha \propto { \omega \beta  \over \SemFactor } e^{ - \tilde u_0 } 
 \ee
 where $\hat \alpha $ is equal to $\alpha$ in \nref{alphsd}, up to a numerical constant.  It is also worth noting that $X_l.X_r \propto (1 + \hat \alpha)^2 $. This implies that the physical distance, which is the 
 logarithm of this quantity increases linearly with $u_0$, as $\ell \sim 2 |\tilde u_0| \sim 4 \pi | u_0/ \beta |$. 
 The standard semiclassical expression discussed in 
 section \ref{SemSLTSym} amounts to expanding \nref{NonLinG} to first order in $\hat \alpha$. 
 Interestingly we find that the generators saturate at an amount of order $\SemFactor $ which is independent of 
 the energy of the particle we have sent in. 
 This might seem surprising, since it naively looks like we are inserting a higher and higher energy state as 
 we take $u_0 \to -\infty $. However, this insertion is moving the dynamic right boundary and is changing the notion of
 momentum. Notice that the bulk Casimir is zero in this limit, see figure \ref{Lorentzian}(b,c).  The saturation of \nref{NonLinG} will be related to the decay of out of time order correlators in section \ref{RelChaos}.

 We could consider a different experiment where we send matter from both sides, see figure \ref{Lorentzian}(c).
  In this case, 
 $\tilde P\sim \tilde B \sim 0$, but $\tilde E $ continues to increase exponentially. 
 
 This computation illustrates how the exact generators \nref{MomInv} and \nref{OtherTwo} can be defined and used, beyond
 the scrambling time.

  \subsection{Other semiclassical expressions for the generators }
 \la{OtherSemExp}

 We have seen that we can get approximate expressions for the \slt generators. These approximate expressions relied purely on the small $\epsilon$ expansion of the boundary trajectories around a given thermofield double state. Here we want to relate these expressions to correlators in the boundary theory. Of course, we have already given an expression of the exact generators in terms of the distances that are probed by boundary correlators, 
 \nref{MomChaXX} \nref{OtherTwoXX}. Here we want to provide simple expressions that give the same answer in the semiclassical limit. 
 
 We have already mentioned one of them. Namely, the boost generator
 can be approximately given in terms of the difference of Hamiltonians 
 \be \la{BoostSum}
 \tilde B \simeq \hat B  = { \beta \over 2 \pi } ( H_r - H_l)
 \ee
 This is also the modular Hamiltonian that arises when we split the system into left and right sides. Note that $\hat B$ is an exact symmetry of the thermofield double state. 
 
 The paper \cite{Maldacena:2018lmt} discussed a coupled system whose
 Hamiltonian could be viewed as the global time translation, $\tilde E$,  in $AdS_2$. This Hamiltonian was defined as 
 \bea  H_{\rm coupled} &=& H_r + H_l - \tilde \eta \sum_j O^j_l O^j_r  \la{Hcoup}
 \\ 
 &\sim & \la{HcoupSch}
H_r + H_l - \eta \left( {t'_l t'_r \over \cosh^2{ t_l + t_r \over2} } \right)^{\Delta} ~,~~~~~ \eta = \tilde \eta N 2^{-2 \Delta}
 \eea
 where we have indicated the approximate expression in the Schwarzian theory in the approximation that the effect of the boundary coupling on the bulk matter is very small. We normalized the operators so that they go like $\langle O_r(u_1) O_r(r_2) \rangle \sim |u_{12}|^{-2 \Delta} $ at short distances.  Then the main effect of
 the coupling is on the Schwarzian variables \cite{Maldacena:2018lmt}.  
 We  expand around a solution of the form 
 \be
 t_l = t_r =  {\SemFactor} u 
 \ee
 The solution that minimizes the energy (and obeys all necessary equations of the two Schwarzian theories) is such that 
 \be \la{etaAnds}
 { {\SemFactor} }^{2 - 2 \Delta} = \Delta \eta ~,~~~~~~~~~~~~  
 \ee
 Since the semiclassical limit involves $\SemFactor \gg 1$, we need that
 $\eta \gg 1$. This can be achieved by having a large number of operators in \nref{Hcoup}. 
 In other words, we take small $\tilde \eta$, but large $N$, so that $\eta$ in \nref{Hcoup} is large. 
 %we vide the last term in \nref{Hcoup} as arising from 
 %\be
 %\tilde \eta \sum_i O_l^i O^i_r \to \tilde \eta \sum_i \langle O_l^i O^i_r
 %\rangle \to \eta
 % \left( {t'_l t'_r \over \cosh^2{ t_ + t_r \over2} } \right)^{\Delta} ~,~~~~~ \eta = \tilde \eta N 
 %\ee
 %where the expectation value is over the matter fields and it depends on the state of the boundary variables $t_l$, $t_r$ as indicated in the last expression. 
 In the construction of \cite{Maldacena:2018lmt} this equation, \nref{etaAnds},  was 
 viewed as determining $ \SemFactor $, or $\beta$,  in terms of $\eta$. 
 The ground state of the system is close to the thermofield double
 at inverse temperature 
 \be \la{sTildeb}
  \beta = { 2 \pi \over  \SemFactor } 
 \ee
 This is not the physical temperature of the coupled system, it is rather the effective 
  temperature of the density matrix of each side on its own.
  %\footnote{This $\tilde \beta$ should not be confused with the $\tilde \beta$ appearing in [CITE].} 
  Finally, the normalized global time translation symmetry is then 
  \bea
  \tilde E  \simeq  \hat E & \equiv & { 1 \over   \SemFactor } \left[ H_{\rm coupled } - 
  \langle H_{\rm coupled} \rangle_{\rm 0} \right] \la{MQEn}
  \\
 \hat E  & \simeq  &  - \SemFactor (  \epsilon_l''' + \epsilon_r''' )
  \eea
  where the last expression agrees with \nref{Chexp}, as expected. Here $\ev{\cdots}_0$ indicates the 
  expectation value in the ground state of the coupled system. 
  
  For the purposes of this paper, we can simply view the TFD state 
  at a given inverse temperature, $\beta$,  as given. And we
   then write \nref{MQEn}, solving
  for $\eta$ in terms of $\SemFactor={ 2 \pi \over \beta}$ via \nref{etaAnds}, and construct $\hat E$ as in
  \nref{MQEn}. 
  The advantage of this procedure is that it gives an approximate expression for 
  $\tilde E$ that is relatively simple, we only need to couple the two sides. 
  
  Finally, we can get a simple expression for $\tilde P$ by taking the 
  commutator of \nref{BoostSum} and \nref{MQEn}, to obtain  
  \bea \la{POper}
  \tilde P \simeq \hat P &\equiv & - i [\hat B , \hat E ] = { -i \over \SemFactor } \left[ H_r - H_l, -\tilde \eta \sum_{j}  O^j_l O^j_r\right] 
  =  { \tilde \eta \over \SemFactor^2 } \sum_{j} ( O^j_l \dot O^j_r - \dot O^j_l O^j_r ) 
  \\
  & \simeq &  { 1 \over \SemFactor^2 }  ( \partial_{u_r} - \partial_{u_l} ) 
 \eta  \left( { \dot t_l \dot t_r \over \cosh^2 { t_l + t_r \over 2 } } \right)^\Delta 
  \cr 
 \hat P & \simeq & \SemFactor (  \epsilon_r'' - \epsilon _l'' )
  \eea
  where we have used \nref{etaAnds}. 
  
  Notice that the generators $\hat B , ~ \hat E, ~\hat P$ are completely well defined if the system has a quantum mechanical dual. For example, 
  they are well defined in the SYK model. However they do not obey an
  exact \slt algebra. In addition, their definition depends on $\beta$ (via 
  $\SemFactor $). This means that they behave as \slt generators only for states close to the thermofield double state with that inverse temperature. The fact that they obey the right algebra for such states comes from their connection to the matter charges in \nref{Chexp}.   Notice that the thermofield double state, or empty wormhole, really comes 
  in a two parameter family, parametrized by the temperature and a relative time shift between the two sides, see \cite{Kuchar:1994zk,Harlow:2018tqv}. 
  Again, these generators act as desired only for a particular 
  synchronization of the two times.
   This is implicit in the above formulas
  when we write left-right correlators  ``at the same time".

  %--------------------------------------------------------------------------------------------------------------

%=======================================
 \subsection{Order from chaos } 
%=======================================
 \la{RelChaos}

We can wonder what happens if we take the generators we defined, which are defined in terms of 
correlators at $u_l=u_r=0$ and we ``evolve'' them with the boost Hamiltonian. 
We then get, in Lorentzian time, 
\bea
e^{ i \tilde u \hat B } \hat E   e^{ - i \tilde u \hat B } &=& { \beta \over 2 \pi } \left[ 
 H_r + H_l - \tilde \eta \sum_j  O^j_l(-u) O^j_r(u) - \langle \cdots \rangle_{\rm TFD} \right]  \simeq 
\cosh \tilde u \hat E - \sinh \tilde u  \hat P  \la{EnTimes}
\cr
e^{ i \tilde u \hat B} \hat P e^{ - i \tilde u \hat B} &=&   \partial_{\tilde u} \left( { \beta \tilde \eta \over 2 \pi } \sum_{j} O^j_l(-u) O^j_r(u) \right)  \simeq - \sinh \tilde u \hat E 
+ \cosh \tilde u \hat P 
\eea
where $\langle \cdots \rangle_{TFD}$ indicates the expectation value of the previous three terms in the TFD state. 
The first equality is what we get from the explicit definition of the hatted generators. The second equality is
expected to hold for states that are close to the thermofield double, and it holds to the extent that we can approximate
the hatted generators by the matter ones in the semiclassical limit, see \nref{Chexp} and to the extent that the hatted operators obey an approximate SL(2) algebra. 
 
%In terms of the boundary variables $\epsilon$ we get 
%\bea
%e^{ i \tilde u \hat B } \hat E   e^{ - i \tilde u \hat B } & \eqsim &  
%  - { 1 \over \SemFactor } \left[ \epsilon_l'''(-\tilde u) + \epsilon_r'''(\tilde u) \right] 
%\cr
%e^{ i \tilde u \hat B} \hat P e^{ - i \tilde u \hat B} & \eqsim &  { 1 \over \SemFactor } \left[ -\epsilon_l''(-\tilde u) + \epsilon_r''(\tilde u) \right] \eea
%where, as before, the primes are derivatives with respect to $\tilde u$. 

 \begin{figure}[ht]
\begin{center}
\includegraphics[scale=.6]{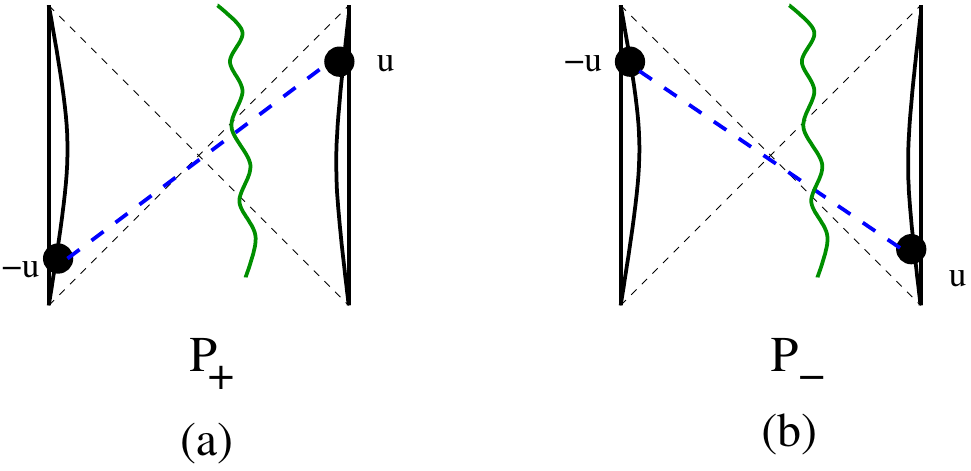}
 \caption{ (a) In order to measure $P_+$ we can consider a correlator in this configuration with large $u$, but smaller than the scrambling time. (b) To measure $P_-$ we consider instead a correlator at these times.  }
\label{Chaos}
\end{center}
\end{figure}

We can think of \nref{EnTimes} as an approximate expression for the approximate symmetries at zero time in terms of operators at other times.  

Notice
that in \nref{EnTimes} we have exponentially growing terms in the right hand side as $\tilde u \to \infty$. Such terms 
can {\it only} come from the term involving $O^j_l( - u) O^j_r(u)$, which indeed can lead to exponential growth. The reason is the following. 
The expectation values of these operators on a state created by acting with operators on the thermofield double is an out of time order correlator. This is an  analytic continuation to Lorentzian time of a 
configuration of operators as in figure  \ref{Charges}(a). In \nref{EnTimes} we are computing the {\it difference}  between this out of time order correlator and the disconnected correlator contained in the thermofield double expectation value $ \langle O^j_l( - u) O^j_r(u)
\rangle_{\rm TFD}$.  The latter is time independent due to the boost symmetry of the empty wormhole or thermofield double. On the other hand the out of time order correlator decays as $u$ increases \cite{Shenker:2013pqa,Kitaev:2014t1}. This initial decay is given by an exponentially growing deviation from the disconnected diagram \cite{Shenker:2013pqa,Kitaev:2014t1}. Since we have a difference between the two correlators in \nref{EnTimes}, we only pick up the correction that is exponentially growing in time.  

We can concentrate on these growing terms and write a simple expression for 
$P_\pm$ at time equal to zero in terms of correlators at other times
\bea
-P_+ &=& {\hat E - \hat P\over 2} = -\lim_{u\to + {\rm large}} 
%\exp\left({  2\pi u\over \beta } \right)\tilde 
e^{-{  2\pi u\over \beta }}
{ \beta \tilde \eta \over 2 \pi }  \sum_j  \left[ O^j_l( - u ) O^j_r ( u) - \langle O^j_l( - u ) O^j_r ( u) \rangle_{\rm TFD} \right]  
\cr
-P_- & = & {\hat E + \hat P\over 2} = - \lim_{u\to   - {\rm large} }  
%\exp\left(-{  2\pi u\over \beta } \right)\tilde 
e^{{  2\pi u\over \beta }}{ \beta \tilde \eta \over 2 \pi } \sum_j  \left[ O^j_l( - u ) O^j_r ( u) - \langle O^j_l( - u ) O^j_r ( u) \rangle_{\rm TFD} \right]  
\eea
where $ \tilde \eta$ is fixed by \nref{HcoupSch} and \nref{etaAnds}. 
The explicit exponential prefactors are decreasing in the corresponding limits and extract the growing pieces of the 
correlator corrections.
% Note that the full expectation values of the correlators {\it decrease} when we have matter. 
%However, here we are considering the {\it correction} to the correlator, which increases exponentially. 
%The $\ev{\cdots}_{TFD}$ means expectation value for the vacuum or for the ``empty'' thermofield double. 
 In these equations when we say ``large'' 
 we mean a large time  but  smaller than the scrambling time. In other words, a time obeying 
  \be
 1 \ll \tilde u \ll \log \SemFactor. % ~,~~~~~~~{\rm or } ~~~~~~~ 1\ll {\rm large} \ll \log \SemFactor
 \ee
 %\be
 %1 \ll \tilde u \ll \log \SemFactor ~,~~~~~~~{\rm or } ~~~~~~~ 1\ll {\rm large} \ll \log \SemFactor
% \ee
  Therefore, these formulas make sense only in the semiclassical limit, where $\SemFactor \gg 1$.

The growing nature of the left-right correlators in the presence of matter is related to chaos \cite{Shenker:2013pqa,kitaevfundamental}. It was found
that this growth is related to gravitational shockwaves which inducing null shifts  of the bulk matter.
 Here we are inverting the logic and
using these growing pieces to {\it define} the action of the generators. 
  In this sense we are getting a symmetry (order) from chaos. 
   
Alternatively, it was shown in \cite{Gao:2016bin} (see also \cite{Maldacena:2017axo}) that the two sided correlators 
induce null displacements of the matter propagating inside, when there is a large relative boost between the two. 
This is related to the phenomenon of quantum teleportation. Here we are using this phenomenon to talk about the 
symmetries. 

In fact, the present discussion suggests that we will be able to use this growth for any non-zero temperature black hole, not just near extremal ones. The only difference will be in the value of the commutator between $[P_+,P_-]$. In our case
this gives $B$. But for a generic black hole we expect that this should be zero, because the symmetry near any horizon 
is just Poincare. 
In fact, even in our case, if we consider excitations that are very close to the horizon, they will have large values
of $P_+ $ and $P_-$ so that is natural to rescale the generators, making them smaller. This in turn  will also rescale the 
commutator.  On the other hand, near any black hole horizon the boost generator has a natural universal normalization 
which is that of the ``modular'' Hamiltonian (conjugate to standard Rindler time).

 Finally, we should remark that by looking at \nref{EnTimes}, which was derived from algebraic and symmetry considerations, we can deduce that 
 the expectation value of 
 $\sum_j O^j_l(-u)O_r^j(u)$, in a perturbed thermofield double state, 
 should contain a term growing exponentially with maximal Lyapunov exponent in order to match the right
 hand side of \nref{EnTimes}. So we can view this as an algebraic derivation of the maximal chaos behavior. Of course, this is essentially the same 
 as the original gravitational derivation using shock waves \cite{Shenker:2013pqa,Kitaev:2014t1} after we use the particular features of nearly AdS$_2$ gravity.  

%Another way to view his is the following. When we evaluate \nref{GenDiffOne} we will need to compute out of time order correlators to be able to evaluate the generator before and after. This leads to the exponential growth. 

\subsection{Generators for the one-sided case}
\la{OneSideGen}

 All of the charges we have been discussing use two-sided operators. To what extent can we define physical matter charges if we only have access to one side? Clearly it is impossible to determine the matter charges for a general state with only one-sided observables. However, it is plausible that we could detect the matter charges on restricted states, where matter is inserted only from one side.

Let us be more concrete. The generators in  section \ref{ExactGenCons}   were of the form $G^A = -e^A_{\,a}  (Q^a_l + Q^a_r)$. Below we will consider choices of $e^A$ that only depend on the right side. Nevertheless, the presence of $Q^a_l$ seems troublesome. 
Now imagine that we start with a state with no matter, e.g., as $u \to \infty$, the matter charge vanishes $Q_m^a = 0$. On such states $ Q^a_l(-\infty) = -Q^a_r(-\infty)$. If we furthermore assume that the left boundary evolves with the standard Hamiltonian $H_l$ with no matter insertions, then $Q^a_l$ is conserved for all times, so we may write
\eqn{Q^a_l = -Q^a_r(-\infty). }
We may also write this as
\eqn{\la{GenDiffOne}\Delta G_{os}^A = -e^A_{\, a}   \Delta Q^a_r, 
~~~~~~~~~ \;\;\; \Delta Q_r^a = Q^a_\text{r,\,after} - Q^a_\text{r,\,before}.}
$\Delta Q_r^a$ measures the change in the right gauge charges before and after the matter insertion. The subscript ``os'' means one sided. 
Note that we are really defining the {\it change} in the generators, not the generators themselves. 

 \begin{figure}[ht]
\begin{center}
 \includegraphics[scale=.75]{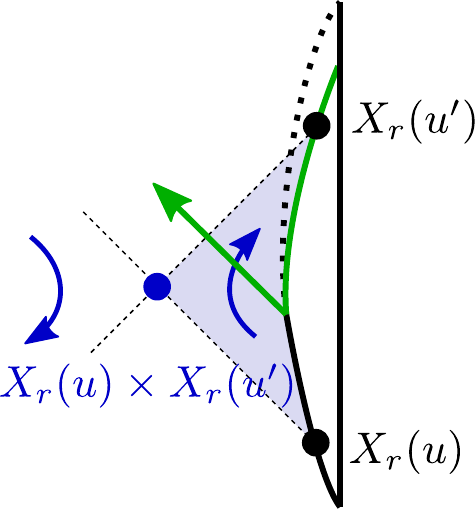}
 \caption{ The one-sided generator. An unknown amount of matter is inserted at some time between $u$ and $u'$. Our task is to detect it using right-sided observables.  The {\color{db} blue} point in the center is proportional to $X_r(u) \times X_r(u')$. The generator corresponding to this point fixes the causal wedge (shaded in light blue).
 } 
\label{OneSideGe}
\end{center}
\end{figure}

We now consider various choices of $e^A$. 
A relatively natural one is to use $X_r^a$ at two different times $X_r(u),X_r(u')$. For the rest of this subsection, all quantities will be on the right side, so we will drop subscripts.
\be \la{OneThreeV}
 e^A_{\, a } = ( e^{-1}_{\, a }, e^+_{\, a} , e^{-}_{\, a } )  = \lp {(X_r \times X_r')_a \over X_r . X_r'}, \sqrt{2}  {X_{r \, a }  (u)\over \sqrt{ X_r . X_r'}}, \sqrt{2}{X_{r \, a } (u')\over \sqrt{X_r . X_r'}} \rp  
 \ee
The generators associated to such vectors have a nice geometric interpretation in terms of causal wedges. 
Namely, if $u< u'$ then we shoot a future directed light ray from $u$ and past directed light ray from $u'$. Then the
``causal wedge'' is defined to be what is enclosed by these light rays, see figure \ref{OneSideGe}. The first vector in 
\nref{OneThreeV} gives a generator that performs a boost around the intersection of the light rays, see figure \ref{OneSideGe}. This boost generator maps points in the causal wedge to points in the causal wedge. 
In the QFT approximation to the bulk physics, ignoring gravity, 
 one can view it as the modular Hamiltonian of the causal wedge\footnote{
 In the full gravity theory the causal wedge should be only an approximate notion that arises when we restrict to simple operators in the boundary theory.}.

 This construction is also closely related to the recent work of \cite{Blommaert:2019hjr}. There, they associate to two boundary times $u$ and $u'$ the point $X_r(u) \times X_r(u')$. They consider a bulk operator at such a point. In other words, they gravitationally dress a bulk matter fields with the gravitational operators $X_r(u) \times X_r(u')$ in order to define a diffeomorphism invariant operator. Here we are dressing the matter charge with the same gravitational operators to obtain $G^{-1}_{os}$. 
     
One problem with the vectors in \nref{OneThreeV}    is that they   fail to commute in the quantum theory. 
To quantify the extent of this problem, let us compute the commutator of the  last two vectors in the semiclassical limit. We can do this by writing the general classical solution
\eqn{X^a %&= {Q \over \sqrt{2H}} + A_\pm e^{\pm \sqrt{2H} u}\\
	&= -{Q^a \over 2H} +  \lp X^a(0)+{Q\over 2H} \rp \cosh(\sqrt{2H}u)+ {\dot{X}^a(0) \over \sqrt{2H} } \sinh (\sqrt{2H} u) }
	We can then compute the Poisson bracket $\{X^a(u), X^b(0)\}_{PB}$. There will be many terms; the important point is that
	%	This implies that at large $u$, the commutator is (FIX factors of 2) and factors of $H
	\eqn{	\{X^a(u),X^b(0)\}_{PB} 
	%q_a q_b  
	= M^{ab}   e^{\sqrt{2H} u} + \cdots ,}
	%{i (X(0) . q)^2 \over 2 \sqrt{ 2H}  } e^{\sqrt{2H} u} + \dot{X} X(0) e^{\lambda u }}
	where $M^{ab}$ is a function of $X(0), \dot{X}(0),H$, but independent of $u$.
	The important point is that there is an exponentially growing contribution to the commutator. On a thermal state, $\lambda = \sqrt{2H} = 2\pi/\beta$; this is exactly the maximal Lyapunov exponent. 
	It is somewhat ironic that the maximal chaos, which we said was useful for constructing the two sided generators, 
	  is also what prevents us from defining good 1-sided charges.

This motivates us to consider $u\to u'$ so that the commutator is smaller. This is equivalent to choosing
\eqn{ \la{DotGen} e^A _{\,a}=  \lp \dot{X}_a, {\ddot{X}_a\over \sqrt{ 2H }},{ (\dot{X} \times \ddot{X})_a \over \sqrt{2H}} \rp.}
Since $\dot{X} . \dot{X} = -1$, the first two vectors are automatically orthogonal $\ddot{X} . \dot{X}  = 0$.
Note, however, that there will be a non-trivial commutator between the different components of $e^A$. For example, the commutator of the first two components will get a contribution from
\eqn{
[\ddot X^a, \dot X^b]= %[Q + \{H,X\}, \dot{X}] \\%= i \epsilon \dot X + [H X, \dot{X}] + [XH, \dot{X}]\\
%=i\epsilon \dot{X} + \{H, [X,\dot{X}]\} + \{ [H,\dot{X}], X\} \\
i \epsilon^{abc} \dot{X}^c + i \{H,  X^a X^b \}-i \{ \ddot{X}^a,X^b\}.
}
Here $\{A,B\} = AB+BA$.
If we contract this expression with the matter charges $\Delta Q^a_m \Delta Q^b_m$ the last two terms  can become large. 
In particular, it grows exponentially as a function of the boundary time when the matter was inserted. So even if we use vectors at the same time, chaos will lead to the breakdown of the one-sided algebra if we wait too long after perturbing the right side.

Finally, let us turn to  the momentum discussed in \cite{Brown:2018kvn}. 
They consider the momentum of a particle thrown in from one side. ``momentum'' here means the variable conjugate to distance. But distance from what? It is most natural to take the distance from the bifurcating surface on the left side, since the bifurcating surface on the right changes when matter is inserted.
The left bifurcating surface sits at a point $Y^a \propto Q_l^a $, which corresponds to a definition of momentum 
\eqn{\tilde P_{\rm os} = Q_{m}.\frac{Q_l\times X_r}{Q_l. X_r} = \frac{\dot X_r. Q_l}{X_r. Q_l} = \frac{d}{ d u_r}\log(-X_r. Q_l).}

From the last line, it is clear that $\tilde P_{\rm os}$ is proportional to the velocity of the right boundary particle relative to the entangling surface. Note, as above, we may replace $Q_l = -Q_r(-\infty)$ to arrive at a purely one-sided quantity.

If we consider semiclassical states with a particle thrown in at some time $u_0<0$ from the right, the one-sided momentum approximates the two-sided momentum $\tilde P_{\rm os} \approx \tilde P$ as long as $|u_r|$ is less than the scrambling time. This is because the geodesic connecting $X_r$ and $X_l$ approximately intercepts the bifurcating surface $Q_l$.

%====================================================
\section{Exploring the bulk}
\la{Explore}
%====================================================

 \subsection{Evolving with the charges}
 \la{EvolvChar}
 
 In section \ref{OtherSemExp} we pointed out that some generators, such a $\tilde E$,  can be approximated, in the
 semiclassical limit, by a simple coupled Hamiltonian \nref{Hcoup} \nref{MQEn}. 
 It is natural to ask whether it is possible to systematically correct this coupled Hamiltonian so that it 
 gives the exact generator $\tilde E$. 
 
 One simple way to think about this is to declare that the full Hamiltonian of the coupled system {\it is } simply 
 \be \la{ChargeEv}
 \tilde H_{\rm coupled} = \tilde E 
 \ee
 This is not the same as \nref{MQEn}, hence the tilde. This seems  a legitimate Hamiltonian from the point of
 view of the gravity theory\footnote{ Non-perturbative corrections that can render the distance between the
 two boundaries ill defined, and therefore \nref{ChargeEv} ill defined. We ignore such corrections here.}.  
 
 This Hamiltonian has a number of differences with \nref{Hcoup} or $\hat E_c$ in \nref{MQEn}. 
 A simple difference is that we do not need to subtract the ground state energy as in \nref{MQEn}. 
 In fact, by construcution, $\tilde E$ annihilates the thermofield double state. A more important difference is
 that $\hat E $ in \nref{MQEn} depends on $ \beta$ (through the temperature dependence on \nref{etaAnds} \nref{sTildeb}. This means that $\hat E$ is close to the generator $\tilde E $ only for states that 
 are close enough to the TFD states with inverse temperature $\beta$. In contrast,  
 the Hamiltonian \nref{ChargeEv} is $\beta$ independent.  TFD states with any temperature and any relative
 synchronization between left and right times are ground states of \nref{ChargeEv}. 
 In \nref{ChargeEv} only states with nontrivial bulk matter have non-zero energy (under the Hamiltonian
  $\tilde H_{\rm coupled }$). 
 
 In the presentation of the charges in \nref{OtherTwo}, we see that $\tilde E$ moves the matter along the global 
 time translation symmetry generator  but leaves the 
 boundaries at the same positions. Up to an $SL(2)_g$ gauge symmetry this is the same as moving the boundaries and 
 keeping the bulk matter fixed.  
 This is close to what we mean by the time evolution of the bulk observer.  
 %
%We now consider acting with $e^{iG^A u_A }$ on some state. 
%The evolution of the boundary particles with respect to these charges is trivial if there is no matter. Said differently, these operators generate a pure gauge transformation.
%We write the charges $G^A = e^A _{\, a } Q^a_m.$ Since $X_l, X_r$ commute with $e^A$, we learn that evolving with $G$ keeps the boundaries fixed, while moving the matter. For example, if we had a clock at rest in the center of N\adsn, by acting with $E$ we could run the clock forward or backward while keeping the boundary particles fixed. 
The 
picture is very similar to the one for the evolution with \nref{Hcoup}, 
 \cite{Maldacena:2018lmt}, where the two physical boundaries move vertically 
in global $AdS_2$. The difference is that these physical boundaries can have any location here, while under 
\nref{Hcoup} they had a preferred location.

In order to explore the relation between $H_{\rm coupled}$ and $\tilde H_{\rm coupled }$ a bit further, it is 
useful to write the expression for $\tilde E$ in terms of the global time $T_l$ and $T_r$. 
In particular, if we act by physical symmetries and choose a special gauge we can classically restrict\footnote{By acting on a general state with the physical symmetries, we can set $\tilde P=\tilde B = 0$ at $u_l = u_r=0$. We then use the \sltg gauge symmetry to set $t_l(0) = t_r(0) = 0$ and  
 $t'_l(0) = t'_r(0)$. So the matter and physical charges align  $e^A_a = \delta^A_a$, at $u_l=u_r=0$. 
 Now the two equations $\tilde P(0,0) =0$, $ \tilde B(0,0)=0$ set $t''_l(0) = t''_r(0)$ and $t'''_l(0) = t'''_r(0)$.
 So at $u_l = u_r = 0$ the coordinates and momenta are equal on both sides. Then using the classical equations of motion, the coordinates and momenta are the same for all times. Hence $t_l(u) = t_r(u)$. Note that this argument could be rerun with the coupled Hamiltonian with almost no modification. } 
 to symmetric configurations $T_l(u) = T_r(u)$. We find from \nref{OtherTwoXX}  
\eqn{ \la{GcE}
\tilde E = 2\lp -T' + {T''^2 \over T'^3} - {T''' \over T'^2}\rp =- 2 e^{-\varphi} ( \varphi'' + e^{ 2 \varphi} ) , \quad ~~~~~ \tilde P =\tilde B =  0\\}
with $\varphi \equiv \log T'$. 
This is also the matter energy in our gauge, and this the same as the gauge constraint $E_m = - Q_l^1(T_l) - Q_r^1(T_r) = \tilde E$. To derive this formulas we can write $X^A_r = ( \cos T_r , \sin T_r, 1 )/T'_r$ and a similar expression for $T_l$. 
%
%Notice that this expression for $\tilde E  $ is exactly just the gauge charge in global coordinates $- \tilde E =  Q^1(T_l) + Q^1(T_r) = E$, 
% (see eqn (B.163) in \cite{Maldacena:2018lmt}). 

We can interpret the full expression for $\tilde{E}$ in \nref{OtherTwoXX} as follows. The prefactor 
\be  \la{Prefac}
(-2X_l . X_r)^{1/2} = 2 e^{ - \varphi } 
\ee
 simply gives a  redshift factor of order  $\beta$ when acting on states near the thermofield double. 
 The first three terms in parentheses of \nref{OtherTwoXX} is precisely the   Hamiltonian in \cite{Maldacena:2018lmt}  with $\Delta = 1$ and $\eta = 1/2$ 
 % \HL{or is it $\eta = 1/2$ ??}.
 % The significance of these parameter values can be seen by looking at the effective action for the variable $\varphi = \log T'$ which is simply twice the regularized geodesic distance between the two sides. For the Maldacena-Qi Hamiltonian $H_\text{coupled} = H_l + H_r - \eta (-X_l . X_r/2)^{-\Delta}$ \HL{Check factors? It seems that we have a factor of 2 mistake somewhere}, we get 
\eqn{ \la{HcoupMQ} 
%H_\text{coupled} = - (2 \ddot \varphi- \dot{\varphi}^2 + e^{2\varphi}) + \eta(2\Delta-1) e^{2\Delta \varphi} 
H_\text{coupled} = - 2 \varphi'' +  \lp \varphi' \rp^2 - e^{2\varphi}  - \eta  e^{2\Delta \varphi}    ~,~~~~~~\eta =\half ~,~~~~~~\Delta =1
}
%for $\Delta = 1$ and $\eta = \half$ we get the first three terms in \nref{OtherTwoXX}. 
The last terms in \nref{OtherTwoXX} give a similar expression which combines to 
\be \la{etem}
\tilde E = ( - 2 X_l . X_r)^{1\over 2} ( -\varphi'' - e^{ 2 \varphi} ) = 2 e^{-\varphi} ( \half e^{\varphi} E_m ) =E_m
\ee
where we used \nref{GcE} viewed as a gauge constraint. 
We see that the dynamical boundary variables have disappeared. So with this Hamiltonian, the boundary has no dynamics. 
This expression, \nref{etem}, looks misleadingly simple because it was written in a special gauge. In order to act with 
$\tilde E$, we do need to know the boundary positions. We need to know the relative synchronization of the two times, for example, and to extract this information we need to measure some left-right correlators. This is a common feature in gauge theories, where an expression in a fixed gauge might look local (here $\tilde E$ appears to involve only one factor in the Hilbert space), but the full gauge invariant operator is not local.  

Before proceeding, let us mention a subtlety in the above discussion. In the above, we wrote expressions which involved derivatives with respect to $u$. These expressions implicitly assumed that $u$-translation was generated by $H_l + H_r$. However, when we imagine evolving with $\tilde{E}$ or with the coupled Hamiltonian $H_\text{coupled}$, our expressions will be modified.
%; for example, the charges $Q$ written in terms of $t(u)$ will depend on $\eta$. 
A better approach is to write this in terms of coordinates and momenta. This is developed in Appendix \ref{appendixCan}. There, we give explicit formulas for the charges and $X^a_{l,r}$ in terms of the coordinates $T, T'$ and their conjugate momenta $p_1, p_2$ (see equation \nref{PQCharges} and \nref{xvec}). Using these formulas, we can express $\tilde{E}$ in terms of coordinates and momenta via the relation \nref{OtherTwo} or via \nref{evec}.

If we denote by $p^r_1, p^r_2$ the momenta conjugate to the global times $T_r(u)$ and $T'_r(u)$, the statement is that when coordinates and momenta for different sides are always equal (up to minus signs), then we get the simple expression
\eqn{T'\tilde{E} = -T'(p_1^l+p_1^r). }
Then evolving by $T' \tilde{E}$ gives the solution $T= -T'u $.

In \cite{Maldacena:2018lmt}, the low energy spectrum of the coupled wormhole was approximately 
 a tensor product of the bulk matter $e^{\varphi} E_m$ and the boundary Schwarzian degrees of freedom whose breathing mode is an anharmonic oscillator. 
While the bulk matter Hamiltonian $e^{\varphi}  E_m$ organizes into SL(2) multiplets, the boundary degrees of freedom does not, since for one thing the oscillator's frequency differs from the AdS$_2$ frequency. 
Viewed as a Hamiltonian $\tilde E$ solves the above problem by subtracting off the kinetic terms and then flattening out the potential energy of the oscillator. Finally, there is an overall factor which removes the redshift factor so that
the Hamiltonian is exactly the bulk energy.  
The result is an energy spectrum of $\tilde E$ that is independent of the Schwarzian modes.
One might also imagine an opposite strategy of achieving an approximate \slt spectrum where instead of removing the energy of the boundary degrees of freedom, one makes the frequency of oscillation so large that the boundary modes are essentially frozen, and we have an effective description that only involves the \slt matter. A preliminary exploration of this idea is given in Appendix \ref{heavyTW}.

While $\tilde H_{\rm coupled}$ makes sense in JT-gravity plus matter, one can question whether we can really 
construct it from a more microscopic theory, such as a full boundary quantum mechanical theory. 
This is of course a question about all $G^A$ generators. In the next section,  we discuss a particular large 
$N$ scaling limit of SYK, where these generators make sense. On the other hand, in a boundary quantum mechanical 
theory with a finite Hilbert space, we should not be able to construct the generators $G^A$ (since they 
generate an infinite number of states). The difficulty lies in measuring distance, as we will discuss in section 
\ref{DistMeas}. 
 %The basic point is that in that limit the correlators are positive,  around low energy states, and we can can raise them to any power or take their logarithms to extract  
% $X_l . X_r$, which are needed for  \nref{OtherTwoXX} (and \nref{MomChaXX}).  
 
% \HL{probably this would be a good place to discuss potential problems with the boundary conditions on matter}

 %WE SHOULD EXPLAIN UNDER WHAT CONDITIONS IT IS REASONABLE TO SET ETA =1 AS ABOVE.... 

 %\HL{I think evolution by $\eta = 1$ is bad because it changes the boundary conditions, but I think it might be reasonable to use an $\eta = 1$ for the purposes of computing expectation values of the global energy}

% ANY REMARKS ABOUT EXTRA FACTORS IN \nref{OtherTwoXX} ? 

 \subsection{Exploring behind the horizon or moving the horizon} 
 \la{ExplBulk} 
 One of our motivations was to understand better how matter moves in the bulk and how that is represented in the 
 boundary theory. 
 The generators we constructed allow us to move matter in the bulk relative to the boundaries, so they allow us 
 to explore the bulk. 
 One would like to be able to explore the region behind the horizon. Indeed these charges allow us to move matter 
 within the Wheeler-de-Witt patch, see figure \ref{SymmetriesWBoundary}. 
 
\begin{figure}[ht]
\begin{center}
\includegraphics[scale=.5, trim = 0 0 0 0 cm, clip]{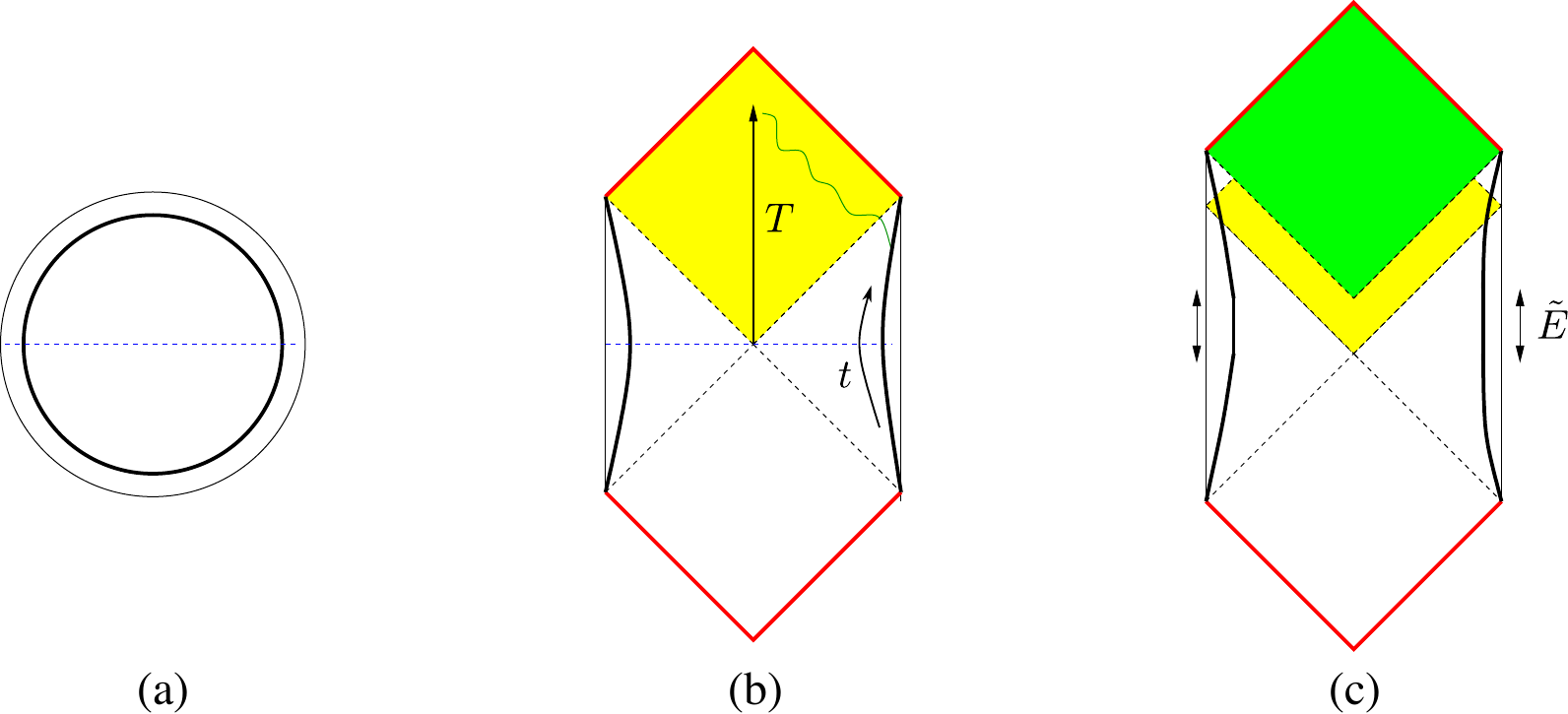}
 \caption{ (a) The standard Euclidean AdS$_2$ picture and its continuation to Lorentzian signature (b). This looks similar to the vacuum decays to AdS studied by Coleman and de Luccia \cite{Coleman:1980aw}. 
 They have shown that the addition of irrelvant operators at the domain wall, which is the boundary for us, leads to a divergence in the bulk at the red line. 
 A bulk observer moving along the central black arrow gets to see a lot of the boundary in a very small proper time. This is the usual blue-shift near the inner horizon. In (c) we evolve for some time using the generator $\tilde E$. This allows us to explore some of the region that was behind the horizon, but the new horizon moves up. The yellow region was behind the horizon and now it is outside.   }
\label{CdL}
\end{center}
\end{figure}

 Actually, it is also important to understand the sense in which we can move matter. When we act with the generator 
 $\tilde E$, for example, we are either moving matter or moving the boundaries (these are two equivalent descriptions). 
 Let us take the point of view that we leave the matter in the bulk as it is but we move the boundaries forwards in time
 along the vertical direction in the Penrose diagram (i.e. by performing a global time translation $T \to T + $constant
 \nref{Coord}). But, if after doing this, we let the boundaries evolve with decoupled Hamiltonians, then we would find now the horizon at a new position, see figure \ref{CdL}(c). 
 So,  we can say   that the   $\tilde E$ generator, allows us to explore the region that would have been behind 
 the horizon if we had done nothing. By the very act of evolving with $\tilde E$, we have brought it out of the horizon (see \cite{Gao:2016bin}). The horizon is a teleological object, and in the quantum theory, it is related to the limitation on the types of experiments we can do. For example, it depends on whether we 
 allow a coupling between the two boundaries. 
We see an important point: a black hole is not just a ``state'', but a state  with some 
 evolution law. Only 
 after specifying the evolution process can we can say that the black hole is ``black'' (it has a horizon). More specifically, if we start with the usual  two boundary wormhole and we do not allow any information exchange between the two boundaries, then we have two black holes. But if we allow information exchange and also allow operators that couple the two boundaries, then we could have an eternal traversable wormhole, as in 
 \cite{Maldacena:2018lmt}. In both of these cases the ``state'' at $u_l = u_r=0$ is the same, or very similar.

\subsection{The inner horizon or Cauchy horizon}

From the point of view of the matter in the bulk, the evolution is given by $E_m$ and it would seem at first sight
that we could continue such evolution ``forever''. 
% 
%However, we can wonder how far we can go into  the region behind the horizon. For example, can we explore the
% region near the singularity (near the red boundary lines in  figure \ref{SymmetriesWBoundary})?.   	
However,  our present discussion does not allow us to move past the inner horizon or Cauchy horizon. The reason is the following. 
We assumed that the matter fields have standard boundary conditions at the 
 AdS$_2$ boundary. These are implied by the boundary conditions at the physical boundaries (curved black lines in figure \ref{SymmetriesWBoundary}). However, beyond the region where the physical boundaries extend, we have no guarantee that the matter boundary conditions are the same as when we had a physical boundary. For this reason we cannot extend the bulk evolution beyond the dotted red lines in figure \nref{SymmetriesWBoundary}. Note that this is also the boundary of the Wheeler de Witt patch when we move both the left and right times to the far future. 
 Of course, it is an important problem to figure out what happens beyond that region! 
 
It has been argued by Penrose that the inner horizon would be generically singular.  (Though it has been demonstrated that classically the   singularity is not too bad \cite{Dafermos:2017dbw} and could be traversed, in some cases.) 
Here we can connect this expectation to the related discussion of  
  vacuum decay into 
AdS that was studied by Coleman and de Luccia \cite{Coleman:1980aw}, see figure \ref{CdL}(a,b).
In the thin wall approximation, the action of the bubble has a surface term and a boundary term, which reproduces JT gravity \nref{JTAct}, \cite{Kitaev:2018wpr,Yang:2018gdb}. In our case,  only the ``true vacuum'' part is present, not the false vacuum. Coleman and de Luccia have argued that in such situations there will be a singularity at the inner horizon. The reason is the following: imagine that we have a scalar field $\phi$ in the bulk and that there is some source for it on the boundary. 
Then, even if the field $\phi$ is massive, and thus corresponds to an irrelevant perturbation 
\cite{Maldacena:2010un}, it is expected to have a non-zero expectation value at the usual horizon. This is translation invariant in the FLRW patch (the yellow patch in figure 
\ref{CdL}(b)). Then the FLRW evolution will generically make it singular along the red line. (For a free bulk field one can avoid the singularity if the corresponding dimension  $\Delta$ is an integer). In the case of the SYK model, we have other operators turned on when we are at finite $\beta {\cal J}$. For a near extremal 4d charged black hole, the fact that the boundary conditions for the fields allow a leakage into the flat space region implies that we have some double trace operators turned on. This could imply that operators such as $\phi \phi $ would get divergent 
expectation values. This is a quantum effect. It is also suppressed in the scaling limit we took, but it seems important if $\beta {\cal J} $ is large and finite or the black hole throat has finite length.

Just to put in some formulas into this discussion we can consider a scalar field $\phi$ and imagine that there is some source on the boundary. We can then use the bulk-to-boundary propagator to compute
\be 
\la{InHo}
 \langle \phi \rangle \propto \int_{\cal C}  du { 1 \over \left[-Y.X(u) \right]^{\Delta }  } \propto 
 \int_{\cal C}  du { 1 \over \left[ \cos T + \sin T \sinh t(u)  \right]^{\Delta } 	}
 \ee
 where we took a point at the center of bulk with $Y^a = ( \cos T, \sin T , 0)$ and a point at the boundary with 
 $X \propto  ( 1, \sinh t , \cosh t )$, see figure \ref{CdL}(b). The inner horizon corresponds to $ T = \pi$, and we see that there the $\sin T$ factor becomes zero and there could be a divergence from the integral over large real times. 
 To analyze this properly we need to specify the contour of integration ${\cal C}$ which goes over the Keldysh contour appropriate for this problem. The regions that contribute are those where $Y$ and $X(u)$ are timelike separated, where the $i \epsilon $ prescription for the forwards and backwards parts Keldysh contour are different and do 
 not cancel if $\Delta$ is not an integer. Formula \nref{InHo} applies also  for   deformations by products of 
 bulk field operators  (``double trace''), where 
 $\Delta$ in \nref{InHo} is the total dimension of the operator.

Notice that this singularity can be moved by evolving the system for some time using the generator $\tilde E$, see 
figure \ref{CdL}(c).   

Note that the yellow region in figure \ref{CdL}(b) looks like a two dimensional Friedman-Lemaitre-Robertson-Walker 
two dimensional cosmology. With the perturbations we discussed, it seems to develop a bulk singularity.
A proper boundary understand of this region from the boundary theory would give us a toy model for a 2d FRLW cosmology. 
This is just the two dimensional version of a general connection between nearly conformal theories on $dS_{D-1}$ and 
negative cosmological constant FLRW cosmologies with $D-1$ hyperbolic slices  \cite{Maldacena:2010un}.

\subsection{Moving operators into the bulk }

When we study gravitational systems with a boundary, one sometimes wants to express operators in the interior in 
terms of operators closer to the boundary. For example, if we had a bulk field $\phi(\tau , x )$ defined in the 
bulk of AdS$_2$, we want to express the operator deep inside in terms of an operator closer to the boundary. 
One way to do it is via the HKLL construction  \cite{Hamilton:2005ju} which involves solving the bulk wave equation and expressing the field at a point in the bulk as an integral of the field near the boundary over a range of times. Here we will provide an alternative construction.

We have constructed an operator $\tilde P$ which performs translations in the bulk, so we could use it 
to translate an operator deep in the interior to an operator closer to the boundary. Roughly we want
\be \la{Oprb}
\phi(\tau , x ) = e^{ i (x -x_{\rm bdy} ) \tilde P   } \phi(\tau, x_{\rm bdy}  ) e^{ - i (x -x_{\rm bdy} ) \tilde P } 
\ee
%where $x$ represents proper distance along  the geodesic that joins the two boundary points $X_l$, $X_r$ (see the 
%FLRW coordinates \nref{Coord}).  
%If we completely fix the two boundary points $X_l$ and $X_r$ then this translation generators gives us a family of
%operators that are exactly local in the bulk, they commute for $x \not = x'$, inheriting the locality properties 
%that we have in the bulk field theory. 
However, this expression is not good enough and we need to   clarify 
  some  subtleties before writing a better expression. 

First, recall that  in the construction of $\tilde P$ we assumed that the actual 
UV boundaries where infinitely far away.  Therefore the point $x_{\rm bdy}$ should still be 
 far away from the boundary,
  but it could sit at a relatively large value of the coordinates $x$ in \nref{Coord}, a value that is large but 
fixed when we take the UV regulator to zero. In other words, we want $x_{\rm bdy}$ to be larger than any value of
$x$ of other operators in the bulk that we want to consider, but finite in the limit that we send the boundaries far away.  
A similar assumption goes into the standard HKLL \cite{Hamilton:2005ju} construction if one wants to use simple AdS wavefunctions. 

A second issue is that the boundaries are dynamical objects and the coordinate points $x$ are not physical by
 themselves. 
%They are also far away, but the actual distance from 
%an coordinate point $x$ is some fixed infinite constant plus $x_r$ or $x_l$, where $x_r$ and $x_l$ are finite order one quantities. 
When we construct the operator relatively close to the boundary we only determine its position relative to the boundary, 
we will call this $\ell_{\rm bdy}$, or $x_{\rm bdy} = \ell_{\rm bdy} + x_r $, where $x_r$ is the position of the right boundary. 
% at a large value of $x_{bdy}$ we will do it relative to the position of the boundary. 
%So the actual position of the operator will be at $x_{bdy} + x_r$ where $x_{bdy}$ is a fixed large value, and independent of $x_r$.
 Such an operator can be constructed in various ways, see e.g. 
\cite{Blommaert:2019hjr}. For an operator with a large value of $\ell_{\rm bdy}$ we expect that quantum fluctuations of the Schwarzian variables are small and the construction will be fairly accurate. 

%
% would be fairly accurate for a large value of $x_{\rm bdy} $. So we need 
%$ 1 \ll x_{bdy} \ll \log \epsilon \sim \log (\beta {\cal J } ) $ (with the last expression in SYK variables)\footnote{
%A similar assumption goes into the standard HKLL \cite{Hamilton:2005ju} construction if one wants to use simple AdS wavefunctions.}. 

With all these caveats,   we can now construct a better expression for 
a bulk operator at some distance $\ell$ from the UV boundary as 
% 
%When we think about the quantum mechanics of the Schwarzian mode, the boundary position is a dynamical variable, so for the above formulas we want to choose $x_{\rm bdy}$ relative to the right boundary particle. 
%$This means that the actual position of the operator \nref{Oprb}. One might worry that this might mean that the final 
% operator would not commute. However,  a more precise form for the bulk operator is 
\bea \la{OprbT}
\Phi(\tau, \ell   ) &=& e^{ i ( \ell - \ell_{\rm bdy} ) \tilde P } \Phi( \tau, \ell_{\rm bdy} )e^{- i ( \ell - \ell_{\rm bdy} ) \tilde P } 
\\ \la{OPnb}
\Phi(\tau, \ell_{bdy} ) &=& \int dx_l dx_r   \phi(\tau, \ell_{\rm bdy} +x_r  )   |x_l,x_r \rangle \langle x_l, x_r |
\eea
The first expression translates the operator from a point closer to the boundary to points deeper into the bulk. 
The second line expresses the operator close to the boundary in a gauge invariant fashion. This operator involves 
an operator $\phi$ acting on the matter Hilbert space. It also involves projection operators onto definite coordinate values for the boundary particles. Thus, it acts on the full Hilbert space of the theory \nref{Constr}. 
  The momentum \sltg gauge generator acts by 
shifting all $x$ coordinates in \nref{OPnb}  by a constant, which can be absorbed by a shift of integration variables. 
We have used a capital $\Phi$ to express the final dressed operators. 

Finally, a more precise description of the bulk operator would also involve the time boundary variables and is 
\be
\Phi(u_l,u_r , \ell) = \int d^2 X_l d^2 X_l  \, 
   \phi[ Y^a( X_l,X_r ; \ell) ]   |X_l , X_r\rangle \langle X_l , X_r  |
\ee
where $Y^a$ is a point determined as follows. First we find the geodesic going between 
$X_l^a $ to $X_r^a$ . 
Then we determine its midpoint. Then we move by a distance $\ell$ to the right along that geodesic to determine $Y^a$.

One can check that   $[ \Phi(\tau, \ell), \Phi(\tau, \ell')] =0$ for $\ell \not = \ell'$. 
This should be compared to the HKLL construction where the construction has to be modified order by order to ensure 
this commutativity \cite{Heemskerk:2012mn,Kabat:2011rz}. 
The construction discussed here includes the full gravitational dressing to all orders in the 
$G_N$ expansion\footnote{Due to topology changing corrections, of order $e^{-S_0}$, this prescription is not 
well defined non-perturbatively, see \cite{Jafferis:2017tiu}.}. Furthermore, 
all matter self interactions have been taken into account.  The prescription is somewhat similar to the prescription of ``shooting a geodesic orthogonal to the boundary'' \cite{Heemskerk:2012np, Kabat:2013wga}. One issue is that, in $AdS$, 
 all spacelike geodesics are orthogonal to the boundary\footnote{One could still select a unique geodesic by choosing more than one point along the boundary, we discuss this in section \nref{OneSideGen}. In our case, this can only work in the semiclassical limit.}. 
 Here we choose a precise geodesic by selecting two boundary points, one on the left and one on the right.

Note that the whole discussion in this subsection is about the bulk theory, not the holographic boundary theory. 
Of course, it is convenient for holography to have the bulk operators written in terms of operators near the boundary.

\section{Connection to SYK and other systems }

%===============================
	\subsection{Generators in SYK}
	%===============================
%	\la{GenInSYK} 
	\la{SYKGen}	
	
	Here we discuss these generators in the context of SYK. 
	%The low energy dynamics of SYK becomes nearly conformal and one can identify an effective degree of freedom that 
	%behaves like the boundary gravitational degree of freedom of the JT theory. 
%	More precisely, the model involves $N$ fermions and there is an energy scale $J$ below which the fermions become strongly coupled. 
	If we consider the system at temperature $\beta$ the effective coupling is $\beta J$. It is convenient to consider the limit
	\be \la{SchLim}
	{\beta J \to \infty} ~,~~~~~~N \to \infty  ~,~~~~~{ N \over \beta J }  = {\rm fixed} 
	\ee
	where the Schwarzian action becomes exact \cite{Stanford:2017thb}. 
%	
%	we get a degree of freedom that is governed by the Schwarzian action. 
	It is important to remark that, in this limit, we can consider quantum mechanical effects in the Schwarzian action.
	%and these effects correctly capture features of the model in this limit \nref{SchLim}. 
	Also, since 
	$S_0 \propto N \to \infty$,  other topologies do not contribute. 
	
	The constructions we discussed above for the charges can be 
	discussed in this model. Whenever we got an expression involving $X_l.X_r$ we could replace it by a fermion operators via 
	\be \la{XXPsiPsi}
	{ 1 \over ( - 2 X_l . X_r )^{\Delta } } \propto { i \over N } 
	 \sum_j \psi_l^j \psi_r^j 
	\ee
	Our expressions for the charges involved functions of 
	$X_l.X_r$. We can then consider functions of these correlators, such as the logarithm or other powers. 
	In the limit \nref{SchLim} these functions are well defined for the low energy states under consideration. 
	Namely, one can be worried that the operator in the right hand side of \nref{XXPsiPsi} has zero or negative 
	eigenvalues. However, in the large $N$
	 limit \nref{SchLim}, we do not access such eigenvalues from low energy states. For such low energy states, and in the limit \nref{SchLim}, 
	  the operator in the right hand side \nref{XXPsiPsi} is a 
	  {\it positive} operator. Therefore we can raise it to arbitrary powers (positive and negative) and we can also take its logarithm in order to construct the 
	  exact generators \nref{MomChaXX} \nref{OtherTwoXX}. 

	  %Alternatively, we could replace operators like $\log \lp i \psi_l^j \psi_r^j \rp$ with $\hf \log \lp i \psi^j_l \psi^j_r \rp^2$. Then the argument of the logarithm is non-negative. In fact, if $N$ is odd, there is no zero eigenvalue of the operator $i \psi_l^j \psi_r^j$, so the operator inside the logarithm is strictly positive, and we have a well-defined operator in the microscopic theory. Of course, just because the operator is well-defined, does not mean that it satisfies an algebra at finite $N$.
	  
	  Now, one can say that in the infinite $N$ limit, \nref{SchLim},  we have an infinite number of fermions anyway so it is not at all surprising that we can find some exact $SL(2)$ algebra. 
	  What is interesting is that the quantum effects of the boundary mode are still finite in this limit, \nref{SchLim}. In particular, the scrambling time for excitations of thermal energies is still finite, in this limit. And also the dynamics of the boundary mode is not conformal invariant.  The non-trivial statement we are making is that, despite these facts, we still have exact \slt generators.

	  Most of our discussion used the language of nearly-AdS$_2$ gravity. However, the SYK model displays the same structure, as reviewed in section \nref{RevSYK}. 
	  %All of our discussion was based on the JT gravity theory. If we analyze the SYK model using the $G, ~\Sigma$ action, we get a structure that is essentially the same as in the JT gravity theory coupled to ``mattter". We can split the excitations of $G$ and $\Sigma$ in to the part involving the ``Schwarzian'' degree of freedom and the 
	 % ``orthogonal ones''. The orthogonal ones are exactly \slt invariant. For the purposes of the computations of this paper they behave exactly like the matter was behaving. In other words, we have some matter gauge covariant charges $Q_m^a$, which should be added to the charges of the left and right system to get zero, etc. 
	  So everything we did in this paper also holds for the 
	  SYK model. Note that we made an important assumption when we discussed the JT gravity theory, right after \nref{JTAct}. We said that the boundary was very far away. This scaling limit of the JT theory, where 
	  $\phi_b \to \infty$, is essentially  the same as \nref{SchLim} in SYK. 
	  Unfortunately, the $G,\Sigma$ action does not give us a simple Hilbert space description, other than \nref{Constr}. In particular, one would like to understand how this emerges from the fermions. Or how the Hilbert spaces in \nref{Constr} are embedded into the full Hilbert space of the fermions. 
	   The analysis of aspects of these symmetries directly in terms of the femions was undertaken in \cite{Qi:2018bje,StreicherPrivate}. The results in this paper provide a ``target'' for such discussions.

	  Of course, an important question is how this structure is broken at finite $N$. We will not discuss that in this paper. However, we will now make the following 
	  simple observation. 
	The action of the generator $\hat E_c$ (or $\hat P_c$) on a state created by a fermion was discussed near \nref{SLAction}. Once we use the expression \nref{MQEn}
	\nref{BoostSum} for the $\hat E_c$ generator, then the computation boils down to a fermion four point function computation of the general form 
	\be \la{EinSYK}
	\SemFactor   { \langle \psi^i(\pi) \psi^j(u_t)  \psi^i(0) \psi^j(u_b) 
	  \rangle \over \langle \psi^i(\pi)    \psi^i(0) \rangle \langle   
	  \psi^j(u_t)    \psi^j(u_b)  
	  \rangle}  ~,~~~~~~~~ \SemFactor \propto { N \over \beta J } 
	 \ee
	and there is no sum over $i$ or $j$.  What we want is that this ratio of correlators is a certain particular order one function of the $\tilde u_i$ variables. 
	Now, the general structure of the four point function is 
	\be
	{ \langle \psi^i(\pi) \psi^j(u_t)  \psi^i(0) \psi^j(u_b) 
	  \rangle \over \langle \psi^i(\pi)    \psi^i(0) \rangle \langle   
	  \psi^j(u_t)    \psi^j(u_b)  
	  \rangle} = 1 +  { \beta J \over N } {\cal F_{\rm enhanced}} + 
	  { 1 \over N } {\cal F_{\rm finite} } 
	\ee
	where ${\cal F}_{\rm enhanced}$ comes from the Schwarzian mode. The $1$ is subtracted with the expectation value in \nref{MQEn}.
	Due to the factor of of $\SemFactor$ in  \nref{EinSYK} we pick up {\it only} the part involving ${\cal F}_{\rm enhanced}$ and the ${\cal F}_{\rm finite}$ part drops out. In fact this last term contains the conformal invariant contribution that features an infinite sequence of composite operators, etc. Such terms depend on something which we can call ``bulk interactions'' \cite{Gross:2017hcz}.   Such terms drop out in the limit \nref{SchLim}. If had not taken that limit, then we see that there are some specific $1/\SemFactor$ corrections that would change the action of $\hat E$ relative to that expected for an 
	\slt generator. We will not discuss here whether this can be fixed up, or to what extent. But it is of course an interesting problem!

\subsection{Relation to ``size''} 
	\label{RelToSize}
	
	 It is worth noting that in SYK language the three generators 
	 in \nref{BoostSum}, \nref{MQEn}, \nref{POper} can be expressed as 
 \begin{eqnarray} \
 \hat B &=& {  \frac{\beta}{2\pi} } ( H_r - H_l )\label{SYKBGen}
 \\
 \hat E &=& { \frac{\beta}{2\pi} } \left[ 
 H_r  + H_l + i\mu  \sum_j\psi^j_l \psi^j_r   - \langle H_r + H_l + i\mu  \sum_j\psi^j_l \psi^j_r \rangle_{\text{TFD}} \right]
 \label{SYKEGen} \\ \label{SYKPGen}
 \hat P & = & -i[ \hat B, \hat E] = -i{ \frac{\beta^2}{4\pi^2}    } 
 \left\{i \mu \psi^i_l  [ H_r , \psi^i_r]    - i\mu[ H_l , \psi^i_l] \psi^i_r \right\}  
 \cr
 & = &  { -i\mu \frac{\beta^2}{4\pi^2}   }  ( \psi^i_l \dot \psi^i_r - \dot \psi^i_l \psi^i_r )
\\
 &~& {\text{ with} } ~~~~ \frac{\mu}{\mathcal{J}} = \frac{4 \alpha_S }{\Delta c_\Delta }\left(\frac{\pi}{\beta\mathcal{J}}\right)^{2-2\Delta} =
   \frac{\alpha_s}{\Delta}\left(\frac{2\pi}{\beta\mathcal{J}}\right)^2\frac{1}{G(\frac{\beta}{2})} \label{muFac}
 \end{eqnarray}
  where $G \lp \frac{\beta}{2}\rp = c_{\Delta}\left(\frac{\pi}{\beta\mathcal{J}}\right)^{2\Delta}
$.
% \begin{align*}
 %	G \lp \frac{\beta}{2}\rp = c_{\Delta}\left(\frac{\pi}{\beta\mathcal{J}}\right)^{2\Delta}
 %\end{align*}

On the other hand, in previous investigations, the concept of the ``size" of an operator 
was of interest \cite{Susskind:2018tei,Roberts:2018mnp,Brown:2018kvn,Qi:2018bje}. Roughly speaking, in SYK, the size of an operator counts the number of fermions that will be affected by applying this operator. One way to characterize this is to consider its commutator with all the fermion operators. \footnote{In the following expressions we assume the operator $\mathcal{O}$ is bosonic. When the operator we are interested in is fermionic, we can multiply it by a fermionic operator from some external system that anticommutes with all $\psi^i$'s.} For an operator normalized so that  $2^{-\frac{N}{2}}\text{tr} \, O^\dagger O = 1$, the operator size is given by \cite{Roberts:2018mnp}:
\begin{align}
\label{size_inf_T}
S_{\infty}(\mathcal{O}) = \hf \sum_{i = 1}^N 2^{-\frac{N}{2}}\text{tr}([\mathcal{O}, \psi^i]^{\dagger}[\mathcal{O}, \psi^i]).
\end{align}
For an operator $\psi(u_r)= \sum_k\sum_{i_1<...<i_k}c_{i_1...i_k}\left(2^{\frac{k}{2}} \psi_{i_1}...\psi_{i_k}\right)$, this expression\footnote{There is a factor of $\sqrt 2$ due to the normalization $(\psi_i)^2 = \frac{1}{2}.$}  gives $S_{\infty} = \sum_k\sum_{i_1<...<i_k}k|\sqrt 2 c_{i_1...i_k}|^2$. This can be written as the expectation value of some ``size" operator defined as
\begin{align}
&\hat S = \sum_j i\psi_l^j \psi_r^j+\frac{N}{2} \\
%&S_{\infty}(\mathcal{O}) = \frac{\bra{I}\mathcal{O}_r^{\dagger}\hat S\mathcal{O}_r\ket{I}}{\bra{I}\mathcal{O}_r^{\dagger}\mathcal{O}_r\ket{I}}
 &S_{\infty}(\mathcal{O}) = \bra{I}\mathcal{O}_r^{\dagger}\hat S\mathcal{O}_r\ket{I},
\end{align}
where $\ket{I}$ is the infinite temperature thermofield double state.\footnote{$\ket{I}$ satisfies $(\psi_L^j+i\psi_R^j)\ket{I} = 0$. As a consequence $\bra{I}\hat S\ket{I} = 0$.} Note that part of the global energy operator $\hat E$ in \eqref{SYKEGen} appears here.

The above notion of size only depends on the operator. A generalization of this notion of size which depends on both the operator and the temperature $1/\beta$ of the system is defined in \cite{Qi:2018bje}, see also \cite{Brown:2018kvn}:
\begin{align}
	S_{\beta}(\mathcal{O}) = \delta_\beta\inv \lp {S_{\infty}(\mathcal{O}\rho^{\frac{1}{2}})-S_{\infty}(\rho^{\frac{1}{2}})} \rp.
\end{align}
Here $\rho$ is the thermal density matrix and $\delta_{\beta}$ is some normalization factor\footnote{$\delta_{\beta} =2G(\frac{\beta}{2})= 2c_{\Delta} \left(\frac{\pi}{\beta \mathcal{J}}\right)^{2\Delta}$, see \cite{Qi:2018bje}.} which fixes $S_{\beta}(\psi^1) = 1$. This overall renormalization also implies $S_{\beta}(\psi^1(u))$ saturates at the value $\frac{N}{2}$. In the limit that $\beta \to \infty$, we recover \nref{size_inf_T}. We may also write this in terms of the size operator $\hat S$:
\begin{align}
\label{size_finite_T}
	S_{\beta}(\mathcal{O}) = \delta_\beta\inv \left({\frac{\langle {TFD}|\mathcal{O}_r^{\dagger} \hat S \mathcal{O}_r |TFD\rangle}{\langle {TFD}|\mathcal{O}_r^{\dagger}  \mathcal{O}_r |TFD\rangle} - \langle {TFD}|\hat S |TFD\rangle 
}\right)
\end{align}
where the expectation value is taken on thermofield double at temperature $1/\beta$.

Here we see that size is related to global energy. Let ${\cal U}_r$ be a unitary operator acting from the right side. %For simplicity, let $U$ be a local (simple) unitary operator; then we will now take the operator ${\cal O}_r$ to be $ {\cal U}_r(u_r) = e^{i H_r u_r} U e^{-i H_r u_r}$, e.g., the simple unitary acting at time $u_r$.  
Then, once
the coefficients are adjusted and the factors of $H_l$ and $H_r$ 
are added as in \eqref{SYKEGen}, \eqref{SYKBGen}, \eqref{muFac}, the size of ${\cal U}_r$ is
\begin{align}
\label{size_symmetry}
S_{\beta}({\cal U}_r) = \frac{\Delta}{2\alpha_s}\frac{\beta \mathcal{J}}{2\pi}\langle TFD| {\cal U}_r^\dagger (\hat E-\hat B) {\cal U}_r|TFD \rangle.
\end{align} 
The fact that ${\cal U}_r$ is unitary is useful because it ensures that 
the term involving $H_l$ drops out from \nref{size_symmetry}. 

There are some interesting consequences of this expression. 
First, the $\hat B$ term is not important if one is only interested in the time dependence of size as we move a given operator insertion to earlier and earlier times (which was the focus of \cite{Brown:2018kvn,Qi:2018bje}), since in that case the energies $H_r$ and $H_l$ are conserved. Let $ {\cal U}_r(u_r) = e^{i H_r u_r} {\cal U}_r e^{-i H_r u_r}$. It was found that its size grows as $|u_r|$ gets large. This growth is the growth of the corrections to the usual out of time ordered correlators if we expand out \eqref{size_finite_T}. 

 Using \eqref{size_symmetry}, we have
\begin{align}
	\pa{S_\beta(\mathcal{U}(u_r)))}{u_r} =-i\frac{\Delta}{2\alpha_s} \frac{\beta \mathcal{J}}{2\pi}\langle{\cal U}_r(u_r)^{\dagger} \lb \frac{2\pi}{\beta}\hat B, \hat E\rb{\cal U}_r(u_r)\rangle =\frac{\Delta}{2\alpha_s} \mathcal{J} \langle{\cal U}_r(u_r)^{\dagger}\hat P{\cal U}_r(u_r)\rangle.
\end{align}
We see that the momentum is related to the time derivative of size. This is also explored in \cite{Susskindpaper}.

The size of such an operator at large $q$ was computed in SYK in \cite{Qi:2018bje}.  To make such a perturbation at $u_r = 0$, we insert an operator of dimension $\Delta$ to the Euclidean circle at $\varphi_b = \frac{2\pi}{\beta}(-\frac{1}{2\mathcal{J}}+iu_r)$, $\varphi_t = \frac{2\pi}{\beta}(\frac{1}{2\mathcal{J}}+iu_r)$. We give a small real part to $\varphi$ to smear the operator by the amount $\sim\frac{1}{\mathcal{J}}$.\footnote{The amount of smearing needed to go from UV to the conformal regime is $\sim\frac{1}{\mathcal{J}}$. We  determined the $\frac{1}{2}$ factor in the real part of $\varphi_t$ and $\varphi_b$, by 
matching the energy to the energy of the exact solution at large $q$  \cite{Maldacena:2016hyu}. The energy can be computed by taking time derivative of the two-point function. %The $\frac{1}{2}$ factor in the real part of $\varphi_t$ and $\varphi_b$ here is chosen such that the size we get from the symmetry generator matches with the result in \cite{Qi:2018bje}. The non-trivial statement is, once we choose this order $1$ coefficient to be $\frac{1}{2}$, the time dependence of size from \eqref{size_symmetry} matches that from \cite{Qi:2018bje}.
} Using \nref{PMel} (and similar formulas for the other two generators obtained through \nref{SLAction})  
we can get the charges of such an excitation.
\begin{equation}
\begin{aligned}
	\left(\hat B, \hat P, \hat E\right) 
	=\ &\left(\Delta\frac{1}{\tan\frac{\varphi_t-\varphi_b}{2}},-i\Delta\frac{\sin\frac{\varphi_t+\varphi_b}{2}}{\sin \frac{\varphi_t-\varphi_b}{2}},\Delta\frac{\cos\frac{\varphi_t+\varphi_b}{2}}{\sin\frac{\varphi_t-\varphi_b}{2}}\right)\\
	\sim \ &\ \Delta\frac{2\beta\mathcal{J}}{2\pi}\left(1, \sinh(\frac{2\pi}{\beta}u_r),\cosh(\frac{2\pi}{\beta}u_r)\right)
\end{aligned}
\end{equation}
where we assumed $|\varphi_t -\varphi_b|\ll 1$. 
Using \eqref{size_symmetry} we find its size
\begin{align}
	S_{\beta}(\psi(u_r)) =\ & \frac{\Delta}{2\alpha_s}\frac{\beta\mathcal{J}}{2\pi}(\hat E-\hat B) = 2\frac{\Delta^2}{\alpha_s}\left(\frac{\beta\mathcal{J}}{2\pi}\right)^2\sinh^2\left(\frac{\pi}{\beta}u_r\right)
	\la{size_check}
\end{align}
This agrees with the explicit fermion counting calculation in \cite{Qi:2018bje} at large $q$.

In the above example, the excitation has boost energy $2\Delta\mathcal{J}$. It starts from near the boundary and falls toward the horizon. A lower energy excitation can be made by creating a particle at rest at a finite proper distance to the horizon $\rho_m$. We denote an operator which creates such an excitation by $\tilde \psi(\rho_m)$.
%$\tilde\psi(\rho_m)$ is the operator creating such an excitation.  
%One expects the operator gets be more complex as it goes deeper into the throat \cite{Susskind:2013aaa,Brown:2018kvn}.
  We can compute the size of such an operator using \eqref{size_symmetry}. To create such an excitation, one can consider insertions at some finite angle $\varphi_t = -\varphi_b$ on the Euclidean circle, see figure \ref{ParticleRest}.
Its distance to the horizon is $\rho_{m} = -\log({\tan(\frac{\varphi_t}{2})})$. We write its charges \eqref{ExpRest} in terms of $\rho_m$: 
\begin{align}
	&\left(\hat B, \hat P, \hat E\right) = \left(\Delta\sinh \rho_m, 0,\Delta\cosh \rho_m\right).
\end{align}
The size of this single-fermion perturbation\footnote{In this formula, as well as in \nref{size_check}, one factor of $\Delta$ originates from the dimension of the operators used to couple the two sides in our expressions for the global energy, whereas the other factor of $\Delta$ is the dimension of the operator inserted. So the size of a more general operator $O$ of dimension $\Delta_O$ would be $S_\beta(O) \propto \Delta \Delta_O$.} is
\begin{align}
	S_\beta(\rho_m) = \frac{\Delta^2}{2\alpha_s}\frac{\beta \mathcal{J}}{2\pi}e^{-\rho_m}.
\end{align}
When the excitation is near the boundary, $\rho_m\sim\log(\beta\mathcal{J})$ and the size is of order $1$. As we move the excitation deeper and deeper into the throat its size increases exponentially and reaches $\sim\beta \mathcal{J}$ before it enters the near-Rindler region.

In fact, one can directly see this exponential growth from the commutation relation $[ P, E-B] = i(E-B)$.\footnote{Here we only consider small excitations on the throat so to a good approximation we can assume the $SL(2)$ algebra holds. See section \ref{OtherSemExp}.} We consider the operators $\tilde\psi(\rho) = e^{-i\hat P(\rho-\rho_b)}\psi e^{i\hat P(\rho-\rho_b)}$.
\begin{align*}
	\frac{d}{d\rho}S_\beta \lp \tilde\psi(\rho)\rp =\ & \frac{\Delta^2}{2\alpha_s}\frac{\beta \mathcal{J}}{2\pi}2\bra{TFD}\tilde\psi(\rho)[i\hat P,\hat E- \hat B]\tilde\psi(\rho)\ket{TFD}\\
	=\ &-S_\beta(\tilde\psi(\rho))
\end{align*}

Part of the message of this paper is that the symmetry structure in \nref{ActThree} \nref{Constr}  is giving us the correct ``target'' generators that should be reproduced by the microscopic analysis. Moreover, to the extent that we have derived the Schwarzian theory from SYK, we have also derived the existence of these generators from the SYK model. 

%{momentum_oneside}In   \eqref{SYKNERN}   the momentum is given by the time derivative of the length of the wormhole.
%The connection of momentum and size/complexity is also explored by Susskind in \cite{Susskindpaper}.
 
  In \cite{Susskindpaper} Susskind discusses the following relation between the momentum of an infalling particle and the complexity of a precursor preparing this perturbation 
\begin{align}
\label{momentum_oneside}
	P \sim \frac{d}{du_r}\mathcal{C}~.
\end{align}
The complexity-volume conjecture  relates the length of the wormhole   with the state complexity \cite{Stanford:2014jda}. Using this we see   that  \nref{momentum_oneside} follows from   \eqref{SYKNERN}. 

 %with \eqref{momentum_oneside} up to the difference in time derivatives. This difference is a difference between two-sided construction and one-sided construction for the momentum operator. See section \ref{OneSideGen}.

	\subsection{Analogy to the Rindler (and AdS-Rindler) decomposition of a higher dimensional field theory }
\la{HigherDimRel} 

In section \ref{OtherSemExp} we discussed the construction of approximate 
symmetry generators. One generator,  $\hat B$ in 
\nref{BoostSum}, is the sum of an operator defined purely on the left and one purely on the right. This generator acts within the entanglement wedges of each of the two sides. On the other hand, the operator \nref{MQEn} is the sum of pieces that act on each side separately plus a term containing a product of left and right operators. This might seem exotic for a symmetry generator. 

Here, we will point out that this type of structure is actually what we get in an ordinary quantum field theory when we use Rindler coordinates. The analogy is most clear if think of a quantum field theory on a spatial sphere. We then divide the sphere in two halves. We then automatically have a generator like $\hat B$.  This is just the   full modular Hamiltonian of the hemisphere (or a solid ball). The full state of the quantum field theory is exactly  the thermofield double for the modular energy. In the bulk, each entanglement wedge looks like the outside of a hyperbolic black hole \cite{Casini:2011kv}. 
 
 Now we can consider a generator such as the global time translation Killing vector of the full system. This generator is exactly related to an integral of the form 
 \be \la{EHigher}
 E = \int_{\rm Sphere} T_{00} = E_l + E_r  + \sum_i \phi^i_l \phi^i_r ~,~~~~~~~
 E_l = \int_{ l-{\rm Hemisphere}} T_{00} 
 \ee
where $E_r$ is the integral over the right hemisphere and $\phi_l \phi_r$ is related to the stress tensor on exactly the boundary between the two hemispheres. In order to make it more manifest, we could regularize the theory using a lattice so that there is no lattice point precisely at the boundary. However, there is a term in the full global Hamiltonian which acts on the sites that are to the left and the right of the boundary and these terms have the structure of an operator on the left and one on the right side . The sum over 
 $i$ is over all fields of the boundary theory\footnote{If we had a gauge theory we can do the splitting by introducing extra boundary charges (in an entangled state)  and we have the same structure \cite{Donnelly:2014gva}.}. 
 \begin{figure}[ht]
\begin{center}
\includegraphics[scale=.4, trim = 0 0 0 0 cm, clip]{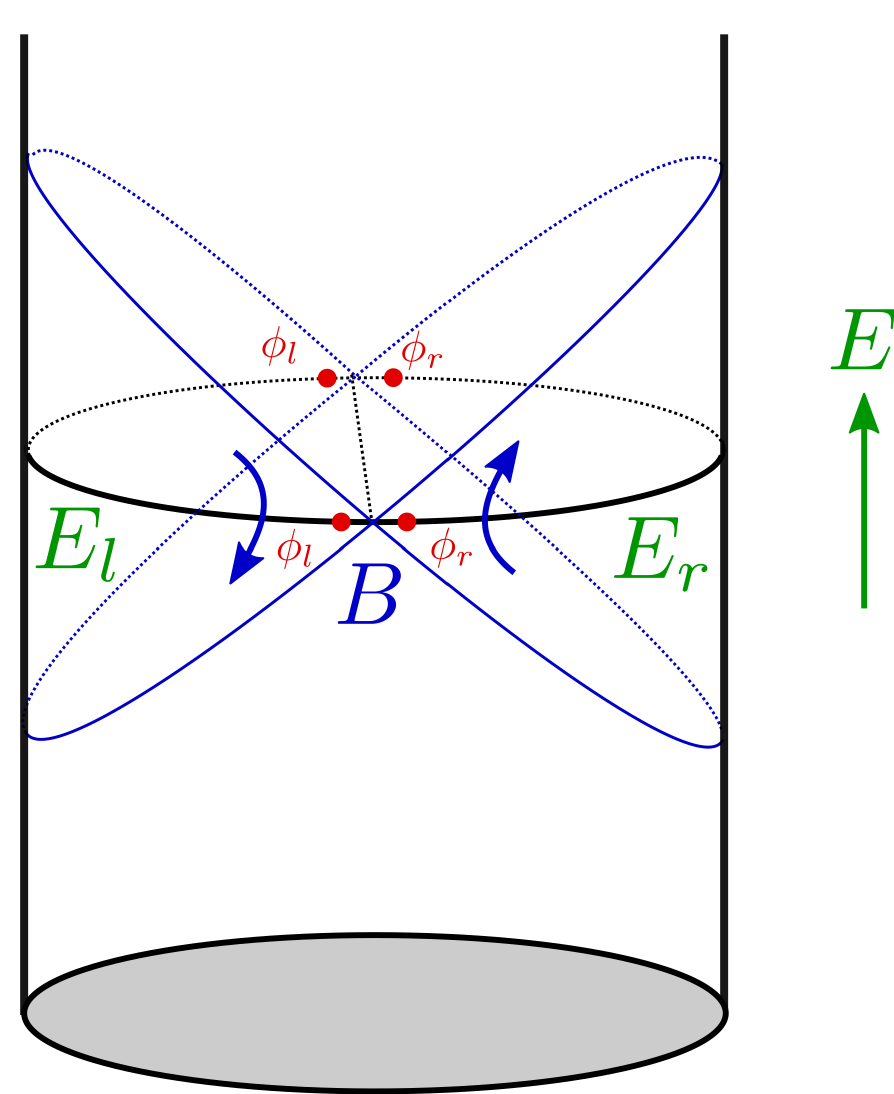}
 \caption{We show a two dimensional boundary theory on a cylinder. We divide the cylinder in two parts, a left and right part. The full modular Hamiltonian corresponds the the boost like generator $B$, which also acts like a boost in  the bulk near horizon region. The global time translation generator $E$ is the sum of three parts. $E_l$ and $E_r$ are completely contained in each region. The third term involves a product of operators in the two regions, depicted here in red.   }
\label{RindlerAdS}
\end{center}
\end{figure}

 So we see that the structure of $\hat E$ in \nref{MQEn} is very reminiscent of what we have in an ordinary QFT when we 
 split it into two parts and we consider the generator $E$. Instead of $H_r + H_l$ (which here would be $B_l + B_r$), 
 we have two other operators that are
 defined purely on the left and the right in \nref{EHigher}. 
 
 The fact that there are terms that act on both the left and the right is crucial to generate the appropriate isometry which can be used to transfer information between the left and right system. In fact, this is the teleportation operator 
 introduced in \cite{Gao:2016bin}. Note also that in \nref{EHigher}   
 we naturally have many operators coupling the left and the right
 because we have many fields in a CFT that is dual to a weakly 
 coupled gravity system. In fact, we can view a particle moving in the gravitational 
 bulk from one side to the other as a particular example of the 
 traversable wormhole discussion in \cite{Gao:2016bin}, but applied to 
 the hyperbolic black holes. 
 
However, one important difference between the higher dimensional
case and  the two dimensional gravity theory is that in higher dimensions we can define an operator $E$, as in 
\nref{EHigher},  that is part of
an  {\it exact} $SO(2,D-1)$ conformal algebra. In our case, we 
only had an approximate expression. 

Notice that in the CFT problem, even if all we cared about was 
the theory on $R \times H_{d-1}$, or the theory on the hemisphere, then when we consider the thermofield double of that system we generate an entangled state which, in the bulk, 
could be extended beyond the union of the two entanglement wedges. This region can be easily explored by evolving the system with the energy generator $E$ \nref{EHigher}.

%=======================================
 \section{Discussion } 
%=======================================
\la{SecDisc}

\subsection{Measuring distance} 
 \la{DistMeas}

In the context of a theory of gravity plus matter the distance between the two boundaries look 
like a reasonable observable (at least if we ignore the effects of topology change, see \cite{Harlow:2018tqv}). 
 
% In the construction of the charges in section XXX, it was crucial to 
% know the distance between the two boundaries. In a gravity theory, this seems a perfectly reasonable observable, at least when we can ignore changes of the spacetime topology. In fact, it is probably not well defined non-perturbatively. 
 
 On the other hand, from the point of view of the holographic boundary theory (the quantum mechanical theory with a finite number of degrees of freedom), this distance is not an obvious operator. 
 We will not attempt to give it a precise meaning in the boundary theory here.  It was proposed in \cite{Susskind:2014rva, Susskind:2014ira, Stanford:2014jda, Brown:2018bms} that this distance is related to ``complexity''.

It will suffice to note that we can define it by considering correlation functions of operators. 
 %The theory has other operators that are indirectly sensitive to this distance. 
 Namely, one can consider correlation functions 
 of operators across the two boundaries, such as 
 \be
 \langle O_l O_r \rangle = { 1 \over (  - X_l . X_r )^{\Delta} }
 \ee
 This is a correlation function on the vacuum and by measuring it, we can indirectly infer the distance \cite{Susskind:2017ney}. However, if we   define the distance in terms of this operator, then we could change it by changing the state of this field. For example, we can    act with an operator that  creates a highly correlated state between left and right values of the field. We can avoid this by imagining 
 that we have many fields, then we can take
 the average of the correlators of many fields. 
 In that case,  changing just one field is not enough to change the 
 distance. But it could be that, if we changed all $N$ of the fields, then we could generate a new geometry. For this to work, the number 
 $N$ has to be comparable to $S_0$, the extremal entropy\footnote{Ideas for measuring distance were mentioned in 
  \cite{Susskind:2017ney}. 
  % had in  mind in the discussion around footnote 8 of \cite{Susskind:2017ney}.
  }.  
 In the SYK we indeed have $N$ fields and an $S_0 \propto N$, so that this method makes sense. This simple method of measuring distances works in the limit \nref{SchLim}. In Appendix \ref{justOne}, we explore how to measure distance in the opposite regime: a bulk theory with only one free scalar field.

 As a qualitative side comment we could say that by increasing the correlations for  just one field, we are
 introducing a small microscopic wormhole, which could become macroscopic when we correlate all $N$ fields. 
 
 A second problem with the definition of distance in terms of correlators is that if we have bulk matter, we can have interactions with matter. If we have such interactions, the correlators could change in the presence of bulk matter and our definition of distance will change. 
 In the particular case of SYK model, these self interactions go like $1/N$, and are suppressed in the limit \nref{SchLim}. 
 Therefore, in this strict limit  this
 definition makes sense.

 %In principle, one could also compute more complicated non-linear quantities like the mutual information of the fields from the left and right side. 
 % In this case, it is likely that we have to mimic the constructions we had in Rindler space for boundary theories in section \nref{HIgherDimeRel}, but now in the bulk. Perhaps we need to carefully construct operators outside the horizon, and then, once we constructed the operators close to the horizon, we could introduce interactions as they would appear in the construction of a bulk stress tensor. 
 
 In more general NAdS$_2$  theories, it is more challenging problem to define this distance, and we will not attempt to do it here. By more general theories we mean, for example, ones with a  small number of matter fields where these fields are self interacting.

	A potentially promising framework for understanding the distance between the two boundaries is the following\footnote{See a related discussion in \cite{Harlow:2015lma}.}. 
	In many body physics, there are some properties that are not defined for single particles but emerge when we have many particles. An example is the phase of a  superconductor  (by a ``superconductor'' here we just mean a system of interacting fermions with a U(1) global symmetry that is spontaneously broken). 
	This phase is good classical variable in the large $N$ limit, but it is not well defined for a small number of fermions. In our case, we have a rather similar variable which is the relative time shift between the two sides of the thermofield double. This relative time shift behaves as a classical variable and it is indeed one of the classical variables of an empty wormhole \cite{Kuchar:1994zk,Harlow:2018tqv}. It can be measured by considering 
	correlation functions, in the same way that the phase of a superconductor can be measured looking at fermion two point functions. 
	One subtle point for our discussion is that in our case this is both spontaneously and explicitly broken. In other
	words, if we evolve by $H_l + H_r$ we increase the value of this time shift. This is analogous to adding a term to the superconductor Hamiltonian that is proportional to the U(1) charge; the phase would then move	linearly with time. 
	We think that we can call this a ``time superfluid" in the sense that the overall time translation symmetry is broken by the wormhole or thermofield double state. In fact, we can think about any state that has a classical time dependence as a ``time superfluid'' in the sense that the time translation symmetry is broken. 
	The wormhole seems special because this time translation is also related to other symmetries of the problem, 
	the approximate SL(2) symmetries of the quantum theory. So another very close to flat direction is the 
	temperature. Namely, states with different temperatures are also related. They are connected by acting with 
	an overall dilation the system. This affects the wormhole by making it longer. In fact, in the wormhole both the time and space directions are related, so the state should be more fittingly called a ``space-time superfluid". 
	We hope to expand on these remarks in a future publication.

\subsection{Conclusions and open questions }

In this paper we have studied the symmetries near the horizon of a black hole. 
The main reason to study them is that these allow us to move into the region behind the horizon, in the sense discussed in section \ref{ExplBulk}. 
So a thorough understanding of these symmetries is crucial for understanding how the interior region emerges in the full quantum theory. 
We have considered near extremal black holes because in this case we have a connection to the SYK model. 
We considered a scaling limit where the temperature becomes very small and the extremal entropy very large, but the
near extremal entropy remains fixed. This physically means that we are keeping the scrambling time fixed in units of 
the temperature and quantum gravity effects are important, but calculable. But topology changing effects are negligible. 
We have defined three SL(2) generators  \nref{MomInv} \nref{OtherTwo}.
 They act on both the matter degrees of freedom and the boundary graviton 
degrees of freedom. If we have a state with fixed values of the boundary positions, the
generators do not move the boundaries but move the matter relative to the boundaries as in figure 
\ref{SymmetriesWBoundary}. 
These generators act on physical  states of the theory grouping them into SL(2) representations. In particular, they
imply that the number of states for the wormhole is infinite. This is not a contradiction because we are 
working in a limit where the extremal entropy is infinite. The non-trivial aspect is that we are keeping the scrambling time finite. So we are making the non-trivial statement that the generators are well defined even after 
the scrambling time. 
These generators do not commute with the Hamiltonian, so we cannot call them ``symmetries'', though from the point of view of a bulk observer made out of matter they act pretty much like symmetries. The total Hamiltonian changes the
generators because it changes the boundary positions. Fortunately the evolution of the boundary positions is a solvable quantum problem so that, in principle, we can undo it. This allows us to define time independent (conserved) 
charges as in \nref{Gzz}. 

We have also expressed the generators in terms of the distance between the left and right boundaries and their time derivatives. In this way we obtain  expressions that depend purely on the boundary and its dynamics. (Of course, 
this is related to matter via the constraints.) These 
formulas also allow us to express the generators in terms of correlation functions of operators, after making some 
extra assumptions.  These assumptions are necessary to relate correlation functions to distances between the boundaries, even in the presence of matter. We have assumed that we have a number of fields scaling like $N$ and that their
interactions with the matter we are probing goes as $1/N$. These conditions are met in the SYK model. 
 
In the semiclassical limit, where $\Delta S = \SemFactor \gg 1$, we can further simplify the expressions for the generators and write them in terms of products of operators on the two boundaries. These are similar to the operators 
that appear in the traversable wormhole discussion \cite{Gao:2016bin}. This is not a coincidence, it is because these
operators generate bulk matter displacements across the horizon. These displacements also play a role in the growth of chaos or in out of time order correlators. 
We have also shown that these operators approximately 
act like SL(2) transformations on the boundary time $u$, so that they can be
viewed as conformal\footnote{We are talking about three generators, and not the infinite dimensional group of reparametrizations.} transformations of the boundary theory. This also allows us to define an operator-state mapping. 
It relates operators inserted during Euclidean evolution by $\beta/2$ to states on the wormhole. These approximate symmetry generators depend explicitly on the temperature of the background wormhole we are expanding around. 

We should emphasize that our discussion is valid both for a nearly-AdS$_2$ gravity theory with a large number of fields 
or the SYK model. In both cases, we can repeat all the steps of this construction. And the discussion is valid in the 
large $N$ limit with finite $N/(\beta J)$. This means that scrambling effects are included, but not effects due to the finite distance to the boundary (finite $\beta J$), or topology changing finite $N$ effects. 
As an example, we have computed the growth of the charges before and  beyond the scrambling time in section \ref{InsertEarly} and found that beyond the scrambling time the momentum or energy of an excitation does not continue to grow. This saturation is related to the fact that out of time order correlators decay to zero. 
In fact, when we evaluate the charges on a state that is created by inserting operators during euclidean evolution, we are computing the same type of correlation functions that appear in out of time order correlators. These out of time order correlators have some pieces that display an exponential growth related to chaos. It is also possible to define the approximate generators in such a way that they depend only on these growing pieces. Of course, this is related to the well known fact that such pieces are related to shock wave scattering, which in this two dimensional context, generate simple bulk displacements \cite{Shenker:2013pqa,Kitaev:2014t1}. Here we are inverting the logic and 
saying that we can use these growing correlators to define symmetry generators. 
It is tempting to say that in any system with maximal chaos we can define translation generators that, together with
the boost generator obey an approximately Poincare algebra near the horizon. For near extremal black holes they obey 
an SL(2) algebra, which is a bit more constraining. The difference is just whether $[E,P] $ is zero or $[E,P] \propto B$. 
Even in the SL(2) case, if we consider excitations very close to the horizon, at distances smaller than the radius of
AdS$_2$, we find that $E$ and $P$ become relatively large, so that after a rescaling we obtain a Poincare looking algebra. It would be nice to understand the conditions under which this structure emerges in a general maximally chaotic large $N$ system.

The approximate generators are related to the ``size'' operator that has been studied recently \cite{Susskind:2018tei,Roberts:2018mnp,Brown:2018kvn,Qi:2018bje}. This connection to a symmetry algebra explains many of its features and helps to clarify  its relation to bulk dynamics in AdS$_2$. A particularly simple relation is that the time derivative of size is the same as the bulk momentum. This is explored further in \cite{Susskindpaper}.  
In \cite{Qi:2018bje} a direct SYK analysis at large $q$ and large $N$ showed how size evolved up to the scrambling time. 
Presumably, similar methods could explain why it saturates after the scrambling time.  In other words, one would like to have a better microscopic picture for these generators in the SYK model, one which does not go through the usual route of the $G,\Sigma$ action but rather constructs them more manifestly in terms of the UV operators of the model, 
as in  \cite{Qi:2018bje}.

There are some straightforward looking generalizations of this discussion, for example one could consider a supersymmetric system and a super-Schwarzian. 

%It would be interesting to generalize our construction to the Super-Schwarzian. The goal would be to construct gauge-invariant generators of the supersymmetry algebra $OSp(1|2)$.

It would be interesting to understand how this story is modified when the length of the throat is not infinite (or 
  $\beta J$ is finite) and eventually covering the case 
 of a generic finite temperature black hole. We know that the chaos correlators are also given in terms of simple displacements in this case too. So we expect that they should also be useful for constructing the local 2d Poincare symmetry near any horizon (the two dimensions are time and the radial direction). 
 We have also recalled the general expectation that the inner horizon would have some kind of singularity. It would be interesting to understand it better, and it is likely that these finite throat length corrections are relevant.

\section*{Acknowledgements}
We thank the anonymous referee, D. Harlow, L. Iliesiu, D. Jafferis, 
 A. Lewkowycz,  X.L. Qi, S. Shenker, D. Stanford, A. Streicher, L. Susskind, Z. Yang, and W. Zhao for discussions.  

H.L. is supported by an NDSEG fellowship.
J.M. is supported in part by U.S. Department of Energy grant
de-sc0009988 and by the Simons Foundation grant 385600. Y.Z. is supported by the Simons foundation through the It from Qubit collaboration.
 We benefited from the workshop ``Chaos and Order: From
strongly correlated systems to black holes" at KITP, supported in part by the National
Science Foundation under Grant No. NSF PHY-1748958 to the KITP.

\appendix
\section{SO(3) analogy}
\la{appendixSO3}
Consider a non-relativistic particle in Euclidean space. We have the position operators and the angular momenta:
\begin{equation}
%<3.4>
[J_i,J_j]=i\varepsilon_{ijk}J_k,\qquad [J_i,X_j]={i}
\varepsilon_{ijk}X_k,\qquad [X_i,X_j]=0,\qquad i,\,j,\,k=1,\,2,\,3.
\end{equation}
This is the Euclidean algebra $E(3)$. (Note that what is normally the momentum generator, $P^i$, is here the position operator). This algebra has two Casimirs,
\begin{equation}
%<3.5>
{ X}^2=r^2,\qquad {J}.{X}=\lambda.
\end{equation}
Our toy model consists of two non-relativistic particles, which we dub ``left'' and ``right,'' each constrained to live on the surface of the sphere $r^2 = 1$. We assume for simplicity that our non-relativistic particles do not carry intrinsic spin, so $\lambda = 0$. In addition, there is some ``matter'' which can carry angular momentum. This matter could for instance be some spinning particle, which lives at the center of the sphere. (Later we will see that the matter has to have integer spin.) Finally, we demand that the overall system has vanishing angular momentum $J_l + J_r + J_m = 0$; e.g., the overall SO(3) symmetry is gauged. Thus the Hilbert space of our system may be denoted
\eqn{\mathcal{H} = (\mathcal{H}_l \otimes \mathcal{H}_r \otimes \mathcal{H}_m)/ SO(3).}
This also means that all physical operators $O$ must commute with the total angular momentum, $[J_l^a + J_r^a + J_m^a, O] = 0$.

\subsection{Exact SO(3) algebra from two copies}
%From now on we set $\zeta=0$ and $r=1$. Now consider two copies of the system. We then have, in addition to the usual casimirs, 
Armed with this Hilbert space, we may consider the physical operators
\eqn{X_l . J_r, \quad X_r . J_l, \quad (X_l \times X_r) . \lp J_l + J_r\rp. }
Note that we can $\lambda = 0 $ and the gauge constraint to rewrite these as
\eqn{G^1 = -X_l . J_m, \quad G^2 \sim -X_r . J_m, \quad G^3  \sim -\lp X_l \times X_r\rp . J_m. \la{gop} } 
This suggests that the 3 operators form an SO(3) algebra. (More precisely, we should replace $X_r$ with the Gram-Schmidt linear combination of $X_l$ and $X_r$ that is orthonormal to $X_l$.) 
We can check this by computing their commutators directly in the 2-body Hilbert space without the gauge constraint. To do so, it is worth introducing a little bit of notation.
%For example,
%\eqn{[X^l_i J^r_i, X^r_k J^l_k] &\\
%				&= X^l_i [J^r_i, X^r_k ]J^l_k +X^r_k[X^l_i, J^l_k]J^r_i \\
%				&= i  X^l_i \epsilon_{ikl} X^r_l J^l_k+ i X^r_k \epsilon_{ikl} X^l_l J^r_i\\
%				&= -i X_l .  (X^r \times J^l) + i J^r . (X_r \times X_l) \\
				%&= - i (X_l \times X^r) . J^l - i J^r . (X_l \times X^r) \\ 
%				&= - i \lp X_l \times X_r\rp  (J_l + J_r) }
%So we have that $[G^1, G^2] = i G^3$. 
%				If we label for example $G^1 = X_l . J_m$, etc., this equations tells us that  $[G^1, G^2] = i G_3$.
%				Now we may compute $[G^2, G^3]$. %This seems a bit more challenging.
				%\eqn{[ J_L J_R ] }
				%Thus, it turns out that in the representation defined by (3.9) the twist
%$\zeta $ can be only integer or half integer.  
%{\bf Theorem}: Let 
Let
$V$ and $W$ transform be vector operators under SO(3) such that
\eqn{
&[V_i,V_j] = [W_i, W_j] = [V_i, W_j] =0\\
&V . V = W. W =1,\\
&V. W = 0.
}
Then defining $e^1 = V, e^2 = W, e^3 = V \times W$, we may write $G^A = - e^A_i J_i$ where $J = J_l + J_r$. Here the capital indices are physical, and the lower indices are the gauge index. 
%&[G^A, G^B] = i \epsilon^{ABC} G_C.}
%Note we have the relations $[V^A_i, V^B_j] = 0$ and 
\eqn{[G^A, G^B] &= [e^A_i J_i, e^B_j J_j]\\
		&= i \epsilon_{ijk} \lp e_i^A e_k^B J_i + e_j^B e_i^A J_k + e_j^B e_k^A J_i \rp \\
		&= - i (e^A \times e^B) . J \\
		&= i \epsilon^{ABC} G_C,}
		where in the last line, we used $e^A \times e^B = \epsilon^{ABC} e_C$. Note that in order to construct this algebra, it was crucial that we had other vector operators that were not just the angular momentum operators. The left and right Hilbert spaces were representations of the Euclidean algebra, and not just the SO(3) algebra, so we are necessarily considering an infinite dimensional Hilbert space.

It is also interesting to construct the Casimir operator
\eqn{C = G^A G^A = e^A_i J_i e^A_k J_k = \delta_{ik} J_i J_k = J^2.}
So we see that the Casimir of the gauge charges is equal to the Casimir of these physical charges.

Note that evolving with one of the $G^A$ leads the $X_l$ and $X_r$ vectors to precess about some axis. Alternatively, if we replace $J_l + J_r$ with $-J_m$, we can view the positions of the particles as fixed under time evolution, but the matter rotates. The invariant statement is that the generators move the left and right particles with respect to the matter.
%\subsection{Dynamics}
%We now consider what happens if we evolve a state by $e^{iGu}$. The equations of motion are
%\eqn{
%\dot{X}_l &= 0,\\
%\dot{X}_r &= X_l \times X_r,\\
%\dot{J}_l &= X_l \times J_r \\
%\dot{J}_r &= -X_l \times J_r.
%}
%The classical solution is $X_l$ constant, with $X_r$ precessing around $X_l$. The total angular momentum $J_l + J_r$ is conserved, as it should be.
%A similar story happens if we set $H = G^2$.
%For $G^3$, let $\vec{N} = X_l \times X_r$. We can check that $[\vec{N},G^3] = 0$, so $\vec{N}$ is constant in time. Then $X_l$ and $X^r$ both precess about $\vec{N}$. To say things more symmetrically, let $N^1 = G_L$, $N^2 =X_r$, and $N^3 = X_l \times X_r$. Then in all cases, both $X_l$ and $X_r$ precess around $N$.
%In addition, we must impose the equations of motion
%\eqn{J_l + J_r + J_m = 0.}

%Note that in the absence of matter time evolution is a pure gauge transformation. In other words, the state at time $u$ is always related to the state at time 0 by a gauge transformation. This is expected: the quantum mechanical operator is exactly 0 on the Hilbert space $Q_l + Q_r = 0$. The dynamics with matter are exactly the same, but the resulting time evolution is no longer a pure gauge transformation since the boundary particles evolve with respect to the matter.

%Finally, note that if we set $X . J = q$ for $q \ne 0$ and used equation (\ref{gop}) as our definition for $G^a$, nothing would really change.

%Almost every aspect of this discussion carries over to the AdS$_2$ case.
\subsection{Action on states/operators}
We would now like to use the physical algebra to organize states and operators in this theory. Let us imagine that the matter sector contains a vector operator $W^a$ which is not $J^a$. For example, if the matter is actually another particle, $W^a$ could be its position vector $Y^a$. 
We may form the gauge invariant operator $W^A = W^a e^A_a.$ Then $[W^A,G^B] = i\epsilon^{ABC} W_C$. %For example, $W^a$ could simply be $X_l^a$. Or it could be some vector operators in the ``matter'' system. 
Furthermore, we can use $W^A$ to construct tensor operators. So we can generate operators which transform in any integer spin representation of the physical $SO(3)$.

Now consider a state $\ket{0}$ that is in a singlet under the gauge SO(3) and the physical SO(3)
\eqn{\ket{0} = \sum_{m,j} \psi(j) (-1)^m \ket{j,m}_l \otimes \ket{j,-m}_r \otimes \ket{0,0}_\text{matter}.}
Here $\psi(j)$ can be any normalizable function. For example, the thermofield double corresponds to $\psi \sim e^{-\beta j^2/4}$.
We may generate states which transform under integer spin by repeatedly applying $W^A$, e.g., $W^A W^B W^C \ket{0}$. We can then organize these into irreps $\ket{j,m}$ of the physical algebra.

%Can we find an operator which transforms like a vector? An obvious candidate would be $W^a$, for example, $W^a = X^a_L$, and construct the state $W^a \ket{0}$. Here $\ket{0}$ denotes any state that is in the singlet rep, e.g.,
%It could be the thermofield double. But clearly $W^a \ket{0}$ is not gauge invariant! We need to instead construct states like 
%\eqn{W^A = W^a e^A_a.}
%Then assuming that $[W^b,e^{A}_a] = 0$, we will have $[W^A,G^B] = i\epsilon^{ABC} W_C$. Notice that there are not many operators which actually satisfy this condition. For example, choosing $W^A = \dot{X}^A$ does not work, because $X$ does not commute with $W$. However, once we find a $W$, we can generate more by evolving $W^A$ by one of the symmetry generators. 

%Now in general, we have states $\ket{j,m}$ that transform under the physical algebra. For example the $\ket{1,m}$ states correspond to things like $X_l^A \ket{0,0}$. We can get higher $j$ states by considering products like $X_l^A X_l^B X_r^C \ket{0,0}$. 

One interesting question is: what matter states are allowed in this gauge theory? Since there are no half-integer states in the two-particle Hilbert space, we conclude that the matter is restricted to be integer spin. For example, the matter cannot be a spin $1/2$ qubit.

\section{Canonical quantization of Schwarzian theory}
\la{appendixCan}
\def\xz{\mathcal{X}^0}
\def\xx{\mathcal{X}^+}
\def\xxb{\mathcal{X}^-}
\def\yy{\mathcal{Y}^+}
\def\yyb{\mathcal{Y}^-}
\def\qq{\mathcal{Q}}
The purpose of this section is to show that the most naive quantization of the Schwarzian theory explicitly realizes the commutation relations needed in our construction of the charges. We will quantize using global time coordinates $T_r(u)$, although one could also use Rindler coordinates related by $\tan (T_r/2) = \tanh (t_r/2)$.
{\bf Friendly warning}: when using global time, we adopt different embedding space conventions from the rest of the paper. Defining $\mathcal{X}^0 = X^1$ and $\mathcal{X}^\pm = X^{-1} \pm i X^0$,  we may write
\eqn{\la{xvec}X_r = (\mathcal{X}^0_r,\xx_r, \xxb_r) &=\lp {1 \over T'_r }, -{e^{iT_r} \over T'_r}, -{e^{-iT_r} \over T'_r} \rp,\\
	X_l &=\lp - {1 \over T'_l }, -{e^{iT_l} \over T'_l}, -{e^{-iT_l} \over T'_l} \rp.
		}
		The metric in these coordinates is $X . Y = \mathcal{X}^0 \mathcal{Y}^0 - \hf \lp \xx \yyb + \xxb \yy \rp$. 
		The advantage of this convention is that the components of $X$ are simple when written in terms of global time, and furthermore \nref{poincare} becomes $[\qq_m,\qq_n] = (m-n) \qq_{m+n}$, where $m,n$ take values from $\{0,+,-\}$. This convention is perhaps more familiar from $d \ge 2$ CFT, where $L_0$ generates dilation, or global time translation along the cylinder.

Let's proceed with the canonical quantization. We start with the Lagrangian on one side, setting $T= T_r(u_r)$:
	\eqn{L = - \{\tan T/2, u\} = \hf {T''^2 \over T'^2} - \hf T'^2 - \lp T'' \over T' \rp'.}
	Dropping the overall derivative, we search for a 4-dimensional phase space with 2 canonical coordinates $T, T'$. The canonical momentum are $p_i = \sum^2_{k=i} (-\pd_u)^{k-i} \pd L/ \pd   T^{(k)} $:
	\eqn{p_1 = -T' + {T''^2 \over T'^3}-{T''' \over T'^2}, \quad p_2 = {T'' \over T'^2}.}

	Note that $p_1$ is just the global $T$ gauge charge $p_1 = \qq^0$. Then the Hamiltonian is
	\eqn{H = \sum p\dot{q} - L = -\hf T'^2 + {3 T''^2 \over 2 T'^2}-{T''' \over T'} = -\{\tan T/2, u\}.}
	We then impose canonical commutation relations
	\eqn{[T,T']=0, \quad [T,p_1] = i,\quad [T',p_2] =i.}
	We would like to check that the gauge charges satisfy an \slt algebra. The other two charges are
	\eqn{\qq^\pm &= e^{\pm iT} \lp \pm i {T'' \over T'}+ {T''^2 \over T'^3} - {T''' \over T'^2}\rp\\
%	\qqb &= e^{-iT} \lp -i {T'' \over T'}+ {T''^2 \over T'^3} - {T''' \over T'^2}\rp.
	}
	Now to compute these commutators, we first rewrite these charges in terms of coordinates and momenta by eliminating $T''', T''$ in favor of $p_1, p_2$. %The correct procedure is to eliminate $T'''$ in favor of $p_1$, and then keep $q_2 = T'$ and $q = t$: 
	\eqn{\qq^\pm = e^{\pm iT} \lp  T' (1 \pm i p_2) + p_1 \rp  ~,~~~~~{\rm and }~~~~~~ \qq^0 = p_1 .\la{PQCharges}}
	In writing these expressions, we have chosen an operator ordering where all the momenta are to the right of the coordinates. 
	Note that these charges generate the expected infinitesimal transformations on the times and satisfy the \sltg algebra:
	\begin{align}\la{Taction}
	 &i[\qq^0,T] = 1, & &i[\qq^0, T'] = 0,\nonumber \\
	 &i[\qq^\pm,T ] = e^{\pm i T},& &i[\qq^\pm, T'] = \pm i e^{\pm i T} T',\\
	 &[\qq^0, \qq^\pm] = \pm  \qq^\pm, & &[\qq^+,\qq^-] = - 2 \qq^0.\nonumber
	\end{align}
	%	This rather simple expression has a nice interpretation if we think of $p_1 = i\pa{}{T}$ and $p_2 = i \pa{}{T'}$. Then $\delta T = \epsilon e^{\pm iT}$ and $\delta T' = \pm i \epsilon e^{\pm i T} T'$ which explains the two factors. 
	%Note also that $Q_\pm ^\dagger = Q_\mp$.
%	\HL{There is some problem with the signs of the commutators! }
	With these gauge charges, $H$ is simply the Casimir
	\eqn{H = -\hf Q. Q. }
	One can check that, e.g., the equation of motion correctly links the two coordinates $i[H,T] = T'$.
	
	Now consider the embedding space vectors in \nref{xvec}. We see that the $X$'s are only a function of the coordinates and not the momenta. So we immediately conclude that all components of $X$ commute amongst themselves. We can also check that $[Q^a,X^b] = i \epsilon^{abc} \eta_{cd} X^d$ transforms like a vector. This verifies that the Schwarzian realizes the Poincare algebra postulated in \nref{poincare}.
	%\eqn{[Q_l^0, X_l^i] = \lp 0, { e^{it_L} \over T'_L}, {- e^{it_L} \over T'_L} \rp = 2  \epsilon^{0ij} g_{jk} X_l^k }

	One advantage of using the Hamiltonian formalism is that expressions for operators written in terms of $T,T'$ and $p_1, p_2$ are fairly agnostic about what time-evolution rule is being used. For example, we may evolve by $H_l + H_r$ or the coupled Hamiltonian $H_l + H_r + \eta (X_l . X_r)^{-\Delta} $. This is not the case when we write expressions in the Lagrangian formalism because the expressions for $Q^a$ change (for example, they depend on $\eta$ in \nref{HcoupSch}).

	As another application of the Hamiltonian formalism, we may check that the approximate construction of the charges discussed in section \ref{OtherSemExp} does not lead to a closed algebra. In other words, the failure of the algebra happens not just at the microscopic level (in terms of the fermions) but even in the Schwarzian limit. This is not too surprising because as discussed in \cite{Maldacena:2018lmt} and in section \ref{EvolvChar}, the spectrum of the eternal traverable wormhole (even in the Schwarzian limit) contains a tensor factor which is not \slt invariant.

\section{Evaluating commutators perturbatively}
	One might wonder whether we can check the commutator computation using standard perturbation theory.
	In this section, we show how this may be done. We focus on checking the commutation relations between the gauge charges; the commutation relations between the physical charges $G^A$ can be checked quite analogously.

	We start by reviewing the linearized Schwarzian action \cite{Maldacena:2016upp}. We use the same variables as in Section \ref{SecLinearized}. We write the Schwarzian as	
	\eqn{I = {\SemFactor \over 2} \int d\tilde{u} \lp \epsilon''^2 + \epsilon'^2 \rp = {\SemFactor \over 2} \int d\tilde{u} \lp -r'^2 - r^2 + q'^2 \rp.}
	The first line is the same expression as in \nref{SchLin}, except now in Lorentzian signature.
	We have rewritten this higher derivative action by introducing the fields $q$ and the ghost field $r$ such that $q=-\epsilon''+\epsilon,$ $r=\epsilon''$. From this action we can read off the commutation relations
	\eqn{[q,q'] ={i \over  \SemFactor }, \quad [r,r'] = -{i\over \SemFactor } .}
	Now as in \nref{Chexp}, we expand $Q$ in powers of $\epsilon$, and then substitute for $q,r$. To quadratic order, we find
	\eqn{
Q^{-1} &\simeq \SemFactor \lp -1-q'-(r^2+r'q'+r'^2 )\rp\\
Q^+ &\simeq  \SemFactor  \lp r-r' +  q(r-r') + r(q'-2r') + 2r'(q'+r')+ 2r^2\rp\\
Q^- &\simeq   \SemFactor \lp -r-r' +  q(r+r') + r(q'+2r') + 2r'( q'+r')+2r^2\rp.
		   }
		   Then we can verify that $[Q^{-1}, Q^\pm] = \pm i Q^\pm$ and $[Q^+,Q^-] = 2i Q^{-1}$ in line with \nref{poincare}. To reproduce the linear terms in $Q^\pm$ on the RHS, it is important to expand to quadratic order the charges in the commutator. 

Note that in order to check the algebra at higher orders, one should not only expand $Q$ to higher powers but also modify the commutation relations due to higher order terms in the action. This quickly gets cumbersome; hence the purpose of the previous section.

\section{Spinor description of the boundary variables}
\la{Spinor}

We have described the boundary motion in terms of a vector $X^a$, with $X^2=0$. Alternatively, we
can use a two component spinor   $\lambda_\alpha$  and construct $X^a$ as
\be
X^a = ( \sigma^a)^{\alpha \beta } \lambda_\alpha \lambda_\beta,
\ee
where $\sigma^a$ are Pauli matrices with one index raised, 
$ ( \sigma^a)^{\alpha \beta } =\epsilon^{\alpha \gamma} ( \sigma^a)_\gamma^{~\beta }$, and  
 $\sigma^{-1} = ( \sigma^{-1})_\gamma^{~\beta }= \hat \sigma_3$, $\sigma^{0} = -\hat \sigma_1$ and $\sigma^1 = i \hat \sigma_2 $  where 
 $\hat \sigma_i$ are the three usual Pauli matrices.  
The inner product between two spinors is antisymmetric $\langle \lambda , \chi \rangle = \epsilon^{\alpha \beta} 
\lambda_\alpha \chi_\beta $. This implies that we automatically obey $X^2=0$, since we cannot build any
non zero $SL(2)$ invariant from a single $\lambda$. 
The evolution equation is then 
\be
\dot \lambda = \slashed{Q} \lambda ~,~~~~~~~ \ddot \lambda = (\slashed{Q})^2 \lambda \propto E \lambda 
\ee
where $\slashed{Q} = Q_a \sigma^a$. 
We could normalize the spinor by a condition of the form  $\langle \dot \lambda , \lambda \rangle = $constant so
that $\dot X^2 =-1$, this also ensures $Q.X=1$. This gives a simple expression for the evolution. 
Also, in terms of left and right spinors then the three generators  have expressions proportional to 
\be \la{ChrSla}
\tilde P \propto
 {\langle \lambda_l , \slashed{Q}_m \lambda_r \rangle \over \langle \lambda_l , \lambda_r \rangle }
~,~~~~~~ \tilde E + \tilde B \propto 
{\langle \lambda_l , \slashed{Q}_m \lambda_l \rangle \over \langle \lambda_l , \lambda_r \rangle }
~,~~~~~~ \tilde E - \tilde B \propto 
{\langle \lambda_r , \slashed{Q}_m \lambda_r \rangle \over \langle \lambda_l , \lambda_r \rangle }
\ee

\subsection{Computation of the charges beyond the scrambling time}
\la{ScramDet}

As an application of this formalism, we will consider the setup in section \nref{InsertEarly}. 
We consider a background solution where 
${\slashed Q}_r = \sigma^3/2$ (we have set $\beta = 2\pi$) and 
\be \la{lamrl}
\lambda_r(0) = ( i , i)/\sqrt{2} ~,~~~~~~~~\lambda_l(0) = (1,-1)/\sqrt{2}
\ee
Then, it is easy to find the time evolution, $\lambda_r(t) = e^{ u \slashed{Q}_r } \lambda_r(0) = 
i ( e^{ u/2} , e^{-u/2} )/\sqrt{2}$, for example. 
We want to insert a perturbation. If we inserted the perturbation at zero time, we would expect it to change $Q_r$. 
The charge will be both rotated and rescaled. The rescaling is negligible in the limit $\omega \to 0$, see 
\nref{alphsd}. So the main effect is a small infinitesimal rotation 
\be
\slashed{Q}_r \to \slashed{Q}'_r =  L \slashed{Q}_r L^{-1} ~,~~~~~~ L \sim 1 + \vec \gamma \vec \sigma
\ee
where $\gamma$ is a vector of size of order $\omega$. 
If we insert this same excitation at some early time, then we should conjugate it by the time evolution, 
\be
L(u_0) = e^{ u_0 \slashed{Q}_r } L e^{- u_0 \slashed{Q}_r } \sim 1 +2 \hat \alpha   \sigma_-
~,~~~~~\sigma_- = \left( \begin{array}{cc} 0 & 0 \cr 1 & 0 \end{array} \right)
\ee
where $\hat \alpha \propto \omega e^{- u_0}$ and have kept only the terms that are finite in the limit 
$\omega \to 0$ and $u_0 \to -\infty$ with $\omega e^{ -u_0}$ fixed. 
This implies that 
\be  \la{MatSl}\slashed{Q}_m = -\delta \slashed{Q}_r = \slashed{Q}_r- L(u_0) \slashed{Q}_r L(u_0)^{-1} = - 2 \hat
\alpha [ \sigma_- , \slashed{Q}_r]
 \ee
Now the new value of $\lambda_r$ at the origin is just given by 
\bea  \la{lampr}
\lambda'_r(0) &=& e^{ -u_0 \slashed{Q}_r'} e^{ u_0 \slashed{Q}_r } \lambda_r(0)  = L(u_0) e^{ - u_0 \slashed{Q}_r } L(u_0)^{-1} e^{ u_0 \slashed{Q}_r } \lambda_r(0) =
\cr 
&=& L(u_0) \lambda_r = (1 + 2 \hat \alpha \sigma_-) \lambda_r(0)
\eea
$\lambda_l(0)$ continues to be as in \nref{lamrl}. Inserting 
\nref{MatSl}, 
\nref{lamrl}, 
\nref{lampr} into 
 (\ref{ChrSla}) we get (\ref{NonLinG}).
 We also find that 
 \be
 \langle \lambda_l , \lambda'_r \rangle \propto (1 + \hat \alpha)  \Longrightarrow   (X_l . X_r') \propto 
 (1 + \hat \alpha)^2 
 \ee

\section{ OTOC correlators and the expectation values of charges } 

Section 6.2 of \cite{Maldacena:2016upp}   considered an OTOC correlator of four operators of dimension $\Delta$. 
It is also possible to do the same computation for the case that two operators have dimension $\Delta_1$ and the 
other two have dimension $\Delta_2$. 
Then the same method gives the answer 
\be \la{GenerE}
{ \langle V_1 W_3 V_2 W_4 \rangle \over \langle V_1 V_2 \rangle \langle W_3 W_4 \rangle } 
= { U( 2 \Delta_1 , 1 + 2\Delta_1 - 2\Delta_2 , { 1 \over z } ) \over z^{ 2 \Delta_1 } } = { 1 \over z^{2\Delta_1} \Gamma(2 \Delta_1) } \int_0^{\infty} dt e^{ - t/z} t^{2\Delta_1-1} (1+t)^{- 2 \Delta_2 } 
\ee
Here, $z = \SemFactor e^{\tilde u}/8$.
Despite appearances this expression is symmetric under $\Delta_1 \leftrightarrow \Delta_2$.  We now want to relate this to the computation in section (4.2.2). 
To evaluate the charges it is necessary to evaluate the distance. 

We can extract the distance from \nref{GenerE} by taking the limit 
 $\Delta_2 \to 0$, since 
\be
{ 1 \over (X_3.X_4)^{\Delta_2} } \sim  1 - \Delta_2  \log (X_3.X_4) = 1 - \Delta_2 \ell 
\ee
where $\ell$ is the distance. By going to higher orders in $\Delta_2$ we could get higher moments for the distance. 
 Expanding (\ref{GenerE}) to first order in $\Delta_2$,  
\be \la{Instr} 
\langle \ell \rangle = 2  {1 \over z^{  2\Delta_1 } \Gamma(2 \Delta_1) } \int_0^\infty  dt e^{ -t/z} t^{ 2\Delta_1-1} \log(1 + t) .
\ee
This has the following expansion for small and large $z$ 
 \be
 \langle \ell \rangle \sim 4 \Delta_1 z ~,~~~~z \ll 1 ~,~~~~~~~~~ \langle \ell \rangle \sim  2 \log z ~,~~~~~~~~ z\gg 1
 \ee
 \nref{Instr} is the general answer. But we are interested in the classical limit, which corresponds to the 
case where $C\gg 1$. We are also interested in computing this for a state that has definite energy $\omega$. 
 
From classical computations, we expected the distance to have the form 
 \be \la{Exp}
\langle \ell \rangle = 2 \log[ 1 + 2 \Delta_1 z ] 
\ee
But this answer is supposed to hold for states with definite energy $\tilde E$. One way to get the energy to be definite is to imagine
that   $\Delta_1$ is large, then we could imagine creating states with definite energy, of order $\Delta_1$ times the appropriate
redshift factors, which would lie inside $z$. 
 In this case we can do the integral \nref{Instr} by saddle point. If we ignore the logarithm, the saddle point and the 
 integral around it gives the gamma function.  Again, ignoring the log, the    saddle point is at 
 \be
 { 1 \over z} = { 2 \Delta_1 \over t_s } \rightarrow t_s = 2 \Delta_1 z 
 \ee
 Then we see that we get the expected answer \nref{Exp} by inserting this value of $t_s$ into the log term,  $ \log(1+t)$,
   in the 
 integral \nref{Instr}. 
 
  It is also clear that, to the extent that we can use this saddle point evaluation of this second term, 
 in  \nref{GenerE}, we will get that the distance behaves classically, with the value \nref{Exp}. 
 To check whether the saddle point is valid, we can look at the second derivative of the exponent at the saddle which gives
 \be
  \partial_t^2 ( - t/z + 2\Delta_1 \log t) = - { 2 \Delta_1 \over t_s^2 }   
  \ee
  So we see that the spread around the saddle is $\delta t \sim  t_s/\sqrt{\Delta_1}$ which is always smaller than $t_s$, which justifies the saddle point approximation. 
 The conclusion is that the large $\Delta_1$ limit of the OTOC computation in 
 \cite{Maldacena:2016upp} reproduces the expression for the distance used in previous sections.
%It is likely that we get a similar answer if we fix the energy by doing a Fourier transform over the positions and 
%we choose $\omega$ to be sufficiently larger than the temperature.  

\section{Stiffer traversable wormholes}
\la{heavyTW}
In this section, we consider other alternative to the global energy generator $\tilde{E}$.

In the Maldacena-Qi wormhole, the spectrum at low energies includes both \slt excitations and excitations of the boundary. Both the energies are $\sim T'$. We would like to modify the eternal wormhole so that the boundary excitations become very heavy. Then we can see the \slt spectrum just by going to low energies.

In the Maldacena-Qi eternal wormhole, one can determine the frequency of oscillation of the boundary graviton by writing down an effective action for the variable $\phi = \log T'$ in the Schwarzian approximation. Assuming no bulk matter, the gauge constraints simplify the action so that $\phi$ becomes the coordinate of a non-relativistic particle in a potential $V(\phi)$: 
\eqn{V(\phi) = e^{2 \phi} - \sum_\Delta \eta_\Delta e^{2 \Delta \phi}.}
The first term is from the Schwarzian, the second term is from the Maldacena-Qi interaction. We are considering a slight generalization of their interaction that could arise, if there are matter fields with different dimensions in the bulk, or by imposing more complicated boundary conditions on the matter fields.
For the SYK model, we could generate this interaction by not just adding a term $\sim i \eta \lp \sum_i \psi^i_l \psi^i_r\rp$, but also adding an interaction $\sim {\rho \over N} \lp \sum_i \psi^i_l \psi^i_r \rp^2 $.
We want to know if the frequency can be made large by a judicious choice of $\eta_\Delta$. Note that the validity of this effective potential requires $\eta_\Delta <\ll 1$ and $\phi_\text{min} \ll 0$. 
Let us first consider a minimal extension:
\eqn{V = e^{2 \phi} - \eta e^{2 \Delta \phi} + \rho e^{2 \tilde{\Delta} \phi}.}
To simplify the algebra a little, we start with $\tilde{\Delta} = 2\Delta$. Now let us ignore the first term (from the Schwarzian) and see if we can find a solution where this is a valid approximation. We have
\eqn{\phi_m =  {1 \over 2\Delta} \log (\eta/2\rho), \;\;\;\;\; T' = \lp{\eta \over 2 \rho}\rp^{1\over 2 \Delta}, \;\;\;\;\;  \lp { \omega \over T'}\rp^2 =  2^{\frac{1}{\Delta }} \Delta ^2 \rho  \left(\frac{\rho}{\eta }\right)^{\frac{1}{\Delta }-2}.}
For some fixed value $0<\Delta<1/2$, taking $\eta/\rho \ll 1$ while keeping $\rho$ small and fixed, we can make $\phi_m \ll 0$ and make the frequency very large. In order for it to be a good approximation to drop the first term, we need 
\eqn{2^{-1/\Delta } \left(\frac{\eta }{\rho }\right)^{1/\Delta } \ll \frac{\eta ^2}{2 \rho }}
Thus in the limit where $\eta/\rho \ll 1$ and $\rho$ is small and held fixed, this condition will be satisfied for $0<\Delta< 1/2$.
In the SYK model, $\tilde{\Delta} = 2/q$ and $\Delta = 1/q$.

Another question we can ask is: how many bound $N_b$ states exist for this spectrum? We can compute this in the WKB approximation, with $p = 2N \dot{\phi}= 2N \sqrt{V}$. 
\eqn{N_b &= {1 \over 2\pi} \oint p \, dq
 = {4 N \over 2\pi} \int_{-\infty}^{\phi_m} d\phi \, \sqrt{\eta e^{2 \Delta \phi} - \rho e^{4 \Delta \phi} }\\
&={2 N \over \pi}  \lp \eta^2 \over 2 \rho \rp^{1/2}  \int_{-\infty}^{0} d\phi \, \sqrt{ e^{2 \Delta \phi} - {1 \over 2} e^{4 \Delta \phi} }\\
&=\lp  N \rho^{1/2} \over \pi \rp  \lp \eta \over  \rho \rp \frac{2+\pi }{4  \Delta }
  }
 So we see that the number of states is large in the classical limit, although it is suppressed now by $\eta/\rho^{1/2}$. So we need this quantity to be small but not too small if we want the classical analysis to be correct.
 
 Now if we add some matter in the bulk with energy $E_\text{bulk}$, then should set the gauge charges of the Schwarzian mode $Q_0 = E_\text{bulk}.$ This changes the effective potential by an amount
 \eqn{V \to V + E_\text{bulk} e^{\phi}/N.}
  Notice that for $\Delta < 1/4$, there is a significant difference between our model and the Malda+Qi wormhole. In particular, suppose we add $N$ particles of energy $\Delta$. In the regime of interest, neither $T'$ nor the frequency will change appreciably! This is completely different from the Maldacena-Qi wormhole, where the $e^\phi$ term will typically dominate over the $e^{2 \phi}$ term in the potential and change both $T'$ and $\omega$.

It is interesting to compute the ground state energy in the presence of matter $q_0$: is the non-linear term in $q_0$ suppressed? The first order correction is $T' E_\text{bulk}/N$. The second order correction is obtained by correcting $\delta \phi = -T'_* E/(N V'')$. Defining the frequency $\omega^2 = V''_*/T'^2$:
\eqn{N \delta V  = ET'_*   - {1 \over 2N} {E^2 \over \omega^2} + O\lp 1\over N^2\rp.  }
This comes from both the contribution of the original potential and the matter piece. We see that when $\omega$ is large, the correction is suppressed.

It would be interesting to analyze this model further, e.g., explore its thermodynamics, study it at higher temperatures in large-$q$ SYK, etc. We leave this to future work.

\section{Measuring distance using a single free field}
\la{justOne}
We suspect that the distance could be measured even if we have a small number of fields in the bulk. As a simple 
example, consider a single {\it free} bosonic field in the bulk with dimension $\Delta$. Then we may measure the distance using considering the operator $d\Delta \sim -\log \phi_l \phi_r$. This operator can be defined by the replica trick, e.g., we consider higher-pt correlation functions and then continue to $m=0$. 
%$d \sim - \lim_{m \to 0} \lp {\lp \phi_l \phi_r\rp^m - 1 \over m}\rp$, 

\begin{figure}[h!]
\begin{center}
\includegraphics[scale = .6, trim = 0pt 0pt 0pt 0pt]{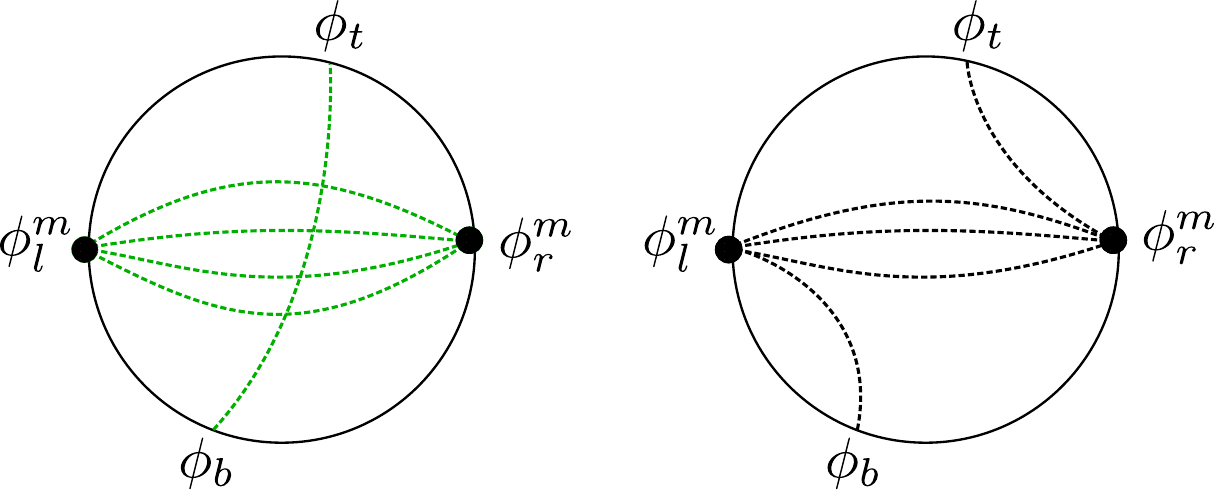}
\caption{
We consider the diagrams contributing to the correlator $\bra{\phi(u_t)}\phi_l^m \phi_r^m \ket{\phi(u_b)} $ in \nref{FreeFeyn}. In {\color{dg} green} is the desired contribution which mimics the action of the charges in Figure \ref{Charges}(a). The ``bad'' diagrams (displayed in black) are morally similar to the ones in Figure \ref{Charges}(b). } 
\label{FreeCharges}
\end{center}
\end{figure}

Now we would like to see how well this operator tracks the distance when there are $\phi$ particles around. We can consider the states defined previously in Section \ref{SemSLTSym},
	\eqn{&\bra{\phi(u_t)}\phi_l^m \phi_r^m \ket{\phi(u_b)} = {\color{dg} m! G_{lr}^m G_{tb}} +  2m^2(m-1)! G_{lr}^{m-1} \lp G_{lt} G_{rb} + G_{rt}G_{lb} \rp \la{FreeFeyn}. }
	Here the first term comes from the ``{\color{dg} good}'' diagrams in Figure \ref{FreeCharges}; the ``bad'' diagrams ones are contractions where one of the $\phi_l$ contract with the matter created from $\phi_b$ or $\phi_t$. 
 The factor of 2 comes from the choice of connecting one of the $\phi_l$ to $\phi_t$ or to $\phi_b$.
 The operators $\phi^m$ should be regulated, for instance by point-splitting. Analytically continuing, 
	\eqn{ d\Delta &\sim - \lim_{m \to 0}{1 \over G_{tb}} \bra{\phi(u_t)} \lp (\phi_l \phi_r) ^m -1\over m \rp \ket{\phi(u_b)} \\
		       &=  - \log\lp G_{lr} \rp + \gamma - 2{G_{lt} G_{rb} + G_{lb} G_{rt} \over G_{lr} G_{tb}}. \la{dist} }
	     In the limit of large distances (e.g., when we push the bulk fields towards the boundary) or when $\Delta$ is large, the leading contribution is the first term, which is precisely the distance.

As a generalization of this calculation\footnote{We thank Wayne Zhao for discussions about the combinatorics.}, we may consider inserting not just a single particle but several particles, e.g., $\phi_b \to \phi_b^k$, $\phi_t \to \phi_t^k$. The analog of \nref{dist} will involve a sum over the number of bonds broken between the left and right. This will generate terms that depend on $k$. The $k$-dependence of the ``bad'' terms means that there is likely a bound on how many particles we can insert before the operator defined above no longer tracks the distance. This is not too surprising, for example, if we considered some coherent state with a large number of quanta, we expect that the expectation value of $\log \phi_l \phi_r$ is determined by the classical field values, which does not have to track the distance.

The preliminary conclusion is that at least for small number of quanta, the operator $-\log (\phi_l \phi_r)$ seems to be a good proxy for the distance between the two sides. We leave to future work a more thorough understanding of the limitations of this distance operator, the effects of interactions, etc.

\mciteSetMidEndSepPunct{}{\ifmciteBstWouldAddEndPunct.\else\fi}{\relax}
\bibliographystyle{utphys}
\bibliography{BigReferencesFile.bib}{}

\end{document}